\documentclass{pasj00}
\usepackage{comment}
\draft
\begin{document}
\SetRunningHead{Author(s) in page-head}{Running Head}
\Received{29-Sep-2014}{}
\Accepted{26-Dec-2014}{}

\title{High Dispersion Spectroscopy of Solar-type Superflare Stars. \\ 
I. Temperature, Surface Gravity, Metallicity, and $v \sin i$}

\author{Yuta \textsc{Notsu}\altaffilmark{1}, Satoshi \textsc{Honda}\altaffilmark{2}, Hiroyuki \textsc{Maehara}\altaffilmark{3,4}, 
Shota \textsc{Notsu}\altaffilmark{1},  Takuya \textsc{Shibayama}\altaffilmark{5}, 
Daisaku \textsc{Nogami}\altaffilmark{1,6}, and Kazunari \textsc{Shibata}\altaffilmark{6}} 
    
\email{ynotsu@kwasan.kyoto-u.ac.jp}    
\affil{\altaffilmark{1}Department of Astronomy, Kyoto University, Kitashirakawa-Oiwake-cho,
Sakyo-ku, Kyoto 606-8502}
\affil{\altaffilmark{2}Center for Astronomy, University of Hyogo, 407-2, Nishigaichi, Sayo-cho, Sayo, Hyogo 679-5313}
\affil{\altaffilmark{3}Kiso Observatory, Institute of Astronomy, School of Science, The University of Tokyo, 10762-30, Mitake, 
Kiso-machi, Kiso-gun, Nagano 397-0101}
\affil{\altaffilmark{4}Okayama Astrophysical Observatory, National Astronomical Observatory of Japan, 
3037-5 Honjo, Kamogata, Asakuchi, Okayama 719-0232}
\affil{\altaffilmark{5}Solar-Terrestrial Environment Laboratory, Nagoya University, Furo-cho, Chikusa-ku, Nagoya, Aichi, 464-8601}
\affil{\altaffilmark{6}Kwasan and Hida Observatories, Kyoto University, Yamashina-ku, Kyoto 607-8471}


\maketitle

\begin{abstract}
We conducted high dispersion spectroscopic observations of 50 superflare stars with Subaru/HDS,
and measured the stellar parameters of them.
These 50 targets were selected from the solar-type  (G-type main sequence) superflare stars 
that we had discovered from the Kepler photometric data.
As a result of these spectroscopic observations, 
we found that more than half (34 stars) of our 50 targets have no evidence of binary system. 
We then estimated effective temperature ($T_{\rm{eff}}$), surface gravity ($\log g$), metallicity ([Fe/H]), 
and projected rotational velocity ($v\sin i$)
of these 34 superflare stars on the basis of our spectroscopic data.
The accuracy of our estimations is higher than that of Kepler Input Catalog (KIC) values, 
and the differences between our values and KIC values ($(\Delta T_{\rm{eff}})_{\rm{rms}} \sim 219$K, $(\Delta \log g)_{\rm{rms}} \sim 0.37$ dex, 
and $(\Delta\rm{[Fe/H]})_{\rm{rms}} \sim 0.46$ dex) are comparable to 
the large uncertainties and systematic differences of KIC values reported by the previous researches. 
We confirmed that the estimated $T_{\rm{eff}}$ and $\log g$ values of the 34 superflare stars are roughly in the range of solar-type stars.
In particular, these parameters and the brightness variation period ($P_{0}$) 
of 9 stars are in the range of ``Sun-like" stars ($5600\leq T_{\rm{eff}}\leq 6000$K, $\log g\geq$4.0, and $P_{0}>$10 days).
Five of the 34 target stars are fast rotators ($v \sin i \geq 10$km s$^{-1}$), while 22 stars have relatively low $v \sin i$ values ($v \sin i<5$km s$^{-1}$). 
These results suggest that stars whose spectroscopic properties similar to the Sun can have superflares, 
and this supports the hypothesis that the Sun might cause a superflare.
\end{abstract}

\bigskip

\section{Introduction}\label{sec:intro}
Flares are energetic explosions in the stellar atmosphere 
and are thought to occur by impulsive releases of 
magnetic energy stored around starspots, like solar flares (e.g., \cite{Shibata2011}).
The total energy released in the largest solar flares is estimated to be of the order of $10^{32}$erg (e.g., \cite{Priest1981}; \cite{Emslie2012}).
Many T Tau stars, RS CVn-type binary stars, and dMe stars have ``superflares," 
which have a total energy of $\sim10^{33-38}$ erg \citep{Schaefer2000}, 
10-$10^6$ times larger than that of the largest solar flares on the Sun.
Such stars generally rotate fast ($P_{\rm{rot}}\sim$a few days and $v \sin i\gtrsim$10km s$^{-1}$)
and magnetic fields of a few kG are considered to be distributed in large regions 
on the stellar surface (\cite{Gershberg2005}; \cite{Shibata1999}, \yearcite{Shibata2002}). 
In contrast, the Sun slowly rotates ($P_{\rm{rot}}\sim$25 days and $v \sin i\sim$2km s$^{-1}$), 
and the mean magnetic field is weak (a few G). 
It had been thought that superflares cannot occur on slowly-rotating G-type main-sequence stars like the Sun 
before our recent discoveries using Kepler described in the following, except for the report by \citet{Schaefer2000}.
\\ \\
\ \ \ \ \ \ \
We recently have analyzed the data obtained with the Kepler space telescope \citep{Koch2010},
and discovered 365 superflare events on 148 solar-type (G-type main-sequence) stars from the data of 83,000 solar-type stars 
for the first 120 days of the Kepler mission \citep{Maehara2012}.
We here define ``$solar$-$type$ $stars$" as stars 
that have a surface temperature of $5100\leq T_{\rm{eff}}\leq 6000$K and a surface gravity of $\log g \geq 4.0$.
Then, we extended the study of \citet{Maehara2012}, and found 1547 superflare events on 279 solar-type stars 
by using Kepler data of a longer period ($\sim$500 days) \citep{Shibayama2013}. 
Kepler is very useful for detecting small increases in the stellar brightness caused by stellar flares, 
because Kepler realized high photometric precision exceeding 0.01\% for moderately bright stars, 
and obtained continuous time-series data of many stars over a long period \citep{Koch2010}.
Kepler data have also been used for stellar flare research on M and K-type stars (\cite{Walkowicz2011}; \cite{Candelaresi2014}) 
and on A and F-type stars \citep{Balona2012}.
\\ \\
\ \ \ \ \ \ \
The analyses of Kepler data enabled us to discuss statistical properties of superflares 
since a large number of flare events were discovered.
Before the discoveries by Kepler, only 9 flare candidates of superflares on solar-type (G-type main sequence) stars were 
reported by \citet{Schaefer2000}\footnote{Some of these 9 events seem doubtful as summarized in Footnote 1 of \citet{Nogami2014}.},
and the number of discovered events were too few to discuss statistical properties. 
\citet{Shibayama2013} confirmed that the occurrence rate ($dN/dE$) of the superflare versus the flare energy ($E$) shows a power-law distribution
of $dN/dE\propto E^{-\alpha}$, where $\alpha\sim 2$, and that this distribution is roughly similar to that for the solar flare.
\citet{Shibayama2013} also estimated that a superflare with an energy of $10^{34-35}$erg occurs once in 800-5000 years in Sun-like stars. 
``$Sun$-$like$ $stars$" are here defined 
as solar-type stars with an effective temperature of $5600\leq T_{\rm{eff}}\leq 6000$K, a surface gravity of $\log g \geq 4.0$, 
and a rotation period ($P$) longer than 10 days.
\\ \\
\ \ \ \ \ \ \
Many of superflare stars show quasi-periodic brightness variations with a typical period of from one day to a few tens of days.
The amplitude of these brightness variations is in the range of 0.1-10\% \citep{Maehara2012}, 
and is much more larger than that of solar brightness variation (0.01-0.1\%; e.g., \cite{Lanza2003}) caused by the existence of sunspots on the rotating solar surface.
\citet{YNotsu2013} showed that these brightness variations of superflare stars can be well explained 
by the rotation of a star with fairly large starspots, taking into account the effects of the inclination angle and the spot latitude.
\citet{YNotsu2013} also clarified that the superflare energy is related to the total coverage of the starspot, 
and that the superflare energy can be explained by the magnetic energy stored around these large starspots.
In addition, \citet{Shibata2013} suggested, on the basis of theoretical estimates, that the Sun can generate large magnetic flux that is sufficient for causing
superflares with an energy of $10^{34}$ erg within one solar cycle ($\sim$11yr).
The results of the superflare researches are becoming extremely important in many fields, 
for example, magnetic activity research in solar/stellar physics (e.g., \cite{Aulanier2013}; \cite{Shibata2013}; \cite{Candelaresi2014}) 
and planetary habitability in astrobiology (e.g., \cite{Segura2010}).
\\ \\
\ \ \ \ \ \ \
The results described above are, however, only based on Kepler monochromatic photometric data. 
We need to spectroscopically investigate whether these brightness variations are explained by the rotation, and whether superflare stars have large starspots.
The stellar parameters and the binarity of the superflare stars are also needed to be investigated by spectroscopic observations,
in order to discuss whether the Sun can really generate superflares.
We have already reported the first results of our spectroscopic observations of three superflare stars 
in \citet{SNotsu2013} and \citet{Nogami2014}.
\citet{SNotsu2013} confirmed that one superflare star KIC6934317 is a G-type main-sequence star.
We investigated its chromospheric activity by using Ca II infrared triplet and H$\alpha$ lines, and these lines of this star suggest high chromospheric activity.
We also measured the projected rotational velocity ($v\sin i$) of this star, and confirmed that this star has a small inclination angle by comparing $v\sin i$ with 
the period of the brightness variation.
\citet{Nogami2014} found that spectroscopic properties ($T_{\rm{eff}}$, $\log g$, [Fe/H], and rotational velocity)
of the two superflare stars KIC9766237 and KIC9944137 are very close to those of the Sun.
This supports the hypothesis that the Sun can cause a superflare.
Apart from our previous spectroscopic studies (\cite{SNotsu2013} and \cite{Nogami2014}), 
\citet{Wichmann2014} performed spectroscopic observations of 11 superflare stars, and 
found several stars of them are young, fast-rotating stars where high levels of stellar activity can be expected. 
For the remaining stars, however, they said that they did not find a straightforward explanation for the occurrence of superflares.
\\ \\
\ \ \ \ \ \ \
We have then performed high-dispersion spectroscopy of more superflares stars (50 stars in total).
In this paper, we first describe the details of these observations, and judge whether the target stars are single or binary stars.
We then estimate the effective temperature ($T_{\rm{eff}}$), surface gravity ($\log g$), 
metallicity ([Fe/H]), and projected rotational velocity ($v\sin i$) of our target stars. 
On the basis of stellar parameters derived in this paper (Paper I), 
we investigate whether the quasi-periodic brightness variation observed by Kepler is explained by the rotation, and whether superflare stars have large starspots 
in \authorcite{PaperII} (\yearcite{PaperII}; hereinafter Paper II).
Ca II infrared triplet and H$\alpha$ lines, which we have already used for investigating chromospheric activity of three superflare stars in \citet{SNotsu2013} and \citet{Nogami2014}, 
are also analyzed in Paper II.
\\ \\
\ \ \ \ \ \ \
We describe the selection of the target stars and the details of our observation in Section \ref{sec:tarobs}.
We discuss the binarity of the targets in Section \ref{subsec:binary}.
In Section \ref{subsec:atmos}$\sim$\ref{subsec:AgeR}, 
we then estimate stellar parameters ($T_{\rm{eff}}$, $\log g$, [Fe/H], $v\sin i$, and stellar radius $R_{\rm{s}}$) of the target stars. 
Finally in Section \ref{sec:discussion}, we comment on the binarity and estimated stellar parameters of superflare stars,
and then perform some analyses in order to check 
whether these spectroscopically derived values are good sources to discuss the actual properties of stars in our next papers (e.g., Paper II).
\\
\\
\section{Targets and Observation}\label{sec:tarobs}
\subsection{Target stars}\label{subsec:tarselc}
We selected 50 superflare stars as target stars. 
The names of these 50 stars and their stellar parameters are listed in Table \ref{tab:shibay}.
Lightcurves of these 50 stars are shown in Supplementary Figure 1. 
Forty-six of these 50 stars were solar-type superflare stars reported in \citet{Shibayama2013}. 
The way that we selected these 46 solar-type superflare stars as targets is summarized 
in the following.
First, we listed all the $solar$-$type$ $stars$ ($5100\leq T_{\rm{eff,KIC}}\leq 6000$K and $(\log g)_{\rm{KIC}} \geq 4.0$) 
having superflare events reported in \citet{Shibayama2013}.
They are 279 stars in total.
We then rejected relatively faint stars ($K_{p}>14.5$mag) from the list of the target stars 
since the target stars should be bright enough to conduct Subaru/HDS observations 
with reasonable exposure time.
We finally selected the 46 stars that have 3 following features with higher priority as target stars.
(1) The brightness variation period reported in \citet{Shibayama2013} is relatively long ($P_{\rm{S}}>10$ days). 
(2) The temperature is similar to the Sun ($5600\leq T_{\rm{eff,KIC}}\leq 6000$K). 
(3) The brightness variation amplitude of the star, which is listed in 
Online Table of \citet{Shibayama2013}, is relatively large ($\gtrsim$0.1\%).
We used the criteria (1) and (2), since we decided to choose stars 
that are relatively similar to the Sun, among the superflare stars we had already discovered.
We used the criteria (3) since such stars are expected to have large starspots.
There was no clear order of priority among the criteria (1), (2), and (3).
\\ \\
\ \ \ \ \ \ \
In addition to the above 46 solar-type superflare stars, 
we also observed the other 4 superflare stars (KIC6934317, KIC7420545, KIC8429280, and KIC11560431).
These 4 superflare stars were not included in \citet{Shibayama2013}, 
since they are not classified as solar-type stars on the basis of Kepler Input Catalog (KIC; \cite{Brown2011}).
We newly confirmed that these 4 stars are also superflare stars by using the same flare detection method 
reported in \citet{Maehara2012} and \citet{Shibayama2013}.
KIC6934317 is a G-type sub-giant star ($(\log g)_{\rm{KIC}} \sim 3.8$) on the basis of KIC data, 
and we have already reported the observational results of this star in \citet{SNotsu2013}, as mentioned in Section \ref{sec:intro}.
KIC7420545 is a K-type sub-giant star  ($(\log g)_{\rm{KIC}} \sim 3.8$), and 
KIC8429280 and KIC11560431 are K-type main sequence stars, respectively, on the basis of KIC.
We observed these three stars (KIC7420545, KIC8429280 and KIC11560431) 
since they are relatively bright ($K_{p}\lesssim10$ mag) among the superflare stars that we had discovered by using Kepler data. 
Adding up these 4 stars and the above 46 solar-type superflare stars, we observed 50 superflare stars in total in this observation.
In addition, five of the 50 target stars can be identified 
with the ROSAT X-ray all-sky survey source (\cite{Voges1999}; \cite{Voges2000}). 
We listed these five stars in Table \ref{tab:ROSAT}.
\\ \\
\ \ \ \ \ \ \
\citet{Shibayama2013} and \citet{YNotsu2013} analyzed the Kepler data \footnote{
As discussed in \citet{Garcia2011}, the analyses of Kepler data needs the pre-processing of the data 
in order to correct the instrumental perturbations (e.g., outliers, jumps and drifts).
We then already used the Kepler data detrended by the PDC-MAP pipeline \citep{Stumpe2012} 
in \citet{Shibayama2013} and \citet{YNotsu2013}.
Through this, many errors such as outliers and temperature drifts are expected to be corrected.
In addition to this, we already conducted the following correction in \citet{Shibayama2013} and \citet{YNotsu2013} 
in order to correct the sudden changes (jumps) in the mean values of the lightcurve at the ``gaps" of the data 
(e.g., data gaps between the quarters).
In this correction process, we calculated the mean flux values of each group of continuous data points between the gaps 
in the lightcurve, and adjusted the flux values of each data group so that the above mean value became equal to each other.
We also removed a linear trend of the data in each quarter.
We applied the above pre-processing processes of the Kepler data 
to the data that were newly analyzed in the following of this paper.
} of Quarter 0$\sim$6 ($\sim$500 days), 
and estimated the period of the brightness variation ($P_{\rm{S}}$) 
that are listed in both Table \ref{tab:shibay} and Supplementary Table 1.
As described in Section 2 of \citet{YNotsu2013}, this variation period ($P_{\rm{S}}$) was calculated 
by choosing the peak from the Fourier power spectrum whose amplitude had the highest ratio 
to the red noise spectrum (e.g., \cite{Press1978}; \cite{Vaughan2005}).
The power spectra used for estimating $P_{\rm{S}}$, which are calculated from the Kepler data of Quarter 0$\sim$6 ($\sim$500 days), are shown in Supplementary Figure 1.
In this paper, we newly estimate the brightness variation period ($P_{1}$) by using the Kepler data of Quarter 2$\sim$16 
($\sim$1500 days) \footnote{
The version of Kepler data used for calculating $P_{1}$ are different from those used for $P_{\rm{S}}$ 
in \citet{Shibayama2013} and \citet{YNotsu2013} (cf. Table 1 of \cite{YNotsu2013}).
We used the latest version of each quarter (Q2$\sim$Q16) data. 
Q2$\sim$Q14 data that we used here were opened to the public in Data release 21 \citep{Thompson2013b}, 
Q15 data were in Data release 20 \citep{Thompson2013a},
and Q16 data were in Data release 22 \citep{Thompson2013c}, respectively.
}.
We did not use the Kepler data of the remaining three quarters (Q0, Q1, and Q17) 
since the data period of these three quarter is short ($\lesssim30$ days) 
compared to those of the other 15 quarters ($\sim$90 days) \citep{Thompson2013d},
and we consider that long-term variations ($P\sim$30 days) cannot well be detected by using the data of these three quarters.
The method of estimating this new period value ($P_{1}$) is the same as that we used in \citet{Shibayama2013} and \citet{YNotsu2013}. 
The power spectra used for calculating $P_{1}$ are also shown in Supplementary Figure 1.
We also estimated other period values $P_{2}$ and $P_{3}$, which correspond to the second and third peak of the power spectra used for 
estimating $P_{1}$, respectively. 
The estimated values of $P_{1}$, $P_{2}$ and $P_{3}$ are listed in Supplementary Table 1. 
\\ \\
\ \ \
We compared these period values ($P_{\rm{S}}$, $P_{1}$, $P_{2}$ and $P_{3}$) in Appendix \ref{sec:periodhik} 
(Figure \ref{fig:periodQ17} (a) and Supplementary Table 1), 
and checked the lightcurves and power spectra by eye shown in Supplementary Figure 1.
We then selected the resultant period value $P_{0}$ from $P_{1}$, $P_{2}$ and $P_{3}$.
The selection method consists of the following two steps.
First, we exclude $P_{1}$, $P_{2}$ and $P_{3}$ value with low signal-to-noize ratio (S/N$<$25).
Next, we select one appropriate period value ($P_{0}$) from $P_{1}$, $P_{2}$ and $P_{3}$ with S/N$\geq$25 by checking the lightcurve and power spectrum of each star by eye. 
In this eye-check process, we confirmed whether the periodicity corresponding to the peak having the highest S/N ratio 
in the power spectrum are also clearly seen in the lightcurve.
The $P_{0}$ values are also listed in Supplementary Table 1 and plotted in Figure \ref{fig:periodQ17} (b). 
In many cases, $P_{1}$ is adopted as $P_{0}$.
In the following, we use only $P_{0}$ as the period of the brightness variation. 
It is because $P_{0}$ is estimated from the longer term data ($\sim1500$ days) than $P_{\rm{S}}$ ($\sim500$ days), 
and some irregular trends, such as effects of starspot evolution and disappearance, are expected to be averaged to some extent.
\\ \\
\subsection{Details of Observations and Data Reduction}\label{subsec:obsred}
Our spectroscopic observations were carried out by using the High Dispersion Spectrograph (HDS: \cite{Noguchi2002}) 
attached at the 8.2-m Subaru telescope during the semester S11B (2011 August 3), S12B (2012 August 6, 7, 8 and September 22, 23, 24, 25), and S13A (2013 June 23, 24).
The spectral coverage was about 6100$\sim$8820\AA. 
This range includes the chromospheric-activity sensitive lines of H$\alpha$ 6563\AA~and 
Ca II infrared triplet 8498, 8542, 8662\AA.
The $2\times2$ on-chip binning mode was adopted.
Spectroscopic resolution ($R=\lambda/\Delta\lambda$) of each observation date is as follows; 
$R\sim100,000$ in S11B (slit width of 0".36),  $R\sim51,000$ in S12B (slit width of 0".7), and $R\sim80,000$ in S13A (slit width of 0".45).
We used image slicer $\#2$ 
\citep{Tajitsu2012}\footnote{Information of Image Slicers is summarized at http://www.naoj.org/Observing/Instruments/HDS/is.html} in S13A observation.
Data reduction (bias subtraction, flat fielding, aperture determination, scattered light subtraction, spectral extraction, wavelength calibration, normalization by the continuum, 
and heliocentric radial velocity correction) was conducted using the ECHELLE package of the 
IRAF\footnote{IRAF is distributed by the National Optical Astronomy Observatories,
which is operated by the Association of Universities for Research in Astronomy, Inc., under cooperate agreement with the National Science Foundation.} software.
\\ \\
\ \ \ \ \ \ \
The observation date of each target star, exposure time of each observation, and obtained signal-to-noise ratio  (S/N) are shown in Supplementary Table 2.
We observed 50 superflare stars as mentioned above, and 19 stars were observed multiple times.
In addition to these 50 superflare stars, we observed 10 bright solar-type stars and the Moon 
as references of solar-type (G-type main sequence) stars.
These 10 stars are relatively bright ($V<8.5$ mag) and easily observable during our observation period with Subaru/HDS,
considering the coordinates of them.
Among these 10 comparison stars, 
8 stars (18 Sco, HD163441, HD173071, HIP100963, HIP71813, HIP76114, HIP77718, and HIP78399) 
are reported as ``solar-twin" stars by the previous studies (\cite{King2005}; \cite{TakedaTajitsu2009}; \cite{Datson2012}).
Stellar parameters of these stars are estimated in detail by these previous studies. 
In Appendix \ref{sec:compConf-ape}, 
we compare these parameters of the previous studies with those derived by us in this paper,
and check whether our result of spectroscopic determination of stellar parameters 
is consistent with that of the previous studies, especially for the stars whose spectrum is similar to the solar one.
In addition to the above 8 ``solar-twin" stars,
we observed two bright solar-type stars 59 Vir and 61 Vir.
These two stars are first used for the above comparison process in Appendix \ref{sec:compConf-ape}, 
as the above 8 ``solar-twin" stars are done.
We also use these two stars as comparison stars 
in the discussion of stellar chromospheric activity in Paper II,
as we have already used in \citet{SNotsu2013}.
59 Vir rotates fast and has the strong average magnetic field ($\sim$500 G), 
while 61 Vir rotates slowly and no magnetic field is detected \citep{Anderson2010}.
In addition to the above 10 comparison stars, the Moon was also observed as a comparison star 
whose spectrum is quite similar to the solar one in this observation, 
as done in many previous studies (e.g., \cite{TakedaTajitsu2009}; \cite{Takeda2010}).
\\ \\
\ \ \ \ \ \ \
The Kepler lightcurves obtained around the periods of S11B and S12B observations are shown in Figure \ref{fig:lcS12B} in Appendix \ref{sec:obslc}. 
There are no lightcurve data for S13A observation since Kepler ended its general observation mode on 2013 May \citep{Thompson2013d}.
No flare events were detected during the observation time.
\\ \\
\section{Analyses and Results}\label{sec:ana}
\subsection{Binarity}\label{subsec:binary}
\begin{figure}[htbp]
 \begin{center}
  \FigureFile(70mm,70mm){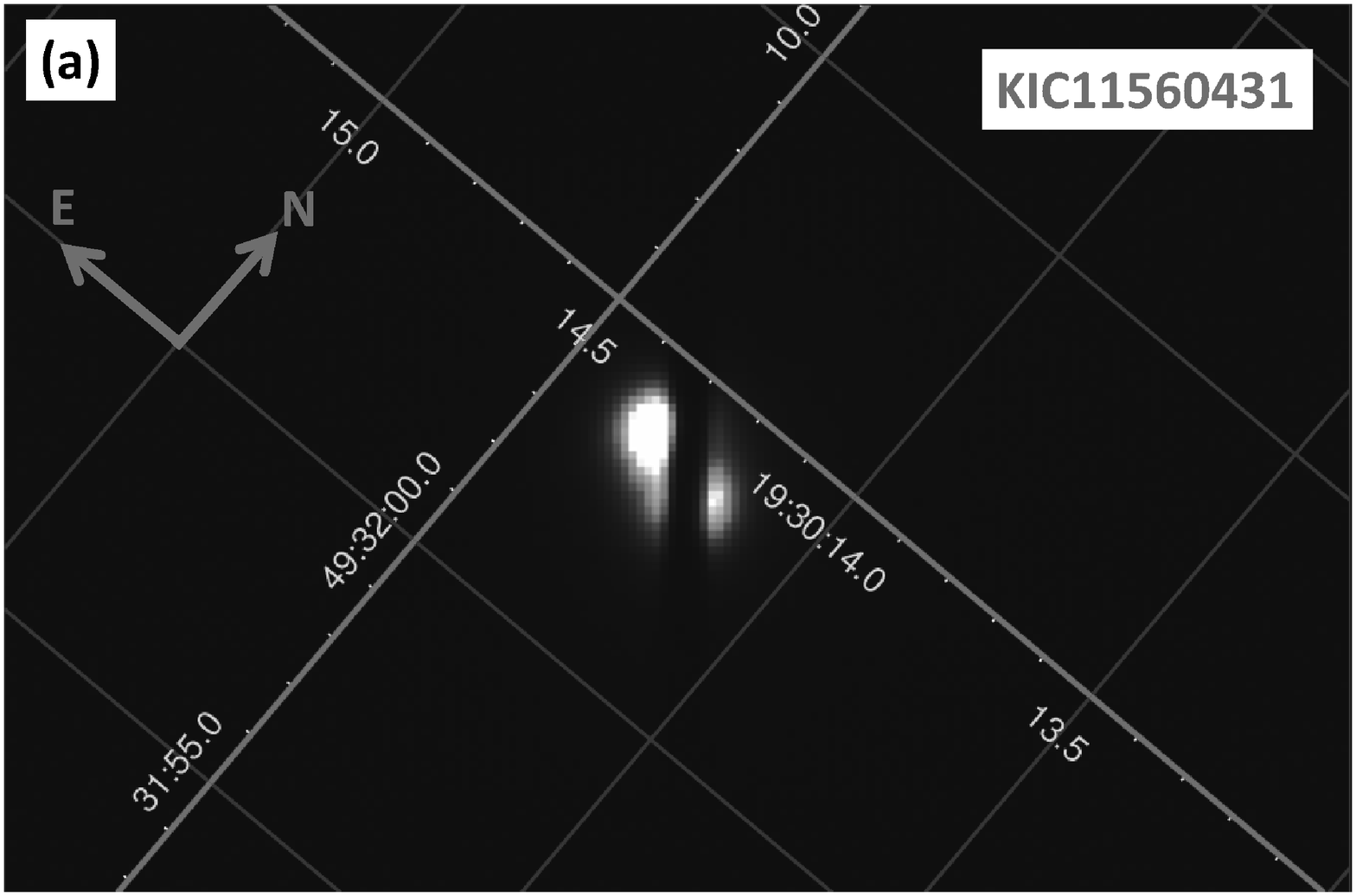}
 \FigureFile(70mm,70mm){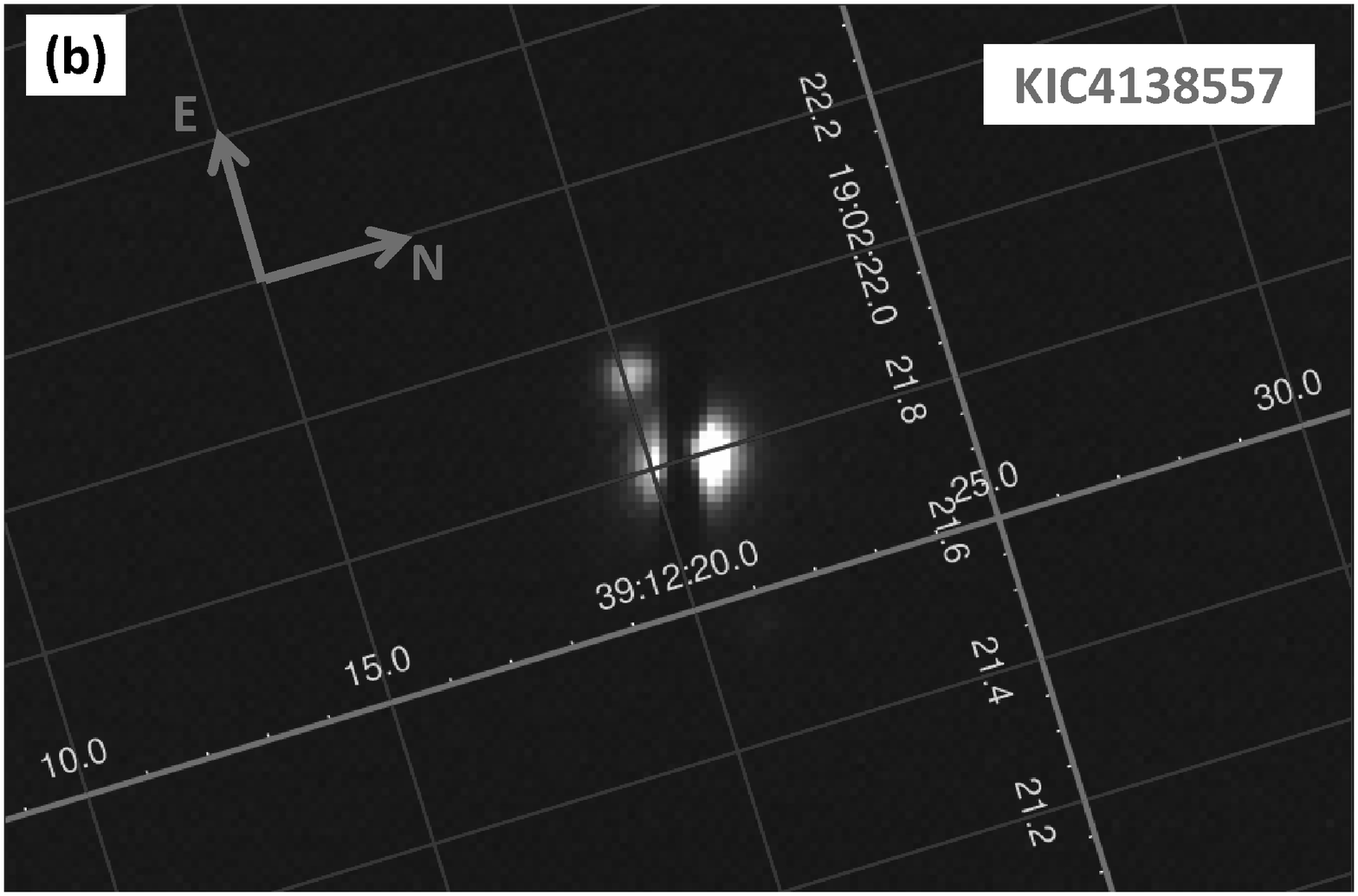}    
  \FigureFile(70mm,70mm){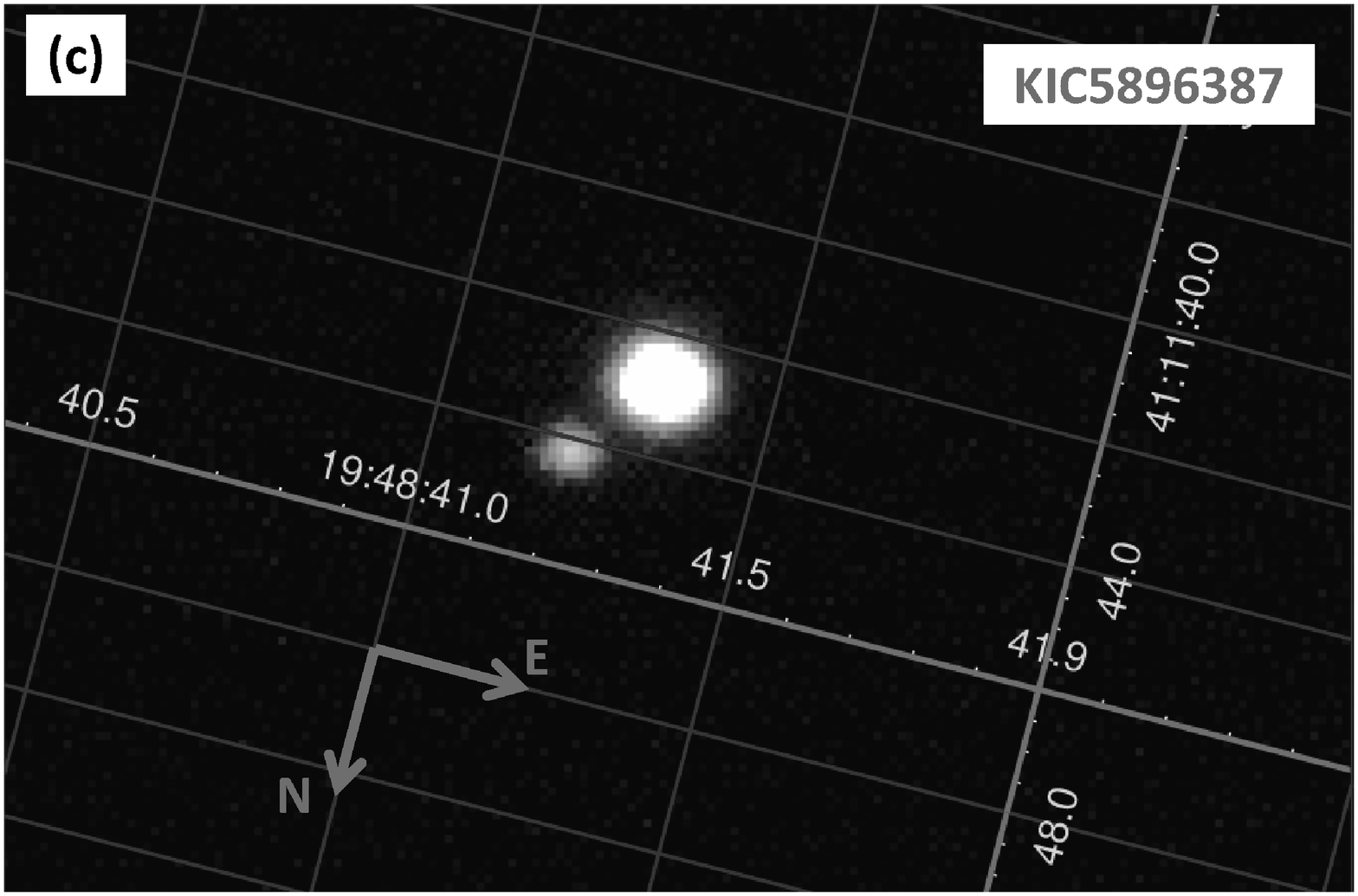}
  \FigureFile(70mm,70mm){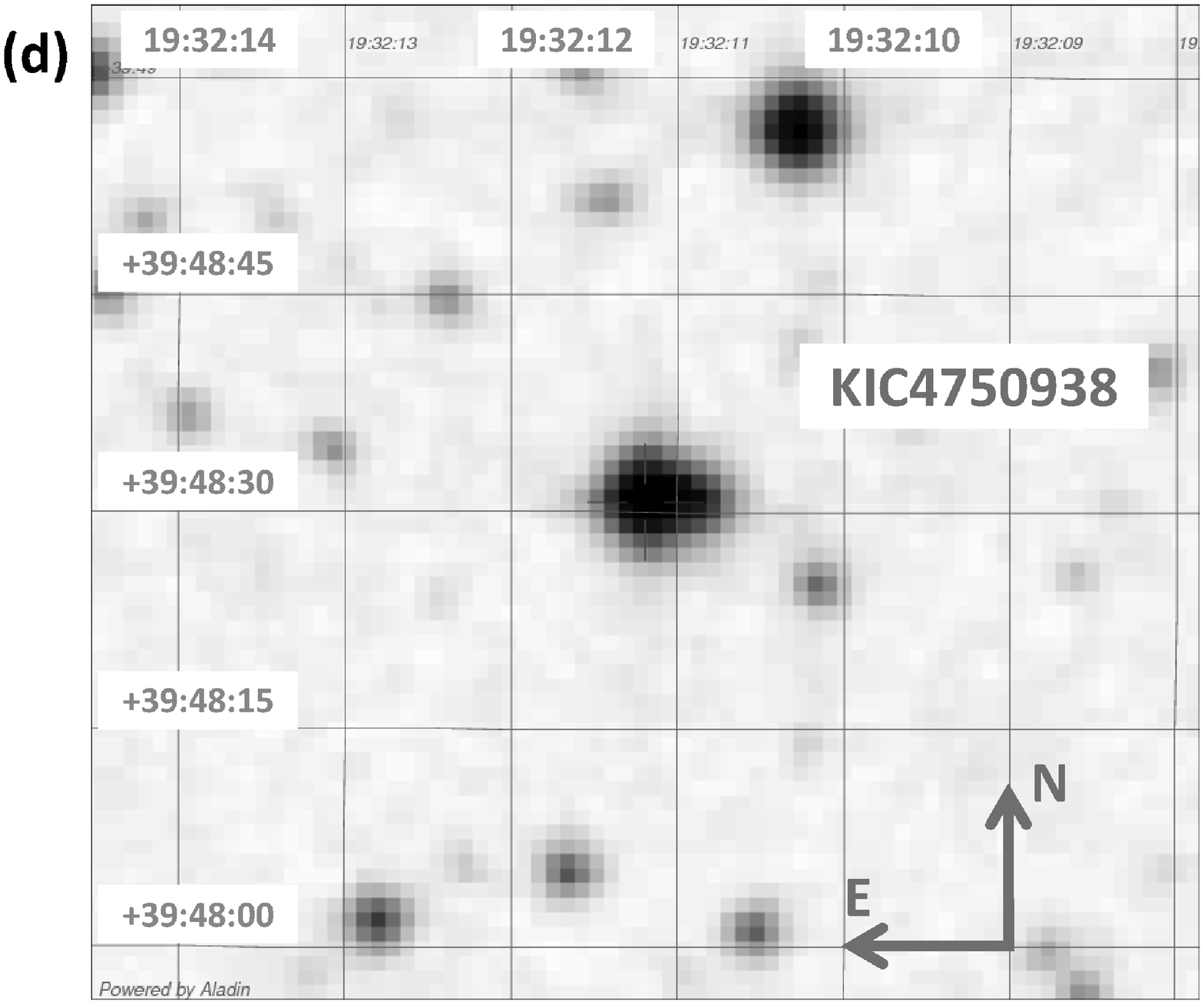}   
 \end{center}
\caption{(a), (b) and (c) : Slit viewer images of KIC11560431, KIC4138557, and KIC5896387, respectively.
These stars were found to have visual companion stars by checking Slit Viewer images of Subaru/HDS.
(d) DSS (Space Telescopes Science Institute Digital Sky Survey) image of KIC4750938. This figure also shows existence of a visual companion star.}\label{fig:sv}
\end{figure}

\begin{figure}[htbp]
 \begin{center}
  \FigureFile(80mm,80mm){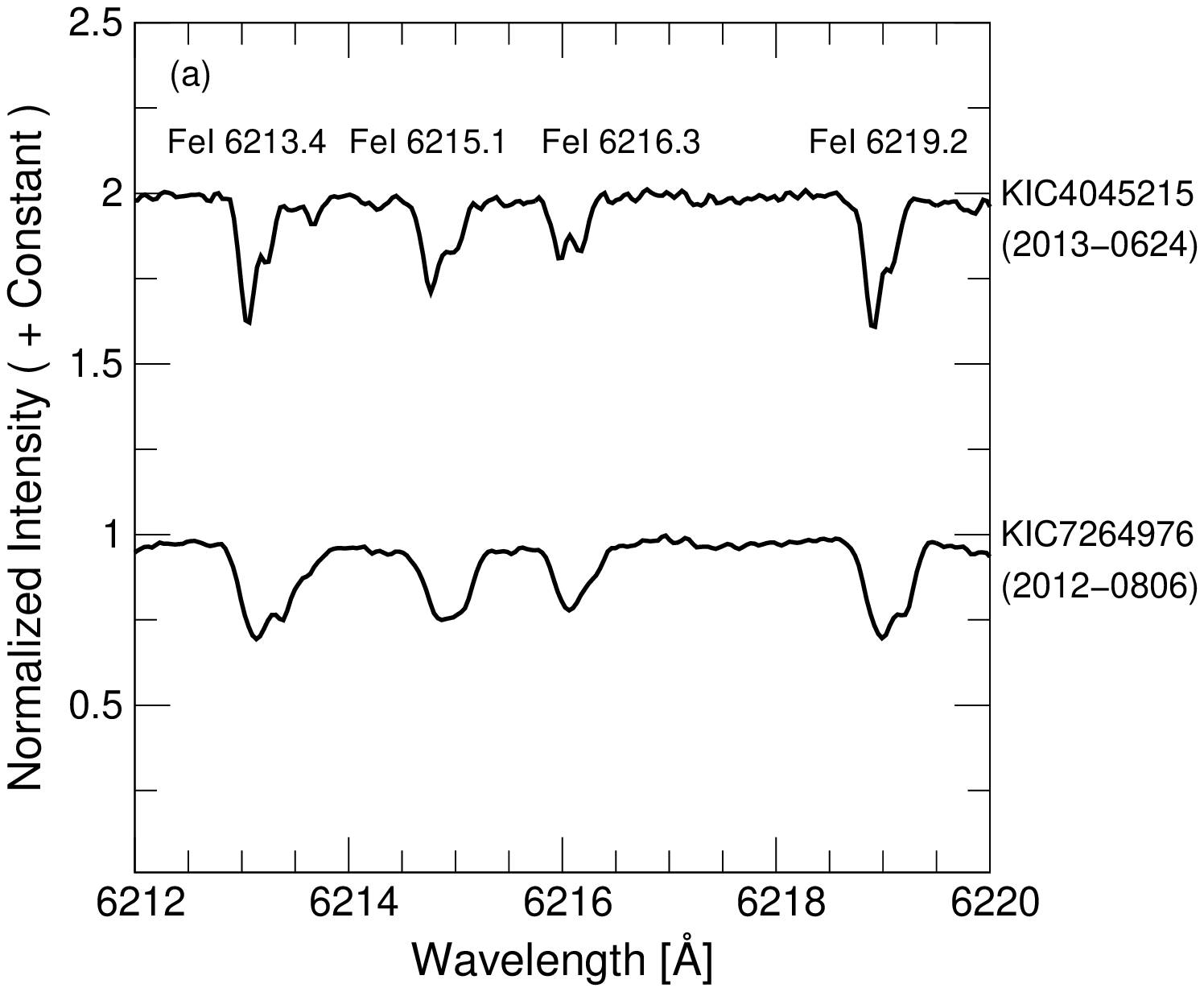} 
  \FigureFile(80mm,80mm){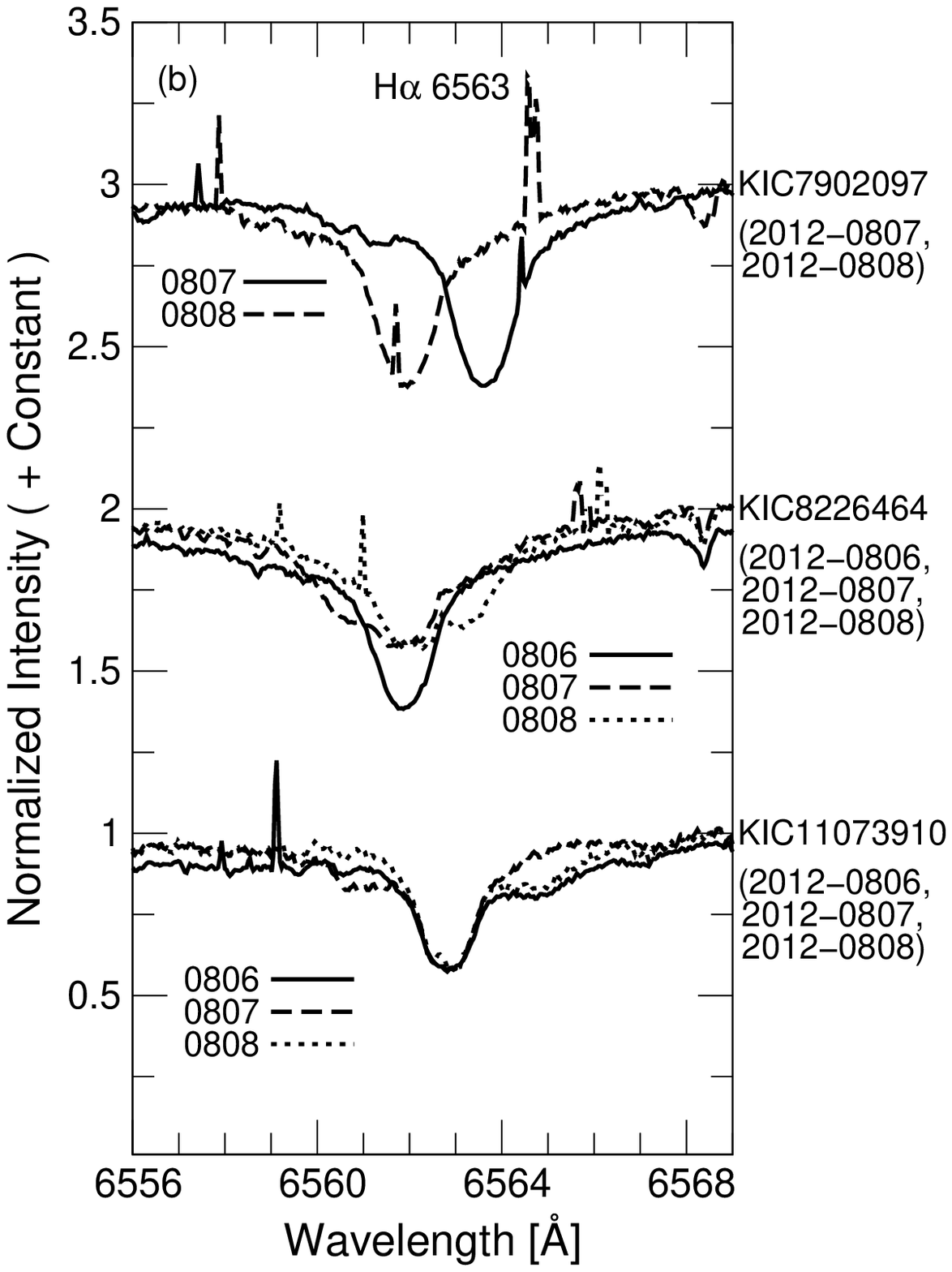}
 \end{center}
\caption{Example of spectra of stars that we consider spectroscopic binary stars. The wavelength scale is adjusted to the heliocentric frame. 
Figures of all spectroscopic binary stars are shown in Supplementary Figure 2. \\
(a) Stars that show double-lined profiles. 
(b) Stars whose spectral lines show time variations between multiple observations, which are expected to be caused by the orbital motion 
in the binary system.}\label{fig:exbi}
\end{figure}

\begin{figure}[htbp]
 \begin{center}
  \FigureFile(70mm,70mm){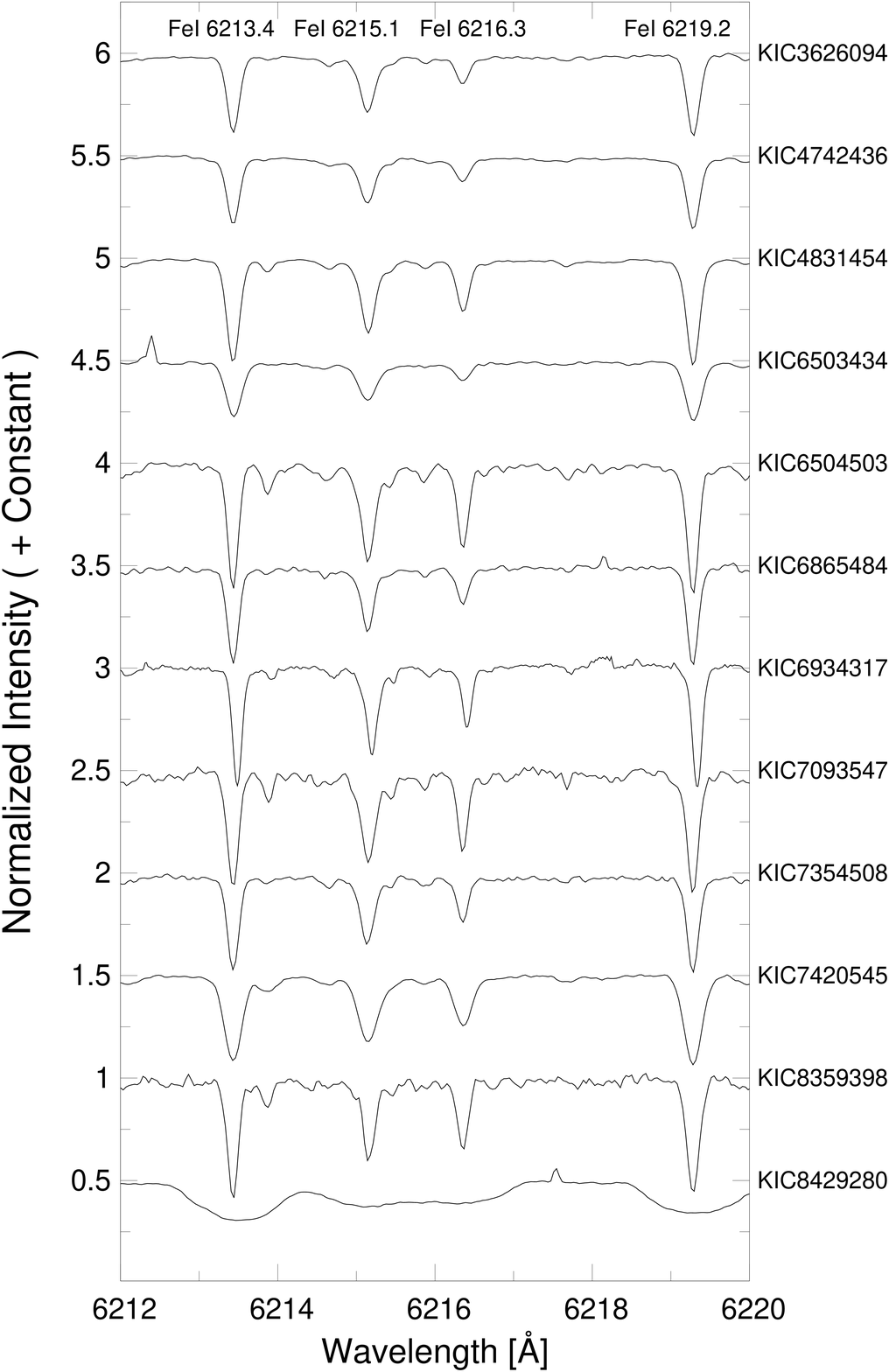}
  \FigureFile(70mm,70mm){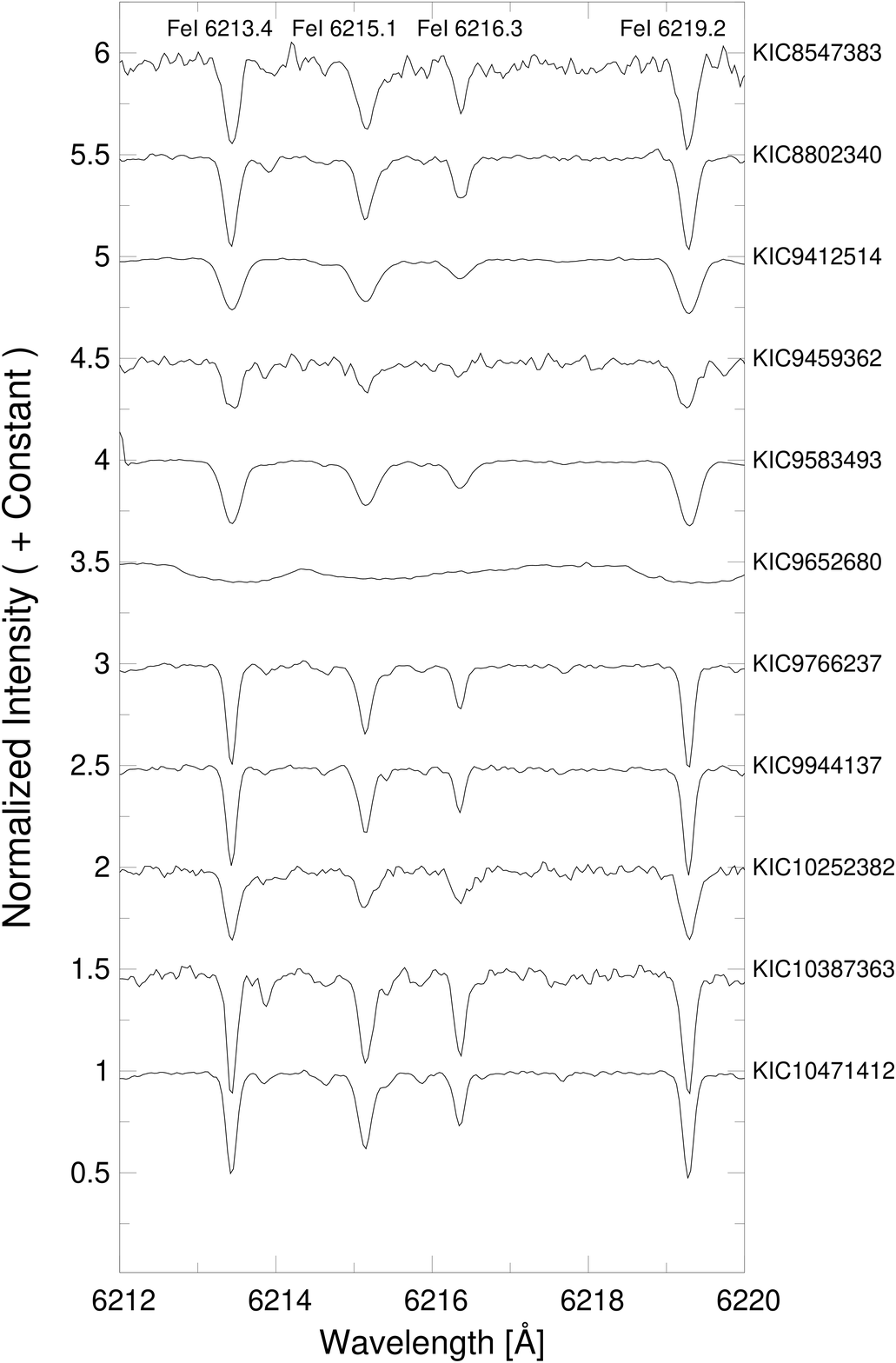}
  \FigureFile(70mm,70mm){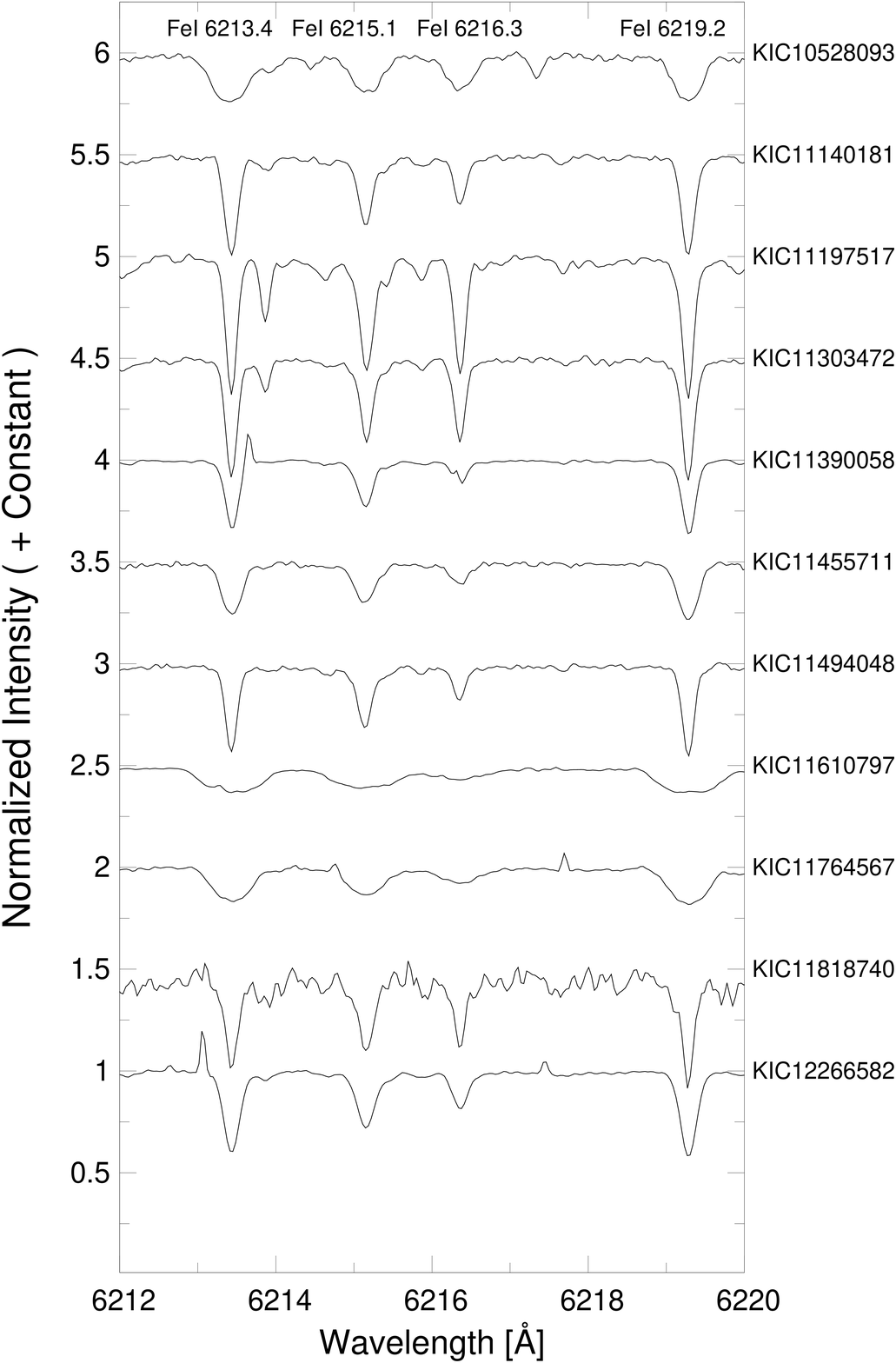}
  \FigureFile(70mm,70mm){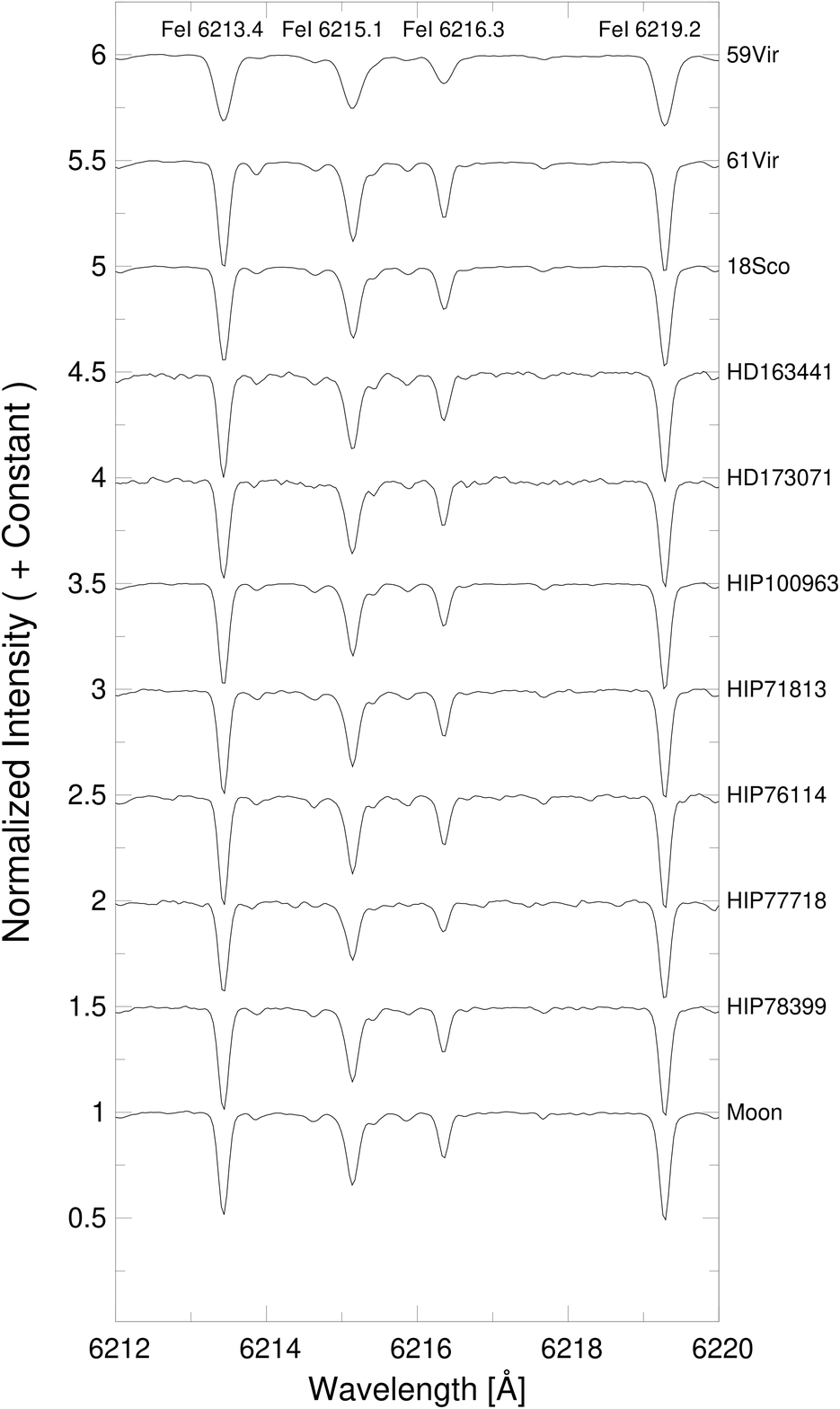}
 \end{center}
\caption{Example of photospheric absorption lines, including Fe I 6213, 6215, 6216, and 6219, of 34 superflare stars that show no evidence of binarity, 10 comparison stars, and Moon. 
The wavelength scale is adjusted to the laboratory frame. Co-added spectra are used here in case that the star was observed multiple times.}\label{fig:sg1Fe}
\end{figure}

For the first step of our analyses, we checked the binarity of each superflare star.
First, we examined slit viewer images of Subaru/HDS of our target stars, and DSS (Space Telescope Science Institute Digital Sky Survey) images of them. 
Four stars (KIC4138557, KIC4750938, KIC5896387, and KIC11560431) have visual companion stars as shown in Figure \ref{fig:sv}. 
These four visual binary stars have ``yes (VB)'' in the 2nd column of Table \ref{tab:shibay}.
Although it is not clear whether these ``visual" companion stars are real binary components of our target stars, 
we did not obtain spectra of them because of the following two reasons: 
(1) we cannot avoid contaminations of the visual companion star in the spectrum of the targets,
and (2) we cannot distinguish which star generated superflares in the Kepler data \footnote{The pixel scale 
of the Kepler CCDs is about 4 arcsec and the typical photometric aperture for a 12 mag star contains about
30 pixels \citep{vanCleve2009}.}.
\\ \\
\ \ \ \ \ \ \
Second, we investigated the line profiles, 
and found that 9 stars (KIC4045215, KIC5445334, KIC7264976, 
KIC8479655, KIC9653110, KIC9764192, KIC9764489, KIC10120296, and KIC10453475)
have double-lined profiles.
In this process, we checked by eye the profile of the main spectral lines (H$\alpha$ 6563, Ca II 8542, 
and 4 Fe I lines in the range of 6212-6220\AA) that are shown in Supplementary Figure 2. 
We also checked by eye the profile of other many ($\sim$50-100) Fe I and II lines of the stars that are not classified as binary stars 
in this paper, simultaneously with measuring equivalent width values of these lines in Section \ref{subsec:atmos}.
The double-lined spectra of KIC4045215 and KIC7264976 are shown in Figure \ref{fig:exbi} (a) as examples.
Figures of all the stars that are considered spectroscopic binary stars are available in Supplementary Figure 2.
Since these double-lined profiles are caused by overlap of the radiation of multiple stars, 
we regard these 9 stars as double-lined spectroscopic binary stars. 
These 9 stars have ``yes (SB2)'' in the 2nd column of Tables \ref{tab:shibay}. 
\\ \\ 
\ \ \ \ \ \ \
Third, we investigated time variations of the line profiles between multiple observations
that are expected to be caused by the orbital motion in the binary system. 
This investigation was for the target stars that we observed multiple times (16 stars).
We measured the radial velocity (RV) of all the target stars that were not classified 
as visual binary stars or double-lined spectroscopic binary stars \footnote{
We do not measure the radial velocities of each component of the double-lined profiles of these stars in this paper, 
since it is not necessary for the following discussions in this paper.
}. 
We used Fe I \& II lines for measuring RV values.
The estimated RVs and the numbers of lines we used here are listed in Supplementary Table 2.
The errors of RVs listed in Supplementary Table 2 are the standard deviations of the RV values that are calculated for individual lines. 
As a result, KIC7902097 shows a large RV change ($>70$km s$^{-1}$) as shown in Figure \ref{fig:exbi} (b).
We consider that this star is in a binary system, and this star has ``yes (RV)'' in the 2nd column of Table \ref{tab:shibay}.
Moreover, the shapes of H$\alpha$ line profile of KIC8226464 and KIC11073910 have large variations between the multiple observations as shown in Figure \ref{fig:exbi} (b), 
though RV values of the main component does not vary significantly. 
We here consider that these variations result from the orbital motion in the binary system, 
and that KIC8226464 and KIC11073910 are also spectroscopic binary stars. 
These stars have ``yes (H$\alpha$-vari)" in the 2nd column of Table \ref{tab:shibay}.
\\ \\
\ \ \ \ \ \ \
In total, we regard 16 superflare stars as binary stars, as summarized in Table \ref{tab:NumBinary} \footnote{
The binary stars are also divided into sub-groups of different period of the brightness variation in Table \ref{tab:NumBinary}.
This classification of binary stars is just for reference in order to confirm how many stars with the brightness variation period 
as long as that of the Sun ($P\gtrsim$20 days) are included among all our target stars.
}.
The remaining 34 superflare stars does not show any evidence of binarity within the limits of our analyses, 
so we treat them as ``single stars'' in this paper. 
These stars have ``no" in the 2nd column of Tables \ref{tab:shibay}.
In the following section, we conduct the detailed analyses only for these 34 stars.
\\ \\
\ \ \ \ \ \ \
Spectra of photospheric lines, including Fe I 6212, 6215, 6216, 6219, of the 34 ``single" superflare stars, 
10 comparison stars, and Moon 
are shown in Figure \ref{fig:sg1Fe}. 
We observed 13 superflare stars multiple times among these 34 superflare stars. 
Three comparison stars (59 Vir, 61 Vir, and 18 Sco) are also observed multiple times.
We made co-added spectra of these 16 stars by conducting the following two steps.
First, we shifted the wavelength value of each spectrum to the laboratory frame on the basis of the radial velocity value of each observation. 
These radial velocity values are listed in Supplementary Table 2. 
Next, we added up these shifted spectra to one co-added spectrum.
These co-added spectra are mentioned as ``S12B-comb" or ``S13A-comb" in Supplementary Table 2.
In Figures \ref{fig:sg1Fe}, co-added spectra of these 16 stars are used. 
Only the co-added spectra are used in the following sections of this paper when we analyze the spectral data of these 16 stars that we observed multiple times.
\\ 
\\
\subsection{Temperature, Surface Gravity, and Metallicity}\label{subsec:atmos}
We estimate the effective temperature $T_{\rm{eff}}$, surface gravity $\log g$, microturbulence $v_{\rm{t}}$, 
and metallicity [Fe/H] of the 34 single superflare stars 10 comparison stars, and Moon (Sun), by measuring the equivalent widths of Fe I and Fe II lines. 
The method is basically the same as the one we used in \citet{SNotsu2013} and \citet{Nogami2014}, 
which is originally based on \authorcite{Takeda2002} (\yearcite{Takeda2002}, \yearcite{Takeda2005}). 
We summarize the method in the following.
\\ \\ 
\ \ \ \ \ \ \
We used Fe I and Fe II lines in the range of  $6120-8370$\AA, 
selected from the line list presented in Online Data of \citet{Takeda2005}.
In the process of measuring equivalent widths, we used the code SPSHOW contained in 
SPTOOL software package\footnote{http://optik2.mtk.nao.ac.jp/$\sim$takeda/sptool/} developed by Y. Takeda, 
which was originally based on Kurucz's ATLAS9/WIDTH9 model atmospheric programs (\cite{Kurucz1993}).
For deriving $T_{\rm{eff}}$, $\log g$, $v_{\rm{t}}$, and [Fe/H] from measured equivalent widths, 
we used TGVIT program\footnote{http://optik2.mtk.nao.ac.jp/$\sim$takeda/tgv/} developed by Y. Takeda.
The procedures adopted in this program are minutely described in \citet{Takeda2002} and \citet{Takeda2005}.
\\ \\ 
\ \ \ \ \ \ \
The resultant atmospheric parameters ($T_{\rm{eff}}$, $\log g$, $v_{\rm{t}}$, and [Fe/H]) of the 34 single superflare stars, 
10 comparison stars, and Moon (Sun) 
are listed in Tables \ref{tab:Sppara} and \ref{tab:compara}, respectively (Table \ref{tab:compara} is in Appendix \ref{sec:compConf-ape}.).
The values of KIC6934317, KIC9766237, and KIC9944137 in Table \ref{tab:Sppara} were already reported in our previous papers.
We reported the values of KIC6934317 in \citet{SNotsu2013}, and the values of KIC9766237 and KIC9944137 in \citet{Nogami2014}. 
In Table \ref{tab:Sppara}, we adopt the atmospheric parameters of KIC8429280 
reported by \citet{Frasca2011}\footnote{
\citet{Frasca2011} provides two sets of the atmospheric parameters: one derived with SYNTHE and the other derived with ROTFIT, 
which agree with each other within 1-sigma error bars. 
In this paper, we adopt the parameters derived with ROTFIT 
since they use the ROTFIT values in the process of estimating Li abundances, and we plan to refer to this value in our future paper. 
} since the rotational velocity of KIC8429280 is so high and 
the spectral lines are too wide to estimate these atmospheric parameters.
For the same reason, we use the atmospheric parameters of KIC9652680 reported by KIC.
\\ \\ 
\ \ \ \ \ \ \
In addition to Table \ref{tab:Sppara}, Figure \ref{fig:atmpa} also shows our $T_{\rm{eff}}$, $\log g$, and [Fe/H] 
of the 34 single superflare stars.
Equivalent width values of all lines of all the stars that we measured in the above process are also listed in Supplementary Data.
\\ 
\\
\subsection{Projected Rotational Velocity}\label{subsec:vsini}
We measure  $v \sin i$ (stellar projected rotational velocity) of the target stars by using the method 
that is basically the same as in \citet{SNotsu2013} and \citet{Nogami2014}. 
The method is originally based on the one described in \citet{Takeda2008}.
We summarize the method in the following.
\\ \\
\ \ \ \ \ \ \ 
We took into account the effects of macroturbulence and instrumental broadening on the basis of \citet{Takeda2008}.
According to \citet{Takeda2008}, there is a simple relationship among the line-broadening parameters, which can be expressed as 
\begin{equation}
 v_{\mathrm{M}}^{2} = v_{\mathrm{ip}}^{2}+v_{\mathrm{rt}}^{2}+v_{\mathrm{mt}}^{2} \ .
\end{equation}
Here, $v_{\rm{M}}$ is $\mathit{e}$-folding width of the Gaussian macrobroadening function 
{$f(v)\propto \exp[-(v/v_{\rm{M}})^{2}]$}, including instrumental broadening ($v_{\rm{ip}}$), 
rotation ($v_{\rm{rt}}$), and macroturbulence ($v_{\rm{mt}}$). 
We derived $v_{\rm{M}}$ by applying an automatic spectrum-fitting technique (e.g \cite{Takeda1995a}; \cite{Takeda2008}), 
assuming the model atmosphere corresponding to the atmospheric parameters estimated in Section \ref{subsec:atmos}.
In this process, we used the MPFIT program contained in the SPTOOL software package. 
\citet{Takeda2008} applied this fitting technique to 6080-6089\AA~region, 
and derived $v\sin i$ values by fitting spectral lines in this region simultaneously. 
We basically adopt their method, but, as we have already mentioned in \citet{SNotsu2013}, 
this region is out of the spectral coverage of our observation (6100$\sim$8820\AA). 
Because of this, we applied the above fitting technique to 6212$\sim$6220\AA \ region, 
which we selected on the basis of the following reasons.
First, this spectral range contains 4 Fe I lines shown in Figure \ref{fig:sg1Fe}, and we can fit these multiple lines simultaneously.
Second, this range does not have so strong lines (Equivalent width values of the 4 Fe I lines are less than 100m\AA), 
and the continuum level is easy to be determined.
Third, this range is expected to have relatively high S/N within the range of this observation (6100$\sim$8820\AA). 
Finally, we have already used this region in \citet{SNotsu2013} and \citet{Nogami2014}, 
and we would like to keep consistency between the results of this paper and those of these papers.
\\ \\
\ \ \ The instrumental broadening velocity $v_{\rm{ip}}$ corresponds to e-folding width of the Gaussian instrumental broadening function. 
This was calculated by using the following relation (\cite{Takeda2008}),
\begin{equation}
 v_{\mathrm{ip}} = \frac{3\times 10^{5}}{2R\sqrt{\rm{ln} 2}} \ ,
\end{equation}
where $R$(=$\lambda/\Delta\lambda$) is the resolving power of the observation.
The macroturbulence velocity $v_{\rm{mt}}$ was estimated by using the relation $v_{\rm{mt}} \sim 0.42\zeta_{\rm{RT}}$~\citep{Takeda2008}.
The term $\zeta_{\rm{RT}}$ is the radial-tangential macroturbulence, and we roughly estimate $\zeta_{\rm{RT}}$ 
by using the relation reported in \citet{Valenti2005},
\begin{equation}\label{eq:macroT}
\zeta_{\rm{RT}} = \left( 3.98 - \frac{T_{\rm{eff}} - 5770 \rm{K}}{650 \rm{K}} \right),
\end{equation}
where $T_{\rm{eff}}$ is the effective temperature of stars.
\citet{Valenti2005} derived the equation (\ref{eq:macroT}) by taking the glowerh boundary of the upper limit of $\zeta_{\rm{RT}}$ 
as a function of $T_{\rm{eff}}$.
Using these equations, we derived $v_{\rm{rt}}$ ($e$-folding width of the Gaussian rotational broadening function), 
and finally $v \sin i$ by using the relation $v_{\rm{rt}}\sim 0.94 v \sin i$ \citep{Gray2005}. 
\\ \\
\ \ \ \ \ \ \
The resultant $v \sin i$ values of the 34 single superflare stars, 10 comparison stars, and Moon (Sun) 
are listed in Tables \ref{tab:Sppara} and \ref{tab:compara}, respectively. 
As mentioned in Section \ref{subsec:atmos},
the $v \sin i$ values of KIC6934317, KIC9766237 and KIC9944137 in Table \ref{tab:Sppara} 
were already reported in our previous papers (\cite{SNotsu2013}; \cite{Nogami2014}).
In Table \ref{tab:Sppara}, we adopt the $v \sin i$ value of KIC8429280 reported by \citet{Frasca2011} since 
we also used the atmospheric parameters reported by them in Section \ref{subsec:atmos}.
The $v \sin i$ value of KIC9652680 is calculated from atmospheric parameters in KIC 
since we used the atmospheric parameters in KIC in Section \ref{subsec:atmos}.
In this process, we assume that microturbulence velocity ($v_{\rm{t}}$) of KIC9652680 is 1 km s$^{-1}$ 
as a typical value of the solar-type stars (see Table 4).
\\ \\
\ \ \ \ \ \ \
\authorcite{Hirano2012} (\yearcite{Hirano2012}, \yearcite{Hirano2014}) 
estimated the systematic uncertainty of $v \sin i$ by changing $\zeta_{\mathrm{RT}}$ by
$\pm$15\% from Equation (\ref{eq:macroT}) for cool stars ($T_{\rm{eff}} \leq $6100K), 
on the basis of observed distribution of $\zeta_{\rm{RT}}$ (See also Figure 3 in \cite{Valenti2005}). 
They explained that the statistical errors in fitting each spectrum are generally smaller 
than the systematic errors arising from different values of $\zeta_{\rm{RT}}$.
We then used this type of error values arising from $\zeta_{\rm{RT}}$ as errors of $v \sin i$ listed in Tables \ref{tab:Sppara} and \ref{tab:compara}.
\\ \\
\ \ \ \ \ \ \
Recently, \citet{Doyle2014} estimated $v \sin i$ values from asteroseismic analysis of 28 main-sequence stars observed by Kepler, and infer macroturblence velocity of them
on the basis of these $v \sin i$ values.
Using these results, they then derived the following new equation between macroturblence velocity, $T_{\rm{eff}}$, and $\log g$:
\begin{eqnarray}\label{eq:Doyle}
\zeta_{\rm{RT}} &=& 3.21 + 2.33 \times 10^{-3} (T_{\rm{eff}} - 5777 ) \nonumber \\
&&+ 2.00 \times  10^{-6} (T_{\rm{eff}} - 5777)^{2} - 2.00 (\log g - 4.44 ) \ .
\end{eqnarray}
Figure 4 of \citet{Doyle2014} shows that there are some differences ($\Delta\zeta_{\rm{RT}}\sim$1 km s$^{-1}$) between 
the $\zeta_{\rm{RT}}$ estimated by using Equation (\ref{eq:macroT}) and those by Equation (\ref{eq:Doyle}).
For comparison, we then derived the new $v \sin i$ value of our targets by using this new equation (Equation (\ref{eq:Doyle}))
in stead of Equation (\ref{eq:macroT}), which we used for estimating the $v \sin i$ value in the above paragraphs.
The resultant value is listed in Supplementary Table 3.
The error value of this new $v \sin i$ is estimated from the errors of Equation (\ref{eq:Doyle}) ($\Delta\zeta_{\rm{RT}}\sim$0.73 km s$^{-1}$) 
reported in \citet{Doyle2014}.
In Figure \ref{fig:vsini-vmac}, we compare this new $v \sin i$ value estimated by using Equation (\ref{eq:Doyle}) with the original $v \sin i$ value by Equation (\ref{eq:macroT}).
As shown in this figure, the difference between these two $v \sin i$ values is not so large ($<$1 km s$^{-1}$) for most of the target stars.
We have already used the values with Equation (\ref{eq:macroT}) in our previous researches (\cite{SNotsu2013}; \cite{Nogami2014}).
Because of these two things, in the following sections of this paper and Paper II, we only use the original $v \sin i$ value 
estimated in the above paragraphs using Equation (\ref{eq:macroT}).
This value with Equation (\ref{eq:macroT}) is plotted in the horizontal axis of Figure \ref{fig:vsini-vmac} and listed in Tables \ref{tab:Sppara} and \ref{tab:compara}.
\\ 
\\
\subsection{Stellar Radius}\label{subsec:AgeR}
Using the stellar atmospheric parameters ($T_{\rm{eff}}$, $\log g$, and [Fe/H]) estimated in Section \ref{subsec:atmos}, 
we roughly estimated the stellar age and stellar mass ($M_{\rm{s}}$) 
for the target stars by applying the PARSEC isochrones \footnote{http://stev.oapd.inaf.it/cgi-bin/cmd} in \citet{Bressan2012}.
In this process, we selected all the data points having possible sets of $T_{\rm{eff}}$, $\log g$, and [Fe/H] from the PARSEC isochrones, 
taking into account the error values of $T_{\rm{eff}}$ and $\log g$ ($\Delta T_{\rm{eff}}$ and $\Delta \log g$, respectively) 
shown in Tables \ref{tab:Sppara} and \ref{tab:compara}. 
There were three stars (KIC8359398, KIC9766237, and KIC10252382) that have no suitable isochrones  
within their original error range of $T_{\rm{eff}}$ and $\log g$.
For these three stars, we then took into account $2\Delta T_{\rm{eff}}$ and $2\Delta\log g$.
We must note that the resultant values of these three stars can have relatively low accuracy. 
\\ \\
\ \ \ \ \ \ \
For each selected data point, we estimated $R_{\rm{s}}$ from the $M_{\rm{s}}$ and $\log g$  
of each data point by using 
\begin{equation}\label{eq:Rs}
\frac{R_{\rm{s}}}{R_{\odot}} =\sqrt{\mathstrut\biggl(\frac{M_{\rm{s}}}{M_{\odot}}\biggl)\bigg/\biggl(\frac{g}{g_{\odot}}\biggl)~} ~.
\end{equation}
We then selected the maximum and minimum $R_{\rm{s}}$ values of each star 
and determined the $R_{\rm{s}}$ value of each target star as a median between these maximum and minimum values.
These $R_{\rm{s}}$ values of the target superflare stars and comparison stars are listed 
in Tables \ref{tab:Sppara} and \ref{tab:compara}, respectively.
The error values of $R_{\rm{s}}$ in these tables correspond to the above maximum and minimum $R_{\rm{s}}$ values of each target star.
The error values of $R_{\rm{s}}$ shown in Table \ref{tab:Sppara} are $\lesssim$20\% for most of the stars.
In addition, values of stellar age on the basis of the above isochrone are shown in Figure \ref{fig:isochrone}.
\\ 
\\ 
\section{Discussion}\label{sec:discussion}
\subsection{Binarity}\label{subsec:dis-binarity}
In Section \ref{subsec:binary}, we described that more than half (34 stars) of 50 target superflare stars have no evidence of binary system. 
We need to remember here that we cannot completely exclude the possibility that some of these 34 ``single" superflare stars have companions, 
only with our limited observations above.
We performed multiple observations only for 13 stars among 34 ``single" stars, 
while the remaining 21 stars were observed only at once, as shown in column 3 of Supplementary Table 2.
Then, we cannot completely exclude the possibility of showing radial velocity shifts especially for these 21 stars.
It is necessary to observe these target stars more repeatedly in order to exclude the possibility of having binary stars as much as possible.
Moreover, we also cannot completely exclude the possibility that the above 34 ``single" stars have 
other faint neighboring stars such as M-dwarfs,
even if no double-lined profiles and no radial velocity shifts were confirmed in this paper.
In the process of investigating whether the target stars have double-lined profiles in Section \ref{subsec:binary},
we only checked by eye the profile of the main spectral lines \footnote{
We also checked by eye the profile of other many ($\sim$50-100) Fe I and II lines of the stars that are not classified as binary stars
in the process of measuring atmospheric parameters in Section \ref{subsec:atmos}.
}
(H$\alpha$ 6563, Ca II 8542 and 4 Fe I lines in the range of 6212-6220\AA) that are shown in Supplementary Figure 2.
This suggests that the classification of double-lined spectroscopic binary stars is not complete.
More detailed analyses such as the cross-correlation of the target and the template spectrum 
are needed to detect the signature of binarity caused by the existence of faint companion stars. 
We must note this point, but we consider that such detailed analyses of binarity are 
not really necessary for the overall discussion of stellar properties of superflare stars in this paper.
We then expect the future detailed observations and analyses.
\\ \\
\ \ \ \ \ \ \
As mentioned in Section \ref{sec:intro}, close binary stars such as RS CVn-type stars have been widely known as active flare stars,
which maintain high rotation rate and high flare activity thanks to the tidal interaction 
between the primary and companion stars (e.g., \cite{Walter1981}).
It is very important whether our superflare stars are such close binary stars or not, 
especially for considering that single stars like the Sun can really have superflares.
The 34 superflare stars show no double-lined profile, as explained above. 
This suggests that many of these 34 stars have no companions whose mass is on the same order of that of the primary star.
This is because if such companions exist, the bolometric intensity of them is comparable to that of the primary star, 
and double-lined profile is expected to be observed in many cases. 
In particular, 13 stars among these 34 stars show no radial velocity shifts between the multiple observations.
It is then possible that many of them are not close binary stars, though more multiple observations are needed, 
as mentioned in the previous paragraph of this section.
\\ \\
\ \ \ \ \ \ \
In addition, the absolute value of radial velocity of the two stars (KIC7354508 and KIC9459362) are relatively large ($>100$km s$^{-1}$) 
compared to the other stars, as listed in Supplementary Table 2. 
This suggests the two possibilities that these stars are close binary stars or high-velocity stars.
We cannot decide which possibilities are right since we observed them only at once and we cannot investigate changes of radial velocities.
In this paper, we then treat them as single stars.
\\ 
\\
\subsection{Estimated Stellar Parameters}\label{subsec:dis-stpara}
\begin{figure}[htbp]
 \begin{center}
  \FigureFile(75mm,75mm){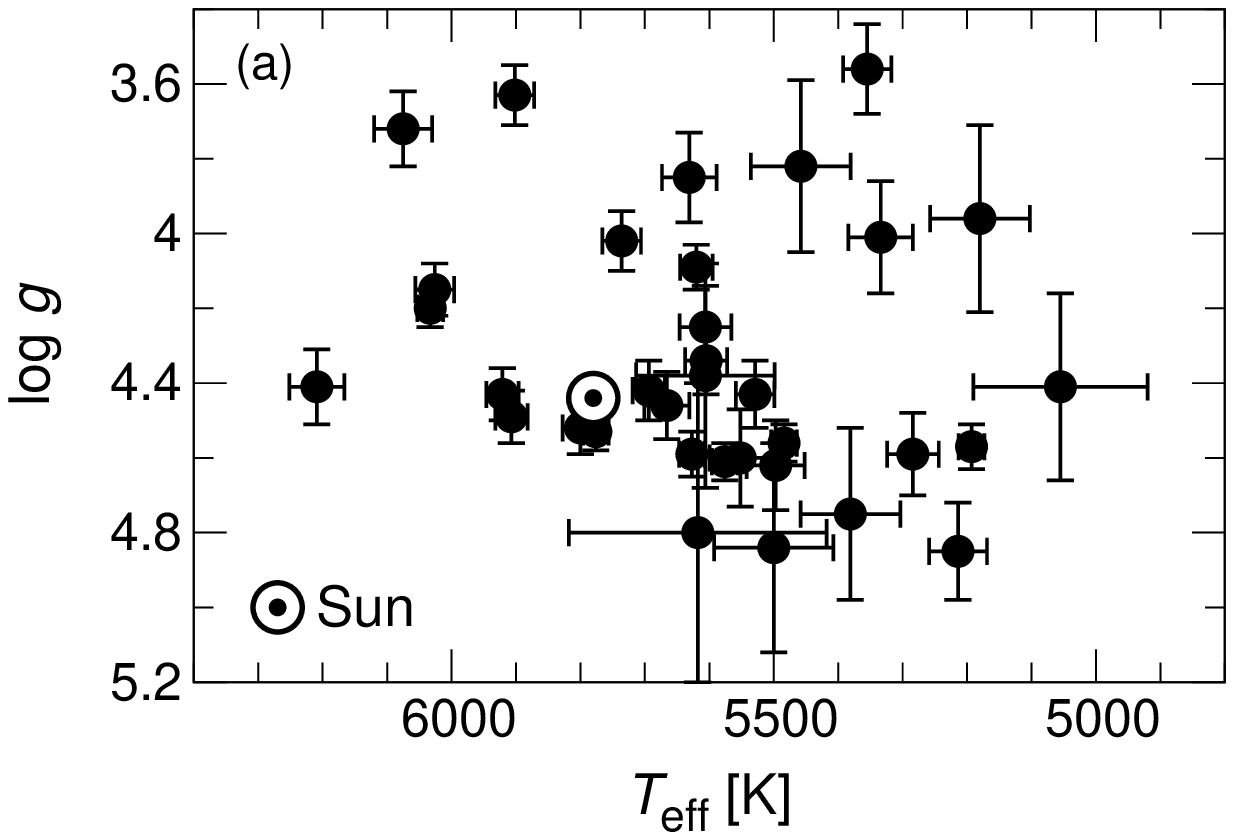}
  \FigureFile(75mm,75mm){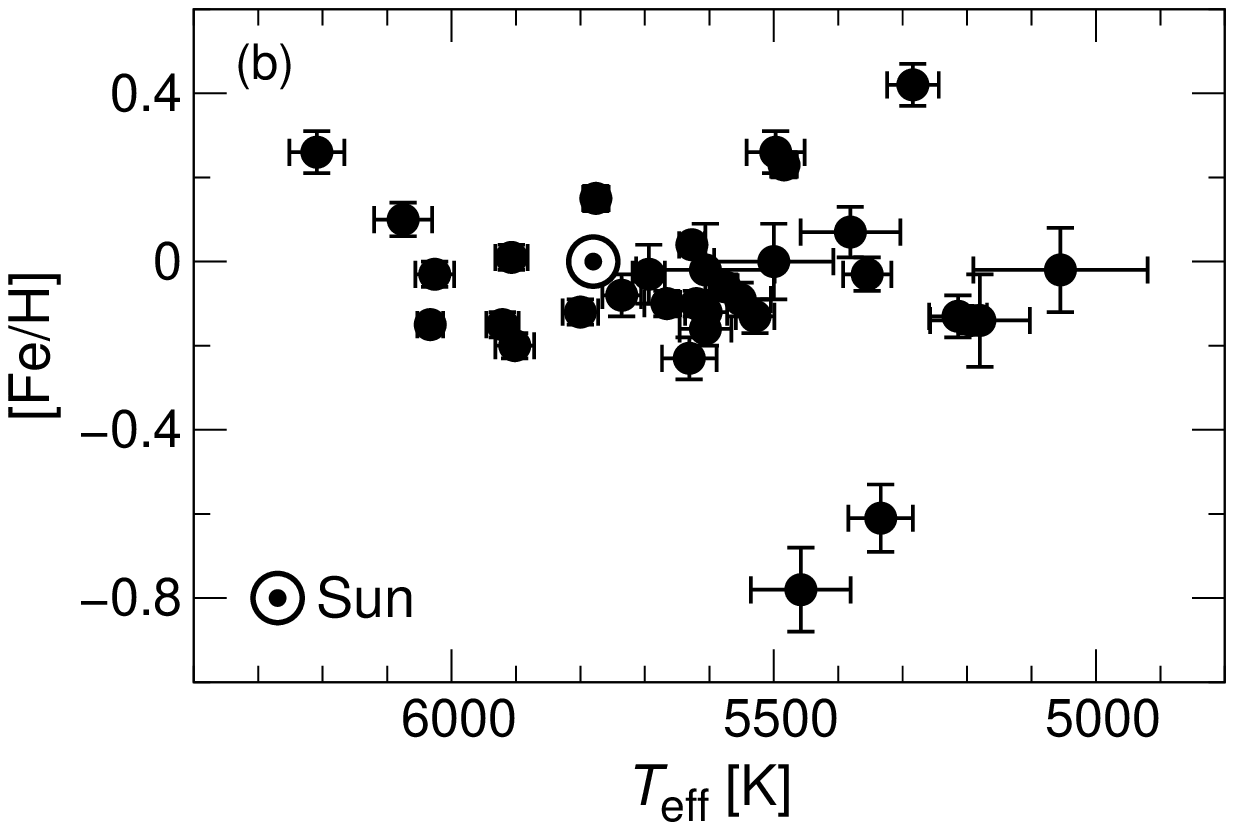}
 \end{center}
\caption{Temperature ($T_{\rm{eff}}$), surface gravity ($\log g$), and metallicity ([Fe/H]) of the target superflare stars. 
These values are estimated by using our spectral data.
The solar value is also plotted using a circled dot point.}\label{fig:atmpa}
\end{figure}

In Section \ref{subsec:atmos}, 
we estimated the atmospheric parameters ($T_{\rm{eff}}$, $\log g$, [Fe/H]) of the 34 superflare stars, 
which are considered as single stars in Section \ref{subsec:binary}. 
According to Figure \ref{fig:atmpa}(a), the measured $T_{\rm{eff}}$ and $\log g$ values of the 34 superflare stars are 
in the range of 5000$\sim$6300K and 3.5$\sim$4.9, respectively.
This means that the stellar parameters of these superflare stars are roughly in the range of solar-type (G-type main sequence stars) stars,
though 6 stars with $\log g<4.0$ are possibly sub-giant G-type stars.
In particular, the temperature, surface gravity, and the brightness variation period ($P_{0}$) 
of 9 stars including two stars (KIC9766237 and KIC9944137) reported in \citet{Nogami2014} 
are in the range of ``Sun-like" stars ($5600\leq T_{\rm{eff}}\leq 6000$K, $\log g\geq$4.0, and $P_{0}>$10 days).
The metallicity ([Fe/H]) of these stars are not so different from solar one as we can see in Figure \ref{fig:atmpa}(b). 
The root-mean-square residual between [Fe/H] of the observed superflare stars and that of the Sun is $\sim$0.23.
No clear ``metal-rich" or ``metal-poor" stars are included in our target stars, though metallicity of the two stars (KIC9459362 and KIC10252382) 
are a bit lower ([Fe/H]$<-0.6$) than that of the other stars.
These two a bit ``metal-poor" stars are plotted  in Figure \ref{fig:isochrone} by using triangle points, 
and this figure suggests the possibility that these two stars are not young, though 
the detailed discussion of stellar age is beyond the scope of this paper, as also mentioned in the next paragraph. 
\\ \\
\ \ \ \ \ \ \
Age of the target stars can roughly estimated by Figure \ref{fig:isochrone} on the basis of the isochrone values. 
Information of the stellar age is important since the stellar activity have deep relation with the stellar age (e.g., \cite{Soderblom1991}),
but the detailed discussion of the stellar age is beyond the scope of this paper.
We also plan to discuss the Li abundances and whether our target superflare stars 
are quite young or not in our future paper (Honda et al. in preparation) 
since the Li abundance is known to provide quite loose constraints 
on the age of G-type stars (e.g., \cite{Soderblom1993}; \cite{Sestito2005}).
\\ \\
\ \ \ \ \ \ \
There are four stars below all the plotted isochrones (100Myr, 1Gyr, and 10Gyr) in Figure \ref{fig:isochrone}.
They are KIC8359398, KIC9652680, KIC10387363 and KIC11818740.
The $\log g$ values of them may be slightly overestimated 
for ordinary G-type main sequence stars.
Regarding KIC9652680, the log g value in KIC is used, as mentioned in Section \ref{subsec:atmos}, 
and can be not so accurate, as we discuss in Section \ref{subsec:dis-pa}.
The S/N ratios of the other three stars are relatively low (S/N$\lesssim$40 around H$\alpha$ 6563\AA) 
among our target stars (see Supplementary Table 2).
This can possibly increase the error values of $\log g$.
\\ \\
\ \ \ \ \ \ \
In Section \ref{subsec:vsini}, we measured the projected rotational velocity ($v\sin i$) of the 34 superflare stars, 
which are considered as single stars in Section \ref{subsec:binary}.
The distribution histogram of $v \sin i$ for these stars are shown in Figure \ref{fig:vsini-bunpu}.
The data of 119 ordinary solar-type stars reported in \citet{Takeda2010} are also plotted in this figure for reference. 
We can roughly regard the data of \citet{Takeda2010} as a random sample of the ordinary solar-type stars,
considering their target selection method explained in \citet{Takeda2007}.
Comparison of these data in Figure \ref{fig:vsini-bunpu} suggests that $v \sin i$ of the observed superflare stars 
tends to be higher than the sample of ordinary solar-type stars, though
we need to remember here that the group of the target stars of our spectroscopic observations 
is not a really random sample of superflare stars, as explained in Section \ref{subsec:tarselc}.
In particular, 5 of the 34 target stars have extremely high $v \sin i$ value ($v \sin i \geq 10$km s$^{-1}$). 
On the other hand,  22 of the 34 target superflare stars have low $v \sin i$ value ($v \sin i < 5$km s$^{-1}$), 
and our target superflare stars include stars rotating as slow as the Sun ($v \sin i\sim$2 km s$^{-1}$).
In Paper II, we will compare this $v \sin i$ value with other stellar properties such as the brightness variation observed by Kepler.
\\ \\
\ \ \ \ \ \ \
Summarizing the above discussions in Section \ref{subsec:dis-binarity} and \ref{subsec:dis-stpara},
more than half (34 stars) of 50 target superflare stars have no evidence of binary system, 
and stellar atmospheric parameters of these stars are basically in the range of ordinary solar-type (G-type main sequence) stars.
Moreover, these 34 stars include stars rotating as slow as the Sun ($v \sin i\sim$2 km s$^{-1}$).
These results suggest that stars whose spectroscopic properties similar to the Sun can have superflares, 
and this supports the hypothesis that the Sun might cause a superflare.
This is consistent with the result of our previous paper \citet{Nogami2014}, 
which found that the spectroscopic properties of 
two superflare stars KIC9766237 and KIC9944137 are very close to those of the Sun.
\\ 
\\
\subsection{Comparison of Our Estimated Stellar Parameters with the Previous Values}\label{subsec:dis-pa}
\begin{figure}[htbp]
 \begin{center}
  \FigureFile(75mm,75mm){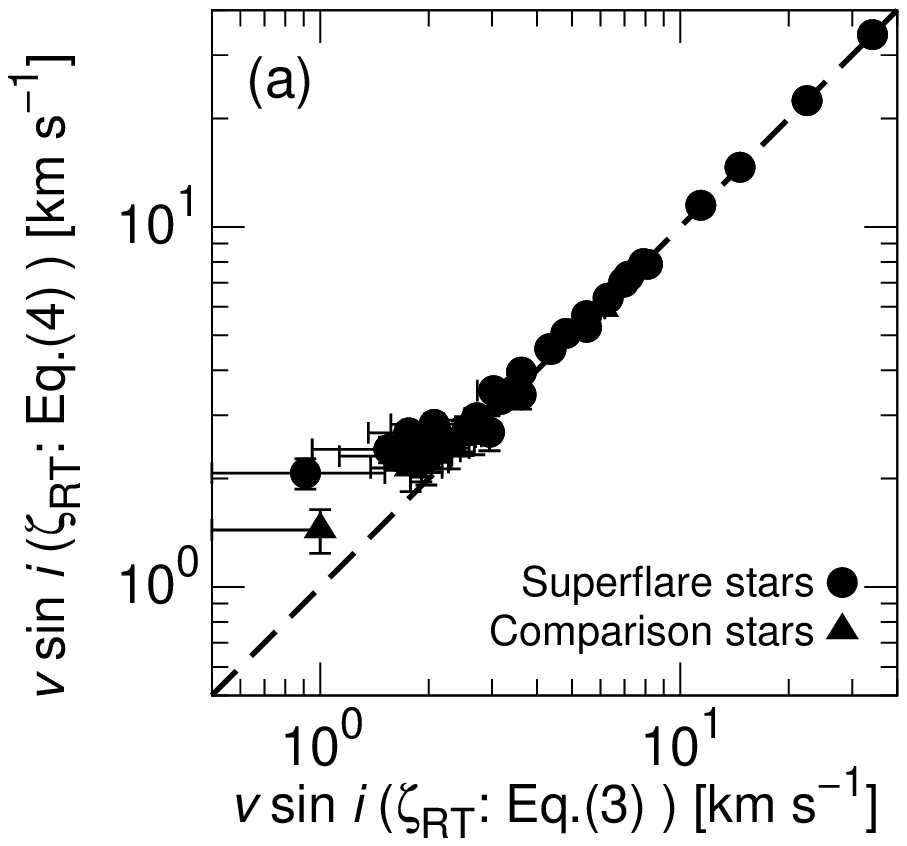}
  \FigureFile(75mm,75mm){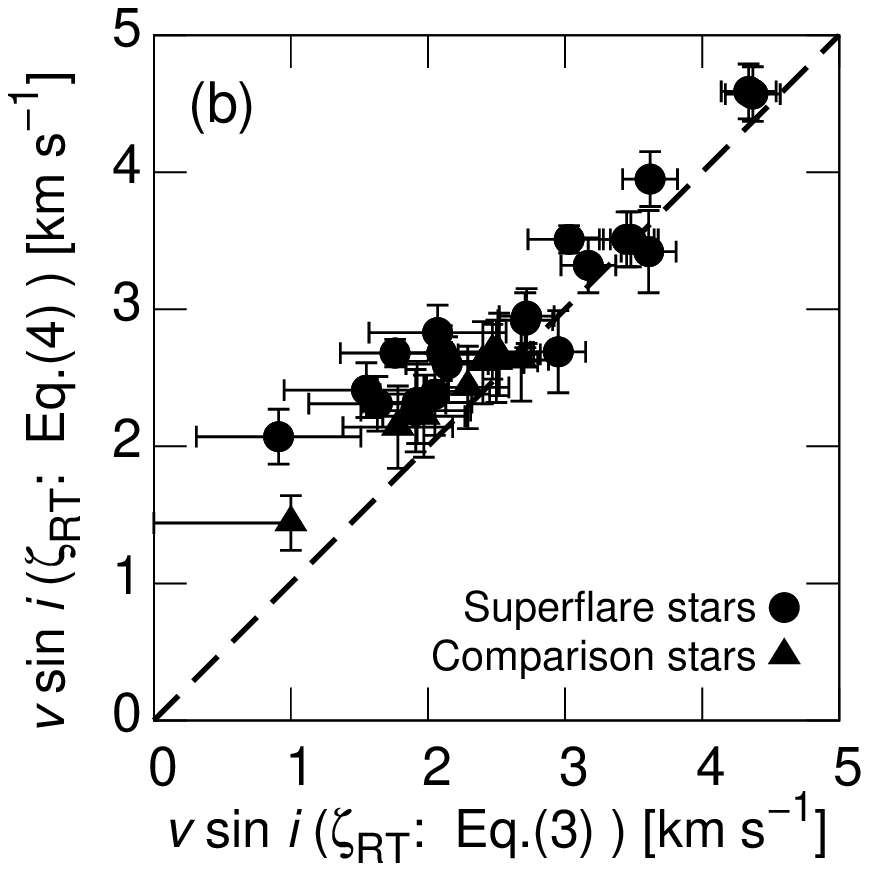}
 \end{center}
\caption{(a) Comparison of two types of $v \sin i$ value using the other equations (Equation (\ref{eq:macroT}) and (\ref{eq:Doyle})) 
for the estimation of macroturblence velocity ($\zeta_{\rm{RT}}$) in the process of $v \sin i$ measurement. 
The black circles correspond to the data of our target superflare stars, while the black triangles are the data of comparison stars.\\
(b) Extended figure of (a). The plot range is limited to 0$\leq v\sin i\leq$5 km s$^{-1}$. }\label{fig:vsini-vmac}
\end{figure}

\begin{figure}[htbp]
 \begin{center}
  \FigureFile(100mm,100mm){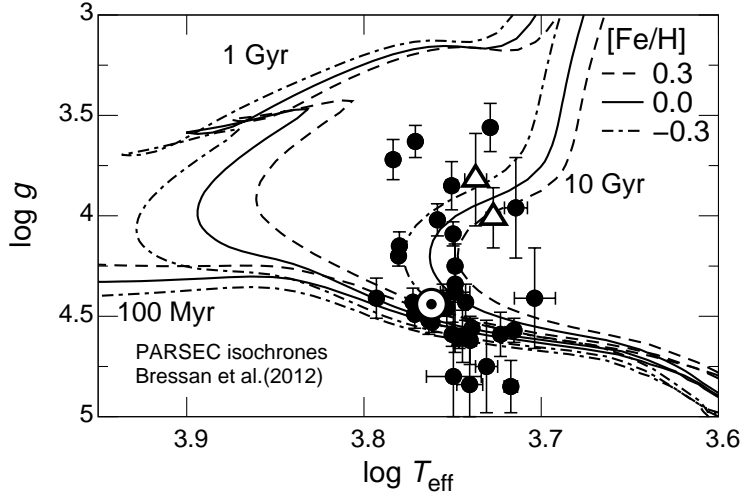}
 \end{center}
\caption{Scatter plot of temperature ($T_{\rm{eff}}$) vs. surface gravity ($\log g$) of the target stars. 
Plotted values are estimated by using our spectral data. 
Two a bit ``metal-poor" stars ([Fe/H]$<-0.6$) KIC 9459362 and KIC 10252382 
among our target stars are plotted by using triangle points.
The overplotted 6 lines show stellar age values of 1 and 10 Gyr for three different metallicity values ([Fe/H]$=0.3, 0.0, -0.3$) 
on the basis of PARSEC isochrones in \citet{Bressan2012}. 
The solar value is also plotted using a circled dot point.
}\label{fig:isochrone}
\end{figure}

\begin{figure}[htbp]
 \begin{center}
  \FigureFile(110mm,110mm){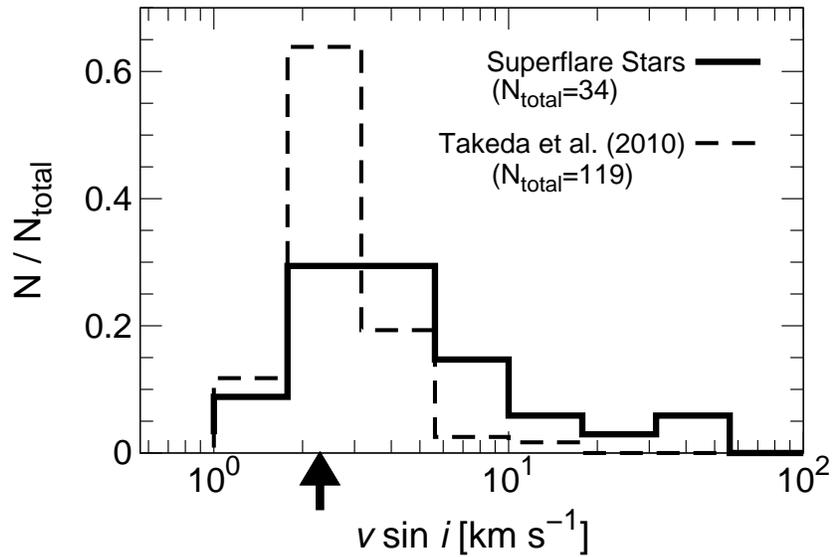}
 \end{center}
\caption{Histograms showing the distribution of $v \sin i$. The filled line shows the data of superflare stars analyzed in this paper, 
while the dotted line shows those of ordinary solar-type stars reported in \citet{Takeda2010}.
The solar $v \sin i$ value is roughly indicated by a black upward arrow.}\label{fig:vsini-bunpu}
\end{figure}

\begin{figure}[htbp]
 \begin{center}
  \FigureFile(70mm,70mm){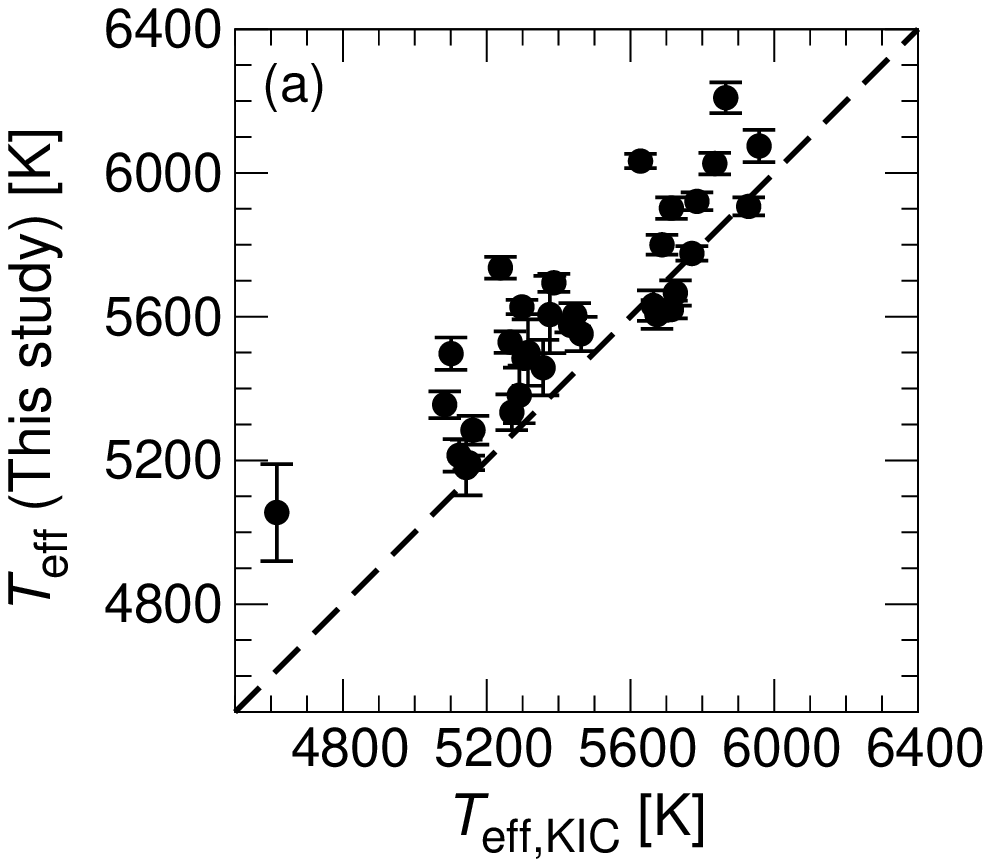}
  \\
  \FigureFile(70mm,70mm){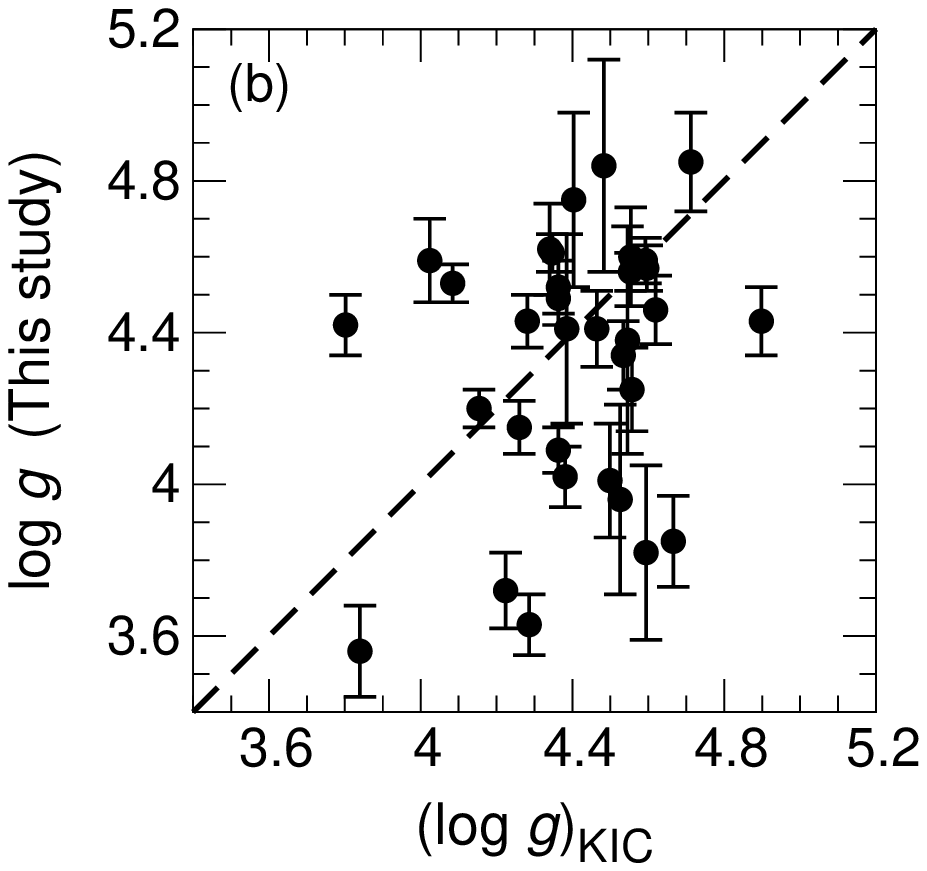}
  \\
  \FigureFile(70mm,70mm){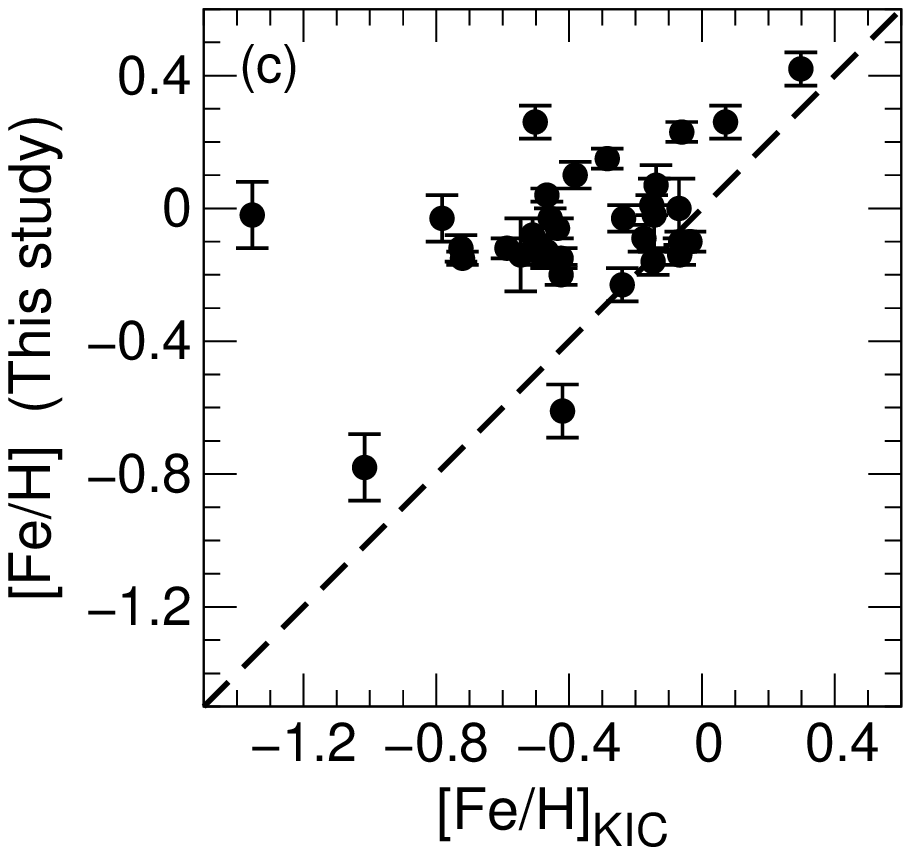}
 \end{center}
\caption{Comparison between the atmospheric parameters ($T_{\rm{eff}}$, $\log g$, and [Fe/H]) that we estimate in this study, 
and those reported in the Kepler Input Catalog, respectively. 
The error bars indicate errors of the values in this study.
KIC9652680 is not included here since the temperature value of KIC9652680 was not estimated from spectroscopic data as described in Section \ref{subsec:atmos}.}
\label{fig:KIChika}
\end{figure}

\begin{figure}[htbp]
 \begin{center}
  \FigureFile(70mm,70mm){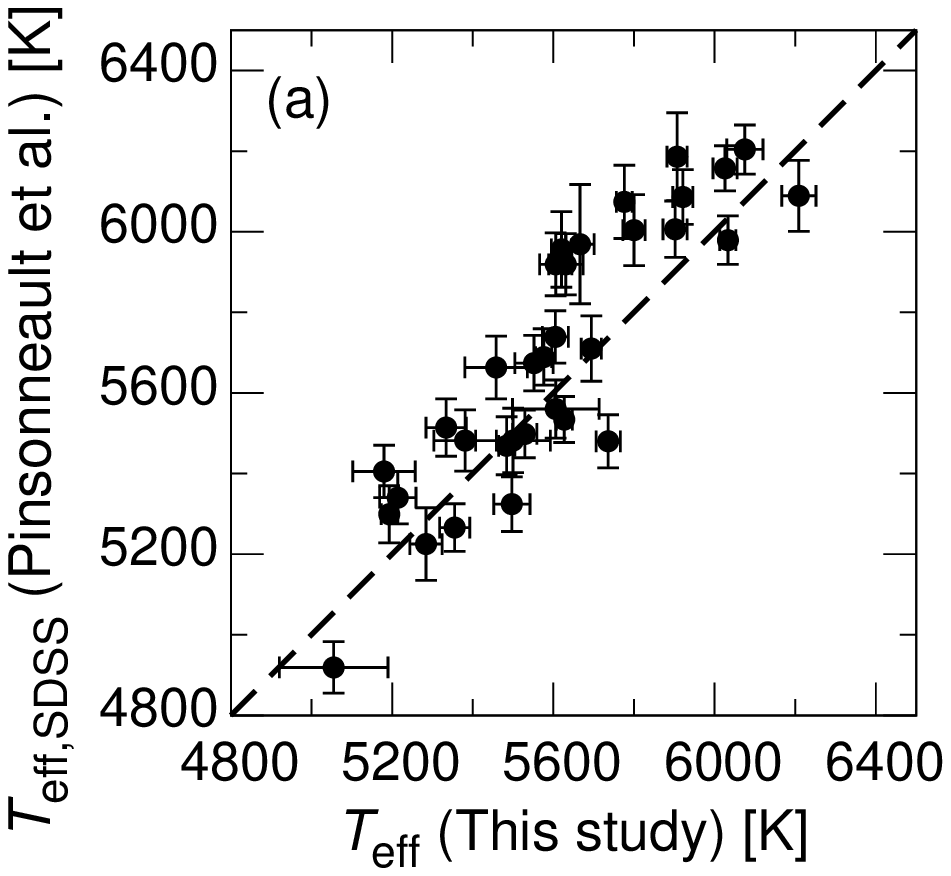}
\\
  \FigureFile(70mm,70mm){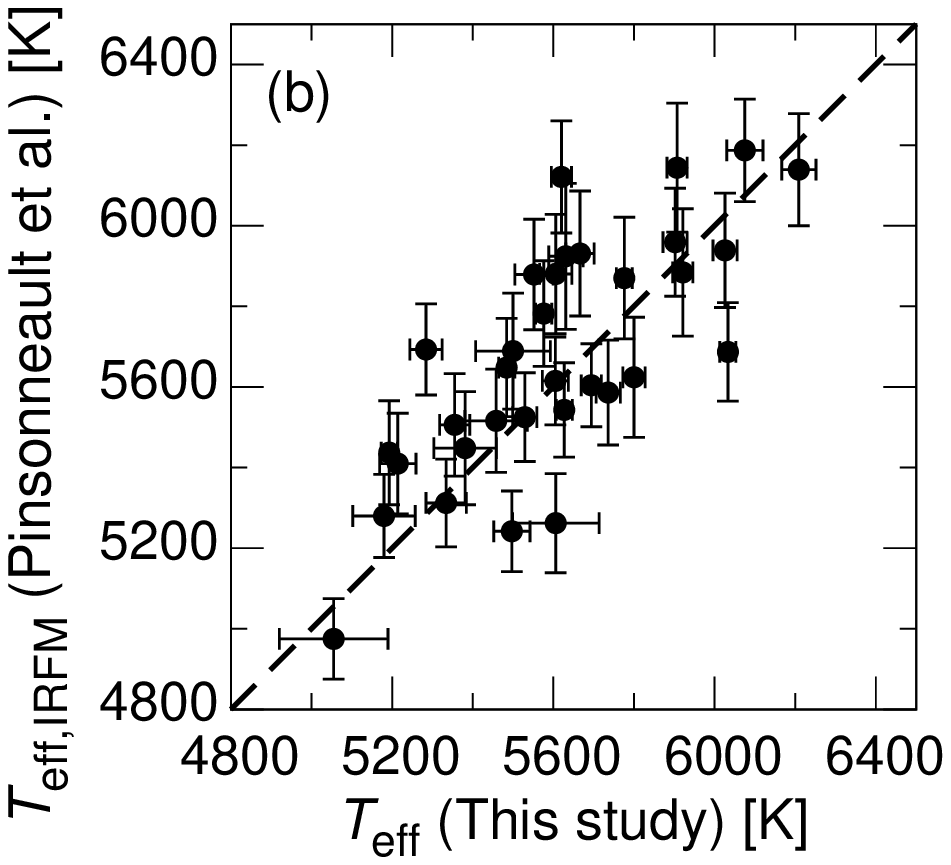}
 \end{center}
\caption{Comparison of the temperature values in this study ($T_{\rm{eff}}$) 
with the value reported in \citet{Pinsonneault2012} ($T_{\rm{eff,SDSS}}$ and $T_{\rm{eff,IRFM}}$). 
KIC9652680 is not included here since we do not have the temperature value of KIC9652680 that was estimated from spectroscopic data as described in Section \ref{subsec:atmos}.}
\label{fig:colhika}
\end{figure}

\begin{figure}[htbp]
 \begin{center}
\FigureFile(67mm,67mm){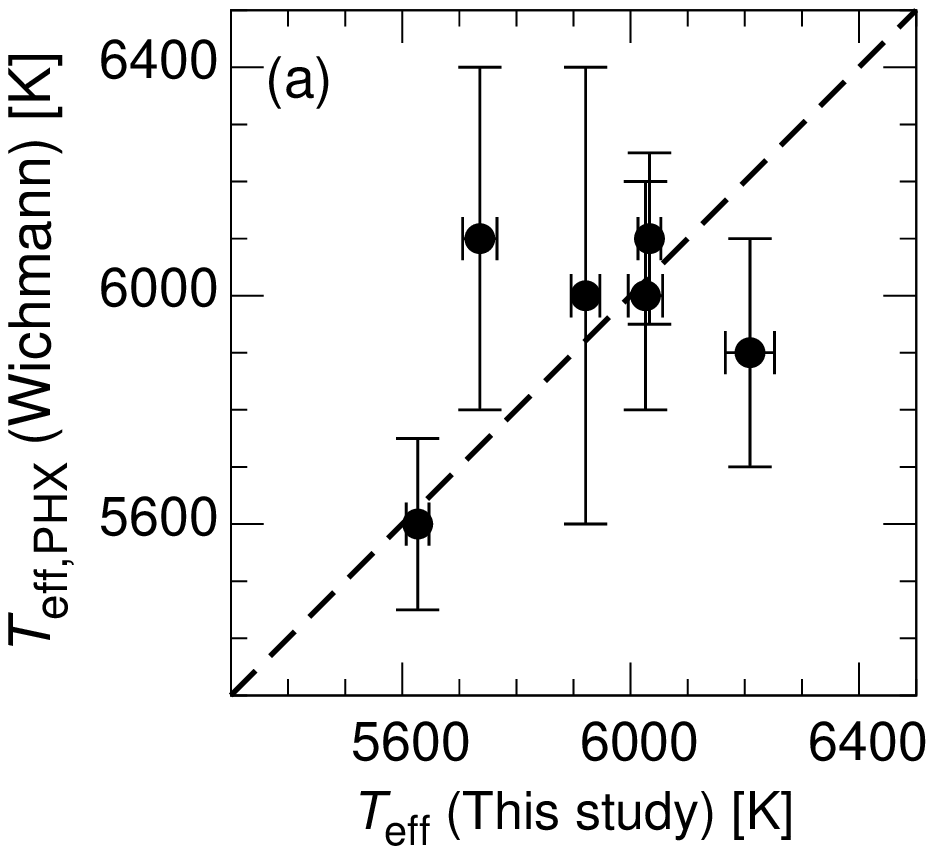}
\FigureFile(67mm,67mm){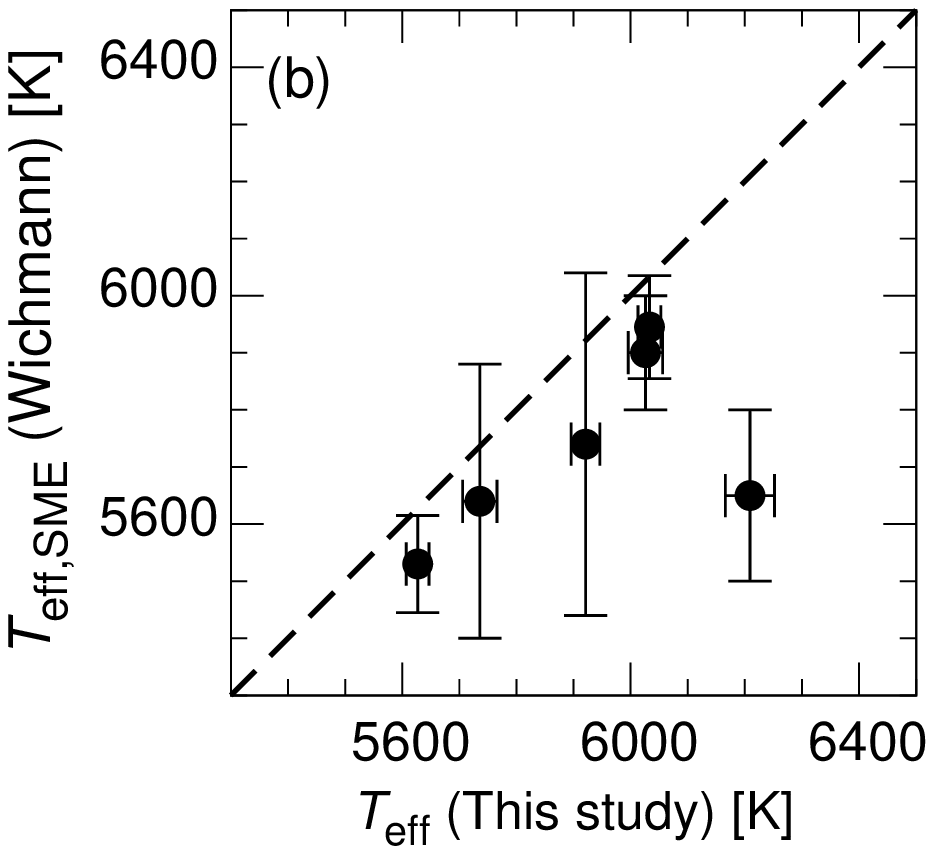}
\FigureFile(67mm,67mm){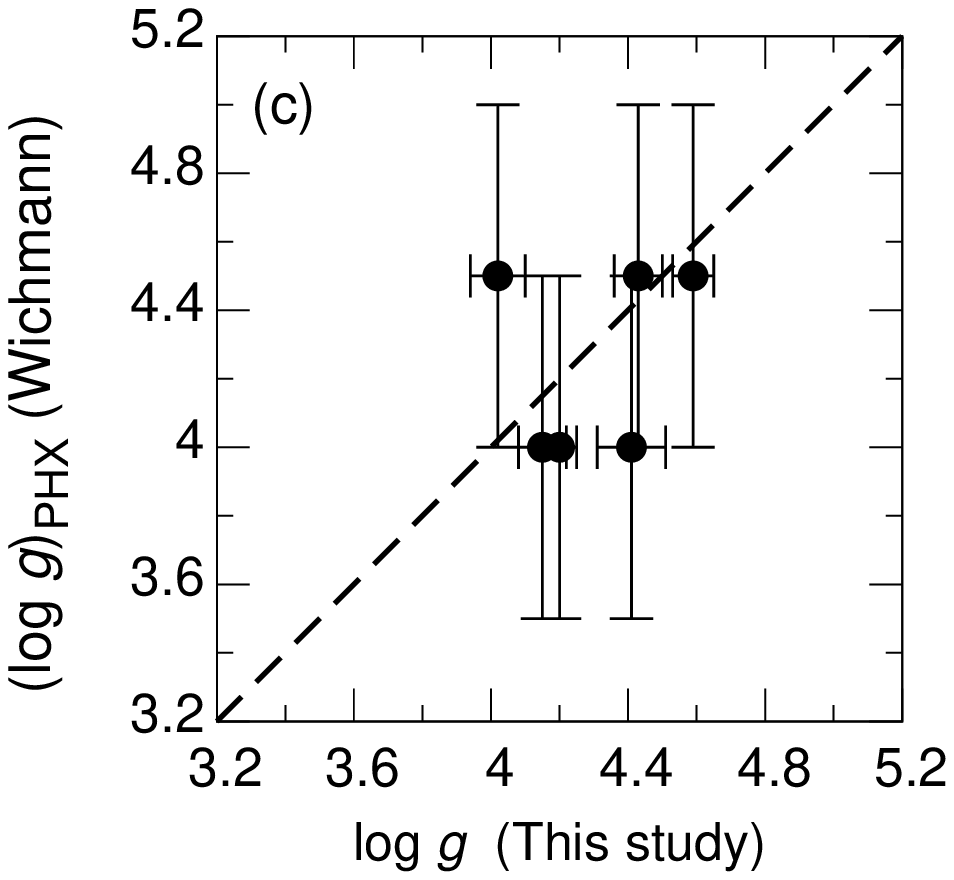}
\FigureFile(67mm,67mm){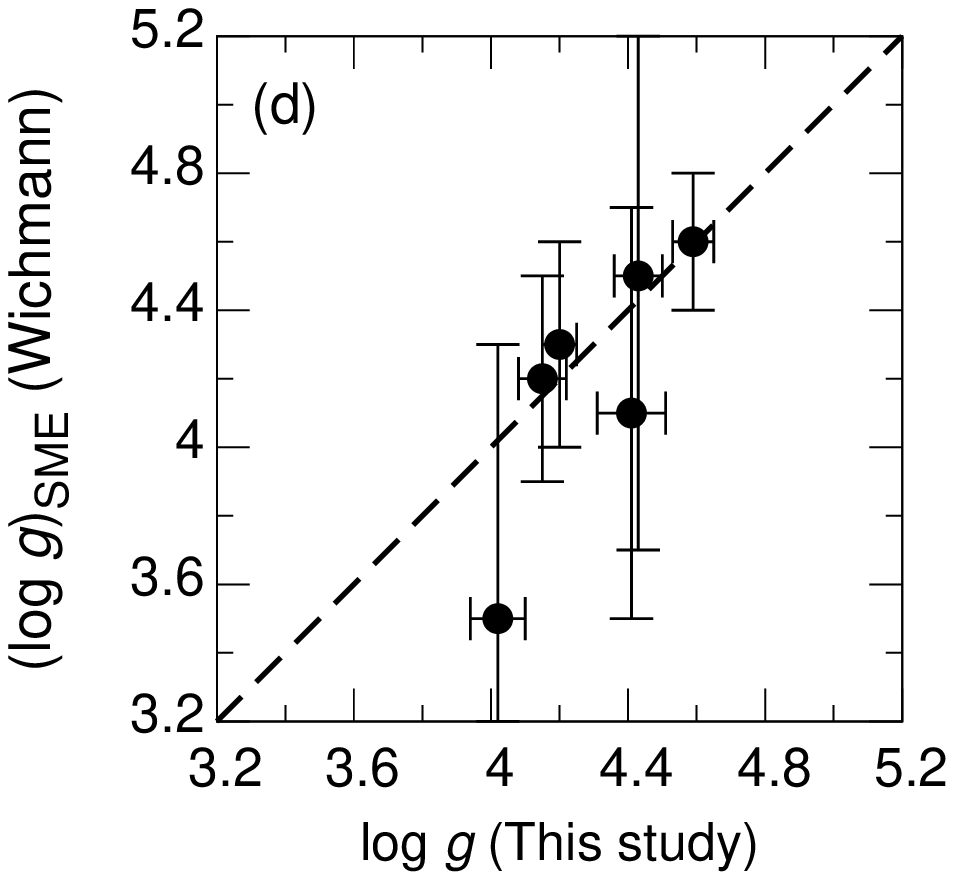}
\FigureFile(67mm,67mm){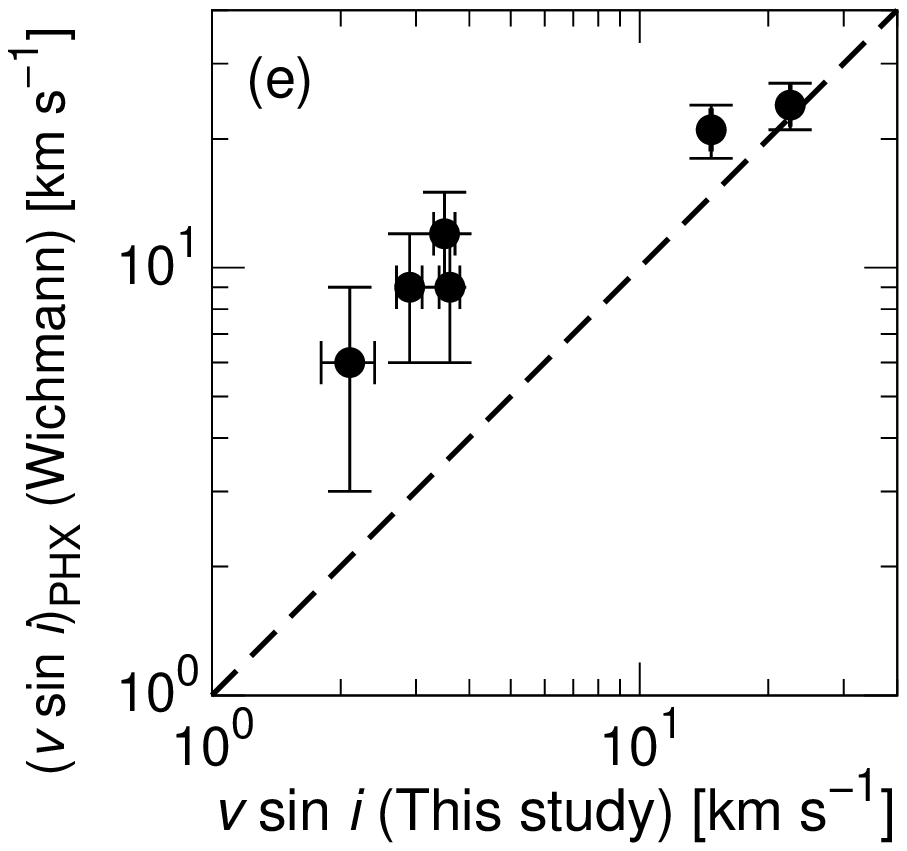}
\FigureFile(67mm,67mm){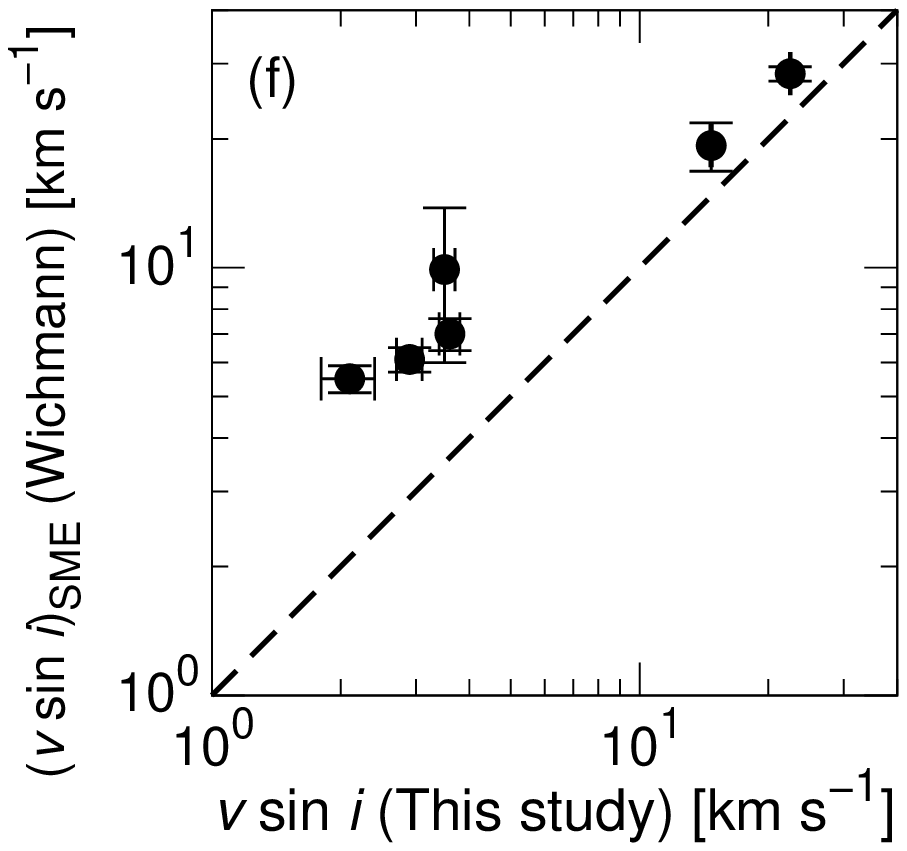}
 \end{center}
\caption{Comparison of the temperature, surface gravity, and projected rotational velocity ($T_{\rm{eff}}$, $\log g$, and $v \sin i$) in this paper
with those reported in Table 3 of \citet{Wichmann2014}. }\label{fig:hik-wichmann}
\end{figure}

In the following, we performed some analyses 
in order to check whether these spectroscopically derived values are good sources to discuss the actual properties of stars.
First, in Appendix \ref{sec:compConf-ape}, we confirmed that our resultant atmospheric parameters and measured equivalent width values of comparison stars 
are comparable to the results of previous researches.
We can say that our result of spectroscopic determination of atmospheric parameters is consistent with that of the previous studies.
In the following, we investigate whether these spectroscopically derived atmospheric parameters of the target superflare stars
are comparable to the parameters estimated from previous photometric catalogs.
\\ \\
\ \ \ \ \ \ \
We compared $T_{\rm{eff}}$, $\log g$ and [Fe/H] with those reported in Kepler Input Catalog (KIC; \cite{Brown2011}), and 
show the results in Figure \ref{fig:KIChika}.
In this Figure, spectroscopically derived values does not seem in so good agreement with KIC values.
Spectroscopic $T_{\rm{eff}}$ and [Fe/H] values tend to be a bit higher than those in KIC, while 
spectroscopic $\log g$ values seem to have much poor correlation with those in KIC.
The root-mean-square residual between the atmospheric parameter values we estimated and those in KIC 
are $(\Delta T_{\rm{eff}})_{\rm{rms}} \sim 219$K, $(\Delta \log g)_{\rm{rms}} \sim 0.37$ dex, 
and $(\Delta\rm{[Fe/H]})_{\rm{rms}} \sim 0.46$ dex, respectively.
However, this result is comparable to the large uncertainties and systematic differences in KIC parameters 
reported by the previous researches (e.g., \cite{MolendaZakowicz2010}; \cite{Brown2011}; \cite{Bruntt2012}; \cite{Thygesen2012}; \cite{Hirano2014}). 
For example, \citet{Brown2011} reported relatively large uncertainties of the temperature and surface gravity in KIC ($\pm$200K for $T_{\rm{eff}}$ and 0.4 dex for $\log g$), 
and also pointed out that the reliability of the metallicity in KIC is especially poor.
\citet{Bruntt2012} compared their spectroscopic values with KIC values, 
and reported that their spectroscopic $T_{\rm{eff}}$ and [Fe/H] values 
are systematically higher by 165K and 0.21 dex than the values in KIC, respectively.
These tendencies are also shown in our Figure \ref{fig:KIChika}(a) and (c).
Summarizing the above points, we have measured stellar atmospheric parameters more accurately compared to KIC values, 
and differences between our values and KIC values ($(\Delta T_{\rm{eff}})_{\rm{rms}} \sim 219$K, $(\Delta \log g)_{\rm{rms}} \sim 0.37$ dex, 
and $(\Delta\rm{[Fe/H]})_{\rm{rms}} \sim 0.46$ dex) are comparable to the large uncertainties and systematic differences 
in KIC values reported by the previous researches. 
\\ \\
\ \ \ \ \ \ \
We also compare our temperature value with the other catalog values for reference.
As we have already explained in detail in Section 3.1 of \citet{SNotsu2013}, 
\citet{Pinsonneault2012} reported a catalog \footnote{This catalog is available at http://vizier.cfa.harvard.edu/viz-bin/VizieR?-source=J/ApJS/199/30 .} 
of revised $T_{\mathrm{eff}}$ for stars in the Kepler Input Catalog (KIC; \cite{Brown2011}). 
They used two methods (``SDSS method" and ``IRFM method") on the basis of stellar color values, 
and derived two revised temperature values ($T_{\rm{eff,SDSS}}$ and $T_{\rm{eff,IRFM}}$), respectively. 
We should note that Pinsonneault et al. (2012) estimate $T_{\rm{eff,SDSS}}$ and $T_{\rm{eff,IRFM}}$ values 
with fixing [Fe/H] values at the mean value of KIC ([Fe/H]=$-0.2$), and the comparison here is rough discussion.  
The $T_{\rm{eff,SDSS}}$ and $T_{\rm{eff,IRFM}}$ values of the 34 single target stars are listed in Supplementary Table 4.
\citet{Pinsonneault2012} argued that these revised temperature values are both about 200K higher than the values of $T_{\mathrm{eff}}$ in the KIC.
In Figure \ref{fig:colhika}, we plotted $T_{\rm{eff,SDSS}}$ and $T_{\rm{eff,IRFM}}$ 
as a function of our temperature values estimated with our spectroscopic data ($T_{\rm{eff}}$). 
The root-mean-square residual between our temperature value and the above two revised temperature values ($T_{\rm{eff,SDSS}}$ and $T_{\rm{eff,IRFM}}$)
are $\sim$177K and $\sim$211K, respectively. 
Comparing Figure \ref{fig:KIChika} (a) and Figure \ref{fig:colhika},
the values we derived spectroscopically ($T_{\rm{eff}}$) seem to be a bit more consistent 
with these revised temperature values ($T_{\rm{eff,SDSS}}$ and $T_{\rm{eff,IRFM}}$) 
than with the temperature values in KIC ($T_{\rm{eff,KIC}}$). 
\\ \\
\ \ \ \ \ \ \
\citet{Wichmann2014} already reported atmospheric parameter values 
of 11 superflare stars on the basis of their spectroscopic data.
Among these 11 stars, 6 stars correspond to 34 single superflare stars that we discuss in this paper. 
For reference, we compare their $T_{\rm{eff}}$, $\log g$, and $v\sin i$ values with the values that we estimated in this paper, 
and show the results in Figure \ref{fig:hik-wichmann}.
The errors of the values in \citet{Wichmann2014} is large, and there are large disagreements especially for 
$v\sin i$ values.
We consider that this large error values are caused by the low S/N ratio of the observation data in \citet{Wichmann2014}\footnote{
The S/N ratio of the observation data in \citet{Wichmann2014} is 17-70 in the range of 5012-6432\AA, 
which are used by them for determining stellar parameters.
}, 
that our spectroscopically derived values are a bit more accurate compared to the values of \citet{Wichmann2014}.
In addition, in \citet{Wichmann2014}, the macroturbulence velocity ($\zeta_{\rm{RT}}$) was set to 3.0 km s$^{-1}$, 
and temperature dependence of the macroturbulence velocity, which are evaluated 
in Equation (\ref{eq:macroT}) or (\ref{eq:Doyle}) was not incorporated.
This can also possibly cause some differences between their $v\sin i$ values and our values.
Because of these things, we do not use the values in \citet{Wichmann2014} in Paper II.
\\ \\
\ \ \ \ \ \ \
On the basis of the discussions above, 
we can confirm that our spectroscopically derived values are better sources to discuss the actual properties of superflare stars compared to KIC values.
We only use the values that we derived spectroscopically in this paper when we discuss stellar properties in Paper II. 
However, we need to remember that  
$v\sin i$ values estimated in Section \ref{subsec:vsini} can be strongly affected by the way of 
estimating macroturbulence ($\zeta_{\rm{RT}}$ in Section \ref{subsec:vsini}). 
We assumed Equation (\ref{eq:macroT}) in this paper, but this equation is an only rough approximation.
The effect of errors in macroturbulence is especially large for slowly-rotating stars ($v\sin i\lesssim2-3$km s$^{-1}$),
as we can see in Figure \ref{fig:vsini-vmac}.
In this figure, the difference between the two $v \sin i$ value, which originally comes from 
the difference in the estimation method of macroturbulence velocity
is a bit larger especially in the range of $v\sin i\lesssim2-3$km s$^{-1}$.
More accurate estimation of macroturbulence is difficult (e.g., \cite{Takeda1995b}), and we think that this is beyond the scope of this paper.
Because of this, we need to remember that the uncertainty of $v\sin i$ is large especially for slowly-rotating stars ($v\sin i\lesssim2-3$km s$^{-1}$).
\\ 
\\
\\
\\
\\
\\
\bigskip

The authors sincerely thank the anonymous referee for his/her very useful and constructive comments. 
This study is based on observational data collected with Subaru Telescope, 
which is operated by the National Astronomical Observatory of Japan. 
We are grateful to Dr. Akito Tajitsu and other staffs of the Subaru Telescope 
for making large contributions in carrying out our observations. 
We would also like to thank Dr. Yoichi Takeda for his many useful advices on the analysis of
our Subaru/HDS data, and for his opening the TGVIT and SPTOOL programs into public. 
Kepler was selected as the tenth Discovery mission. 
Funding for this mission is provided by the NASA Science Mission Directorate. 
The Kepler data presented in this paper were obtained from the
Multimission Archive at STScI. 
This work was supported by the Grant-in-Aids from the Ministry of Education, 
Culture, Sports, Science and Technology of Japan (No. 25287039, 26400231, and 26800096).

\clearpage

\clearpage

\clearpage

\begin{center}
\begin{longtable}{lcccccccc}
\caption{Targets of our observations and their stellar parameters reported in the Online Data of \citet{Shibayama2013}\footnotemark[a].}\label{tab:shibay}
\hline
Starname & binary\footnotemark[b] & $T_{\rm{eff,KIC}}$\footnotemark[c] & $(\log g)_{\rm{KIC}}$\footnotemark[d] 
& $R_{\rm{KIC}}$\footnotemark[e] & $K_{p}$mag\footnotemark[f]  & $P_{\rm{S}}$\footnotemark[g]  & (BVAmp)$_{\rm{LF}}$\footnotemark[h] 
& Max $E_{\rm{flare}}$\footnotemark[i] \\ 
 &  & [K] & [cm s$^{-2}$] & [$R_{\odot}$] & [mag] & [day] & [\%] & [erg]  \\ 
\hline
\endhead
\hline
\endfoot
\hline
\multicolumn{9}{l}{\hbox to 0pt{\parbox{160mm}{\footnotesize
    \footnotemark[a] KIC7420545, KIC6934317, KIC8429280, and KIC11560431 are not reported in \citet{Shibayama2013} 
since they are not solar-type stars on the basis of KIC parameters, as mentioned in Section \ref{subsec:tarselc}. 
We newly analyzed the Kepler photometric data of these stars in the same way as in \citet{Shibayama2013}.}}}  
\\ 
\multicolumn{9}{l}{\hbox to 0pt{\parbox{160mm}{\footnotesize
    \footnotemark[b] For the details of this column, see Section\ref{subsec:binary}. 
``no" means that the star show no evidence of binary system. 
``yes (SB2)" correspond to stars that have double-lined profile.
``yes (RV)" means that the star show radial velocity changes, 
while ``yes (H$\alpha$-vari)" correspond to stars whose H$\alpha$ line profile show changes between the multiple observations. 
In addition, ``yes (VB)" means that the star has a visual companion star.}}}
\\
\multicolumn{9}{l}{\hbox to 0pt{\parbox{160mm}{\footnotesize
    \footnotemark[c] Effective temperature reported in Kepler Input Catalog (KIC; \cite{Brown2011}).}}}
\\
\multicolumn{9}{l}{\hbox to 0pt{\parbox{160mm}{\footnotesize
    \footnotemark[d] Surface gravity reported in KIC.}}}
\\
\multicolumn{9}{l}{\hbox to 0pt{\parbox{160mm}{\footnotesize
    \footnotemark[e] Stellar radius in units of solar radius, which is reported in KIC.}}}
\\
\multicolumn{9}{l}{\hbox to 0pt{\parbox{160mm}{\footnotesize
    \footnotemark[f] Kepler band magnitude in KIC.}}}    
\\
\multicolumn{9}{l}{\hbox to 0pt{\parbox{160mm}{\footnotesize
    \footnotemark[g] Period of the stellar brightness variation calculated in \citet{Shibayama2013} from the Kepler Q0$\sim$6 data.}}}     
\\
\multicolumn{9}{l}{\hbox to 0pt{\parbox{160mm}{\footnotesize
    \footnotemark[h] Amplitude of the stellar brightness variation (\%). 
The amplitude values listed here are estimated using Kepler data of the specific Quarter 
in which the largest superflare event of each star is detected. 
(The amplitude values estimated from the data of each Quarter are listed in the Supplementary Data of Paper II.)}}}     
\\
\multicolumn{9}{l}{\hbox to 0pt{\parbox{160mm}{\footnotesize
    \footnotemark[i] Energy of the largest superflare event of each star reported in \citet{Shibayama2013}.}}}
\\
\multicolumn{9}{l}{\hbox to 0pt{\parbox{160mm}{\footnotesize
    \footnotemark[j] We have already reported the results of our spectroscopic observation of KIC6934317 in \citet{SNotsu2013}, 
and those of KIC9766237 and KIC9944137 in \citet{Nogami2014}, respectively.}}}    
\endlastfoot
KIC3626094 & no & 5835 & 4.3 & 1.3 & 11.1 & 0.7 & 0.12 & $1.4 \times 10^{35}$ \\ 
KIC4742436 & no & 5628 & 4.2 & 1.5 & 10.6 & 2.3 & 0.28 & $2.0 \times 10^{35}$ \\ 
KIC4831454 & no & 5298 & 4.6 & 0.8 & 10.7 & 5.2 & 2.80 & $2.8 \times 10^{34}$ \\ 
KIC6503434 & no & 5714 & 4.3 & 1.2 & 12.6 & 3.9 & 0.21 & $7.5 \times 10^{34}$ \\ 
KIC6504503 & no & 5304 & 4.6 & 0.9 & 12.8 & 30.5 & 0.13 & $4.3 \times 10^{33}$ \\ 
KIC6865484 & no & 5688 & 4.4 & 1.1 & 13.8 & 11.2 & 0.68 & $9.9 \times 10^{34}$ \\ 
KIC6934317 \footnotemark[j] & no & 5387 & 3.8 & 2.3 & 12.3 & 2.5 \footnotemark[a] & 0.10 \footnotemark[a] & $8.4 \times 10^{35}$ \footnotemark[a] \\ 
KIC7093547 & no & 5101 & 4.3 & 1.1 & 13.8 & 14.3 & 0.13 & $1.4 \times 10^{34}$ \\ 
KIC7354508 & no & 5714 & 4.4 & 1.1 & 13.4 & 17.0 & 0.80 & $7.7 \times 10^{33}$ \\ 
KIC7420545 & no & 5083 & 3.8 & 2.2 & 10.0 & 36.2 \footnotemark[a] & 1.13 \footnotemark[a] & $3.3 \times 10^{35}$ \footnotemark[a] \\ 
KIC8359398 & no & 5123 & 4.7 & 0.7 & 14.1 & 13.5 & 1.66 & $3.1 \times 10^{34}$ \\ 
KIC8429280 & no & 4616 & 4.4 & 1.0 & 9.6 & 0.6 \footnotemark[a] & 3.52 \footnotemark[a] & $9.1 \times 10^{34}$ \footnotemark[a] \\ 
KIC8547383 & no & 5376 & 4.5 & 0.9 & 14.2 & 14.8 & 0.76 & $1.5 \times 10^{35}$ \\ 
KIC8802340 & no & 5265 & 4.9 & 0.6 & 13.0 & 10.6 & 1.95 & $6.6 \times 10^{33}$ \\ 
KIC9412514 & no & 5958 & 4.2 & 1.4 & 11.4 & 3.7 & 0.03 & $8.6 \times 10^{33}$ \\ 
KIC9459362 & no & 5357 & 4.6 & 0.8 & 14.1 & 12.3 & 1.46 & $1.0 \times 10^{35}$ \\ 
KIC9583493 & no & 5445 & 4.5 & 0.9 & 12.7 & 5.3 & 1.48 & $5.6 \times 10^{34}$ \\ 
KIC9652680 & no & 5618 & 4.8 & 0.7 & 11.2 & 1.4 & 4.54 & $4.1 \times 10^{34}$ \\ 
KIC9766237 \footnotemark[j] & no & 5674 & 4.6 & 0.9 & 13.9 & 21.8 & 0.14 & $1.1 \times 10^{34}$ \\ 
KIC9944137 \footnotemark[j] & no & 5725 & 4.6 & 0.8 & 13.8 & 25.3 & 0.05 & $9.9 \times 10^{33}$ \\ 
KIC10252382 & no & 5270 & 4.5 & 0.9 & 14.0 & 17.8 & 2.57 & $5.5 \times 10^{34}$ \\ 
KIC10387363 & no & 5291 & 4.4 & 1.0 & 14.2 & 27.4 & 0.27 & $1.6 \times 10^{34}$ \\ 
KIC10471412 & no & 5771 & 4.1 & 1.6 & 13.4 & 15.1 & 0.20 & $5.2 \times 10^{35}$ \\ 
KIC10528093 & no & 5143 & 4.5 & 0.9 & 13.6 & 12.5 & 2.56 & $1.7 \times 10^{35}$ \\ 
KIC11140181 & no & 5463 & 4.6 & 0.9 & 13.3 & 11.2 & 1.32 & $1.5 \times 10^{34}$ \\ 
KIC11197517 & no & 5162 & 4.0 & 1.7 & 12.6 & 19.1 & 0.29 & $1.1 \times 10^{34}$ \\ 
KIC11303472 & no & 5150 & 4.6 & 0.8 & 13.7 & 13.8 & 1.67 & $5.7 \times 10^{33}$ \\ 
KIC11390058 & no & 5785 & 4.3 & 1.3 & 12.6 & 12.1 & 0.47 & $1.7 \times 10^{34}$ \\ 
KIC11455711 & no & 5664 & 4.7 & 0.8 & 14.0 & 13.9 & 1.74 & $1.0 \times 10^{34}$ \\ 
KIC11494048 & no & 5929 & 4.4 & 1.1 & 13.4 & 14.9 & 0.42 & $2.8 \times 10^{34}$ \\ 
KIC11610797 & no & 5865 & 4.5 & 1.0 & 11.5 & 1.6 & 2.47 & $3.5 \times 10^{35}$ \\ 
KIC11764567 & no & 5238 & 4.4 & 1.1 & 13.2 & 22.2 & 1.90 & $1.1 \times 10^{35}$ \\ 
KIC11818740 & no & 5315 & 4.5 & 0.9 & 14.2 & 15.2 & 1.35 & $3.3 \times 10^{34}$ \\ 
KIC12266582 & no & 5434 & 4.3 & 1.1 & 13.0 & 6.8 & 1.21 & $4.0 \times 10^{34}$ \\ 
KIC7902097 & yes (RV) & 5626 & 4.6 & 0.8 & 12.3 & 3.9 & 0.88 & $1.3 \times 10^{34}$ \\ 
KIC8226464 & yes (H$\alpha$-vari) & 5754 & 4.1 & 1.7 & 11.5 & 3.1 & 1.55 & $5.3 \times 10^{35}$ \\ 
KIC11073910 & yes (H$\alpha$-vari) & 5381 & 4.6 & 0.8 & 11.7 & 2.1 & 1.12 & $3.2 \times 10^{34}$ \\ 
KIC4045215 & yes (SB2) & 5229 & 4.5 & 1.0 & 13.6 & 12.0 & 0.12 & $1.8 \times 10^{34}$ \\ 
KIC5445334 & yes (SB2) & 5137 & 4.7 & 0.7 & 12.8 & 7.7 & 0.55 & $1.6 \times 10^{34}$ \\ 
KIC7264976 & yes (SB2) & 5184 & 4.1 & 1.7 & 12.0 & 12.6 & 3.08 & $9.8 \times 10^{35}$ \\ 
KIC8479655 & yes (SB2) & 5126 & 4.6 & 0.8 & 12.8 & 19.3 & 3.09 & $3.8 \times 10^{35}$ \\ 
KIC9653110 & yes (SB2) & 5223 & 4.4 & 1.0 & 12.9 & 1.6 & 3.27 & $7.4 \times 10^{34}$ \\ 
KIC9764192 & yes (SB2) & 5551 & 4.6 & 0.8 & 12.9 & 3.5 & 2.76 & $3.2 \times 10^{34}$ \\ 
KIC9764489 & yes (SB2) & 5447 & 4.7 & 0.7 & 14.1 & 10.7 & 2.25 & $8.7 \times 10^{33}$ \\ 
KIC10120296 & yes (SB2) & 5490 & 4.4 & 1.1 & 12.9 & 3.9 & 1.83 & $1.1 \times 10^{36}$ \\ 
KIC10453475 & yes (SB2) & 5202 & 4.5 & 1.0 & 14.2 & 15.2 & 5.59 & $6.4 \times 10^{35}$ \\ 
KIC4138557 & yes (VB) & 5675 & 4.5 & 1.0 & 12.0 & 1.0 & 0.34 & $5.4 \times 10^{34}$ \\ 
KIC4750938 & yes (VB) & 5804 & 4.3 & 1.2 & 12.9 & 2.1 & 1.19 & $9.8 \times 10^{34}$ \\ 
KIC5896387 & yes (VB) & 5560 & 4.4 & 1.1 & 13.2 & 11.3 & 0.43 & $1.1 \times 10^{35}$ \\ 
KIC11560431 & yes (VB) & 5094 & 4.5 & 0.9 & 9.7 & 3.1 \footnotemark[a]  & 1.80 \footnotemark[a] & $6.1 \times 10^{34}$ \footnotemark[a] \\ 
\hline
\end{longtable}
\end{center}

\begin{table}[htbp]
  \caption{Target superflare stars that can be identified with the ROSAT X-ray source.}\label{tab:ROSAT}
  \begin{center}
    \begin{tabular}{lccc}
      \hline
       Starname & ROSAT ID & count rate & position error \\
        & & [s$^{-1}$] & [arcsec] \\
      \hline 
KIC4742436	&	1RXS J192149.3$+$395017	&	0.0122$\pm$0.0063 	&	14	\\
KIC4831454	&	1RXS J192200.0$+$395957	&	0.0166$\pm$0.0070 	&	17	\\
KIC8429280	&	1RXS J192502.2$+$442948	&	0.2423$\pm$0.0218 	&	8	\\
KIC11610797	&	1RXS J192737.8$+$493949	&	0.0186$\pm$0.0063 	&	27	\\
KIC11560431	&	1RXS J193016.8$+$493156	&	0.1660$\pm$0.0153 	&	9	\\
      \hline     
    \end{tabular}
  \end{center}
\end{table}

\begin{table}[htbp]
  \caption{Number of the single and binary stars.  \footnotemark[a]}\label{tab:NumBinary}
  \begin{center}
    \begin{tabular}{*{4}{c}}
      \hline
       -- & Single & Binary \footnotemark[b] & Sum. \\
      \hline
      $P_{0}\geq 20$day & 4 & 0 & 4 \\
      $10\leq P_{0}< 20$day & 19 & 5 & 24 \\
      $P_{0}< 10$day & 11 & 11(4) & 22 \\
       Sum. & 34 & 16(4) & 50 \\
      \hline
      \hline
      $P_{\rm{S}}\geq 20$day & 6 & 0 & 6 \\
      $10\leq P_{\rm{S}}< 20$day & 17 & 6(1) & 23 \\
      $P_{\rm{S}}< 10$day & 11 & 10(3) & 21 \\
       Sum. & 34 & 16(4) & 50 \\
      \hline      
     \multicolumn{4}{l}{\hbox to 0pt{\parbox{145mm}{\footnotesize
    \footnotemark[a] The values of $P_{0}$ and $P_{\rm{S}}$ are listed in Supplementary Table 1.}}} 
    \\  
    \multicolumn{4}{l}{\hbox to 0pt{\parbox{145mm}{\footnotesize
    \footnotemark[b] Numbers in parentheses correspond to visual binary stars.}}}
    \end{tabular}
  \end{center}
\end{table}

\begin{center}
\begin{longtable}{lccccccc}
\caption{Stellar paremeters of the superflare stars estimated from our spectroscopic results.}\label{tab:Sppara}
\\ \\ 
\hline
Starname &$ v \sin i$ & $T_{\rm{eff}}$ & $ \log g$ & $v_{\rm{t}}$ &[Fe/H]& $R_{\rm{s}}$~\footnotemark[a, b] & $P_{0}$~\footnotemark[c]  \\ 
 & [km s$^{-1}$] & [K] & [cm s$^{-2}$]& [km s$^{-1}$] & & [$R_{\odot}$] & [days] \\ 
\hline
\endhead
\hline
\endfoot
\hline
\multicolumn{8}{l}{\hbox to 0pt{\parbox{170mm}{\footnotesize
    \footnotemark[a] The resultant stellar radius ($R_{\rm{s}}$) value in this column 
is a median between the maximum and minimum values among all the possible $R_{\rm{s}}$ values selected from the isochrone data.
The error value in this column corresponds to these maximum and minimum values. 
}}}
\\
\multicolumn{8}{l}{\hbox to 0pt{\parbox{170mm}{\footnotesize
    \footnotemark[b] When we estimate $R_{\rm{s}}$ values, 
KIC8359398, KIC9766237, and KIC10252382 have no suitable isochrones within their original error range of $T_{\rm{eff}}$ and $\log g$.
For these three stars, we then took into account $2\Delta T_{\rm{eff}}$ and $2\Delta\log g$.
We must note that the resultant values of these three stars can have relatively low accuracy. 
}}}
\\
\multicolumn{8}{l}{\hbox to 0pt{\parbox{170mm}{\footnotesize
    \footnotemark[c] Period value estimated from the Kepler Quarter 2$\sim$16 data, which is explained in Section \ref{subsec:tarselc}.}}}
\\
\multicolumn{8}{l}{\hbox to 0pt{\parbox{170mm}{\footnotesize
    \footnotemark[d] We already reported the values of KIC6934317 in \citet{SNotsu2013}. }}}
\\
\multicolumn{8}{l}{\hbox to 0pt{\parbox{170mm}{\footnotesize
    \footnotemark[e] We list the values of KIC8429280 reported in \citet{Frasca2011} since the rotational velocity is so high and 
the spectral lines are too wide to estimate atmosphric parameters ($T_{\rm{eff}}$, $\log g$, and [Fe/H]) 
in our way using equivalent width of Fe I/II lines (See Section \ref{subsec:atmos} for the details).}}}
\\    
\multicolumn{8}{l}{\hbox to 0pt{\parbox{170mm}{\footnotesize
    \footnotemark[f] KIC values are listed and used for investigating $v \sin i$ and $R_{\rm{s}}$ since the rotational velocity of KIC9652680 is so high 
and the spectral lines of these stars are too wide to estimate atmosphric parameters ($T_{\rm{eff}}$, $\log g$, and [Fe/H]) 
in our way using equivalent width of Fe I/II lines (See Section \ref{subsec:atmos} for the details). 
Microturbulence velocity ($v_{\rm{t}}$) is assumed to be 1 km s$^{-1}$ when we estimated $v \sin i$.}}}
\\    
\multicolumn{8}{l}{\hbox to 0pt{\parbox{170mm}{\footnotesize
    \footnotemark[g] We already reported the values of KIC9766237 and KIC9944137 in \citet{Nogami2014}. }}}    
\endlastfoot 
KIC3626094 & $2.9\pm 0.2$ & $6026\pm 30$ & $4.15\pm 0.07$ & $1.17\pm 0.12$ & $-0.03\pm 0.03$ & $1.44\pm 0.13 $ & 0.7    \\ 
KIC4742436 & $3.6\pm 0.2$ & $6033\pm 20$ & $4.20\pm 0.05$ & $1.09\pm 0.17$ & $-0.15\pm 0.02$ & $1.31\pm 0.08 $ & 2.3    \\ 
KIC4831454 & $2.1\pm 0.3$ & $5627\pm 20$ & $4.59\pm 0.06$ & $1.13\pm 0.10$ & $0.04\pm 0.02$ & $0.88\pm 0.01 $ & 5.2    \\ 
KIC6503434 & $5.5\pm 0.1$ & $5902\pm 30$ & $3.63\pm 0.08$ & $0.96\pm 0.13$ & $-0.20\pm 0.03$ & $3.01\pm 0.13 $ & 3.9    \\ 
KIC6504503 & $1.6\pm 0.5$ & $5484\pm 20$ & $4.56\pm 0.05$ & $0.77\pm 0.16$ & $0.23\pm 0.03$ & $0.90\pm 0.02 $ & 31.8    \\ 
KIC6865484 & $2.7\pm 0.2$ & $5800\pm 28$ & $4.52\pm 0.07$ & $1.11\pm 0.13$ & $-0.12\pm 0.03$ & $0.91\pm 0.04 $ & 10.3    \\ 
KIC6934317 & $1.9\pm 0.4$ \footnotemark[d] & $5694\pm 25$ \footnotemark[d] & $4.42\pm 0.08$ \footnotemark[d] & $0.87\pm 0.14$ \footnotemark[d] & $-0.03\pm 0.07$ \footnotemark[d] & $0.99\pm 0.08 $ & 2.5    \\ 
KIC7093547 & $2.1\pm 0.4$ & $5497\pm 45$ & $4.62\pm 0.12$ & $0.18\pm 0.44$ & $0.26\pm 0.05$ & $0.91\pm 0.02 $ & 14.2    \\ 
KIC7354508 & $3.2\pm 0.2$ & $5620\pm 25$ & $4.09\pm 0.06$ & $0.92\pm 0.11$ & $-0.10\pm 0.03$ & $1.48\pm 0.02 $ & 16.8    \\ 
KIC7420545 & $6.3\pm 0.2$ & $5355\pm 38$ & $3.56\pm 0.12$ & $1.30\pm 0.14$ & $-0.03\pm 0.04$ & $3.45\pm 0.58 $ & 36.2    \\ 
KIC8359398 & $1.8\pm 0.4$ & $5214\pm 45$ & $4.85\pm 0.13$ & $0.75\pm 0.40$ & $-0.13\pm 0.05$ & $0.73\pm 0.02 $ \footnotemark[b]  & 12.7    \\ 
KIC8429280 & $37.1\pm 3$ \footnotemark[e] & $5055\pm 135$ \footnotemark[e] & $4.41\pm 0.25$ \footnotemark[e] & $-$ & $-0.02\pm 0.10$ \footnotemark[e] & $0.76\pm 0.06 $ & 1.2    \\ 
KIC8547383 & $4.4\pm 0.2$ & $5606\pm 108$ & $4.38\pm 0.30$ & $0.83\pm 0.35$ & $-0.02\pm 0.11$ & $1.16\pm 0.33 $ & 14.8    \\ 
KIC8802340 & $4.3\pm 0.2$ & $5529\pm 30$ & $4.43\pm 0.09$ & $1.25\pm 0.18$ & $-0.13\pm 0.04$ & $0.89\pm 0.05 $ & 10.3    \\ 
KIC9412514 & $8.1\pm 0.1$ & $6075\pm 45$ & $3.72\pm 0.10$ & $1.55\pm 0.16$ & $0.10\pm 0.04$ & $2.85\pm 0.38 $ & 3.7    \\ 
KIC9459362 & $7.0\pm 0.1$ & $5458\pm 78$ & $3.82\pm 0.23$ & $1.39\pm 0.47$ & $-0.78\pm 0.10$ & $2.31\pm 0.28 $ & 12.6    \\ 
KIC9583493 & $7.2\pm 0.1$ & $5605\pm 33$ & $4.34\pm 0.09$ & $1.28\pm 0.18$ & $-0.12\pm 0.04$ & $0.98\pm 0.04 $ & 5.5    \\ 
KIC9652680 & $34.2\pm 0.1$ \footnotemark[f] & $5618\pm 200$ \footnotemark[f] & $4.80\pm 0.40$ \footnotemark[f] & 1 \footnotemark[f] & $-0.30$ \footnotemark[f] & $0.84\pm 0.12 $ & 1.5    \\ 
KIC9766237 & $2.1\pm 0.4$ \footnotemark[g] & $5606\pm 40$ \footnotemark[g] & $4.25\pm 0.11$ \footnotemark[g] & $0.88\pm 0.17$ \footnotemark[g] & $-0.16\pm 0.04$ \footnotemark[g] & $1.19\pm 0.31 $ \footnotemark[b]  & 14.2    \\ 
KIC9944137 & $1.9\pm 0.4$ \footnotemark[g] & $5666\pm 35$ \footnotemark[g] & $4.46\pm 0.09$ \footnotemark[g] & $0.93\pm 0.13$ \footnotemark[g] & $-0.10\pm 0.03$ \footnotemark[g] & $0.93\pm 0.09 $ & 12.6    \\ 
KIC10252382 & $5.5\pm 0.2$ & $5334\pm 50$ & $4.01\pm 0.15$ & $1.40\pm 0.42$ & $-0.61\pm 0.08$ & $2.04\pm 0.11 $ \footnotemark[b]  & 16.8    \\ 
KIC10387363 & $1.5\pm 0.6$ & $5381\pm 78$ & $4.75\pm 0.23$ & $0.65\pm 0.39$ & $0.07\pm 0.06$ & $0.84\pm 0.03 $ & 29.9    \\ 
KIC10471412 & $2.7\pm 0.2$ & $5776\pm 20$ & $4.53\pm 0.05$ & $0.91\pm 0.15$ & $0.15\pm 0.03$ & $0.97\pm 0.02 $ & 15.2    \\ 
KIC10528093 & $11.4\pm 0.1$ & $5180\pm 78$ & $3.96\pm 0.25$ & $0.71\pm 0.44$ & $-0.14\pm 0.11$ & $2.06\pm 0.30 $ & 12.2    \\ 
KIC11140181 & $3.6\pm 0.2$ & $5552\pm 48$ & $4.60\pm 0.13$ & $1.25\pm 0.16$ & $-0.09\pm 0.04$ & $0.85\pm 0.05 $ & 11.5    \\ 
KIC11197517 & $0.9\pm 0.6$ & $5284\pm 40$ & $4.59\pm 0.11$ & $0.62\pm 0.35$ & $0.42\pm 0.05$ & $0.90\pm 0.04 $ & 14.3    \\ 
KIC11303472 & $2.1\pm 0.5$ & $5193\pm 20$ & $4.57\pm 0.06$ & $0.93\pm 0.16$ & $-0.14\pm 0.03$ & $0.76\pm 0.04 $ & 13.5    \\ 
KIC11390058 & $3.5\pm 0.2$ & $5921\pm 25$ & $4.43\pm 0.07$ & $1.11\pm 0.13$ & $-0.15\pm 0.03$ & $1.00\pm 0.08 $ & 12.0    \\ 
KIC11455711 & $7.9\pm 0.1$ & $5631\pm 43$ & $3.85\pm 0.12$ & $1.16\pm 0.14$ & $-0.23\pm 0.05$ & $2.09\pm 0.38 $ & 13.9    \\ 
KIC11494048 & $3.4\pm 0.2$ & $5907\pm 25$ & $4.49\pm 0.07$ & $1.03\pm 0.17$ & $0.01\pm 0.03$ & $0.98\pm 0.05 $ & 14.8    \\ 
KIC11610797 & $22.5\pm 0.1$ & $6209\pm 43$ & $4.41\pm 0.10$ & $1.70\pm 0.20$ & $0.26\pm 0.05$ & $1.23\pm 0.07 $ & 1.6    \\ 
KIC11764567 & $14.7\pm 0.1$ & $5736\pm 30$ & $4.02\pm 0.08$ & $1.71\pm 0.26$ & $-0.08\pm 0.05$ & $1.60\pm 0.16 $ & 22.4    \\ 
KIC11818740 & $3.0\pm 0.3$ & $5500\pm 93$ & $4.84\pm 0.28$ & $0.80\pm 0.47$ & $0.00\pm 0.09$ & $0.83\pm 0.02 $ & 15.4    \\ 
KIC12266582 & $4.8\pm 0.2$ & $5576\pm 20$ & $4.61\pm 0.05$ & $1.20\pm 0.12$ & $-0.06\pm 0.03$ & $0.83\pm 0.01 $ & 6.9    \\ 
\hline
\end{longtable}
\end{center}

\clearpage
\appendix
\section{Comparison of period values}\label{sec:periodhik}
\begin{figure}[htbp]
 \begin{center}
  \FigureFile(75mm,75mm){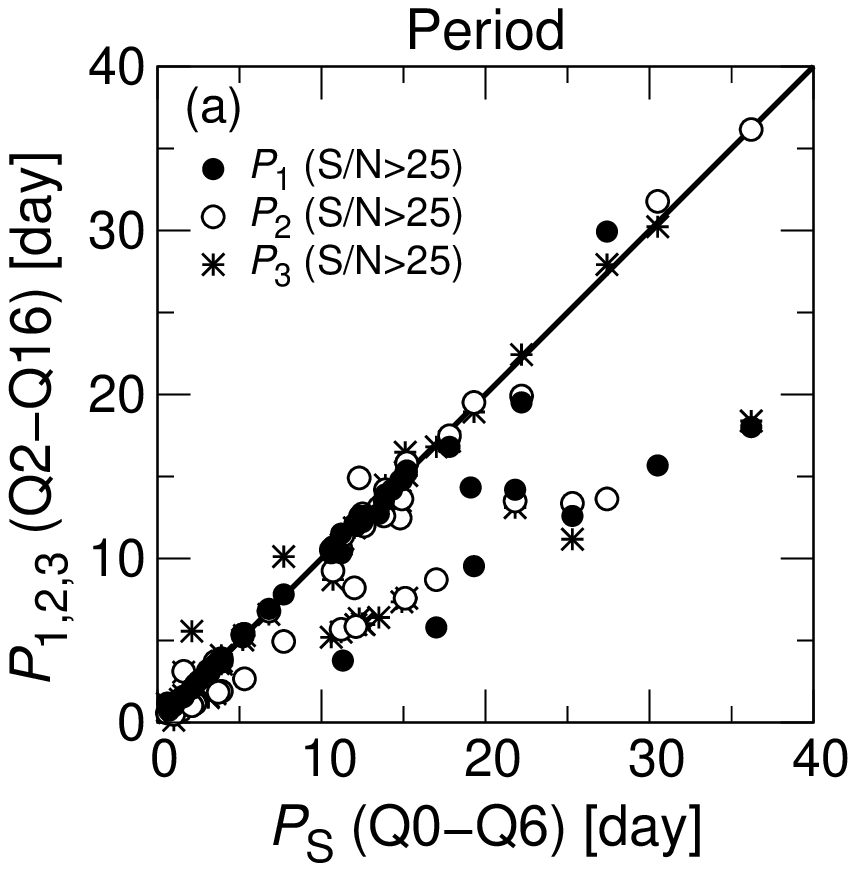} 
  \FigureFile(75mm,75mm){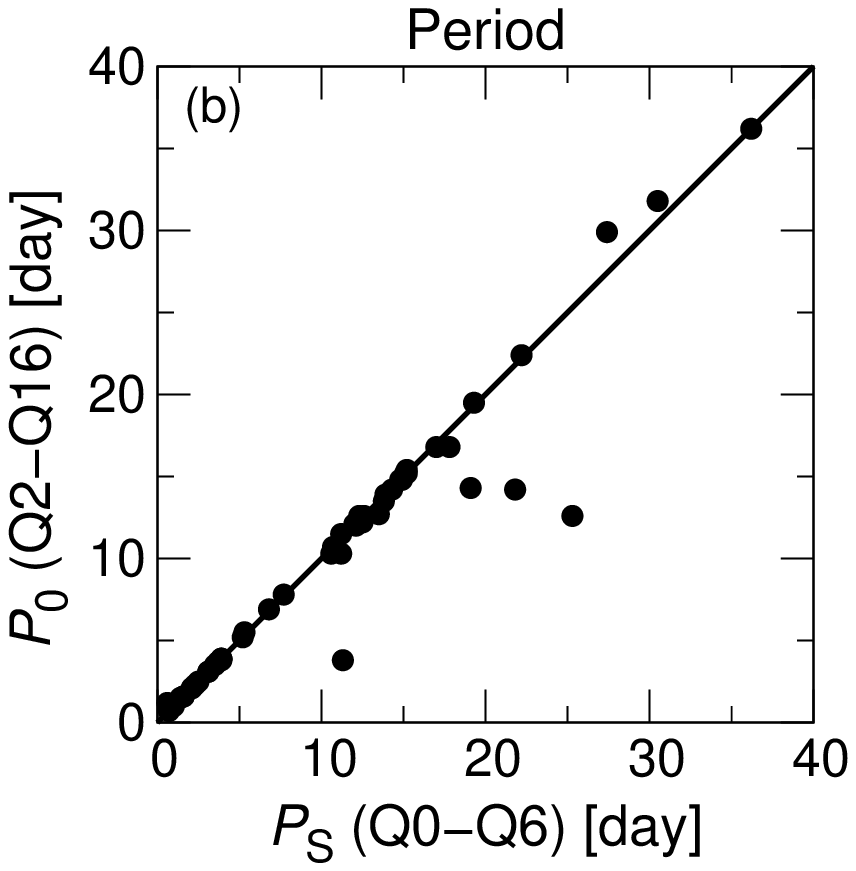} 
 \end{center}
\caption{
(a) Comparison of the period value we newly estimated using Kepler Quarter 2$\sim$16 data ($P_{1}$) 
with that in \citet{Shibayama2013} using Kepler Quarter 0$\sim$6 data ($P_{\rm{S}}$). 
In addition to $P_{1}$, the $P_{2}$ and $P_{3}$ values, 
which are the second and third peak of the power spectra of each star, are also plotted as a function of $P_{\rm{S}}$.  
(b) Comparison of the resultant period value $P_{0}$ selected from $P_{1}$, $P_{2}$ and $P_{3}$
with that in \citet{Shibayama2013} using Kepler Quarter 0$\sim$6 data ($P_{\rm{S}}$).
}\label{fig:periodQ17}
\end{figure}

\begin{figure}[htbp]
 \begin{center}
  \FigureFile(75mm,75mm){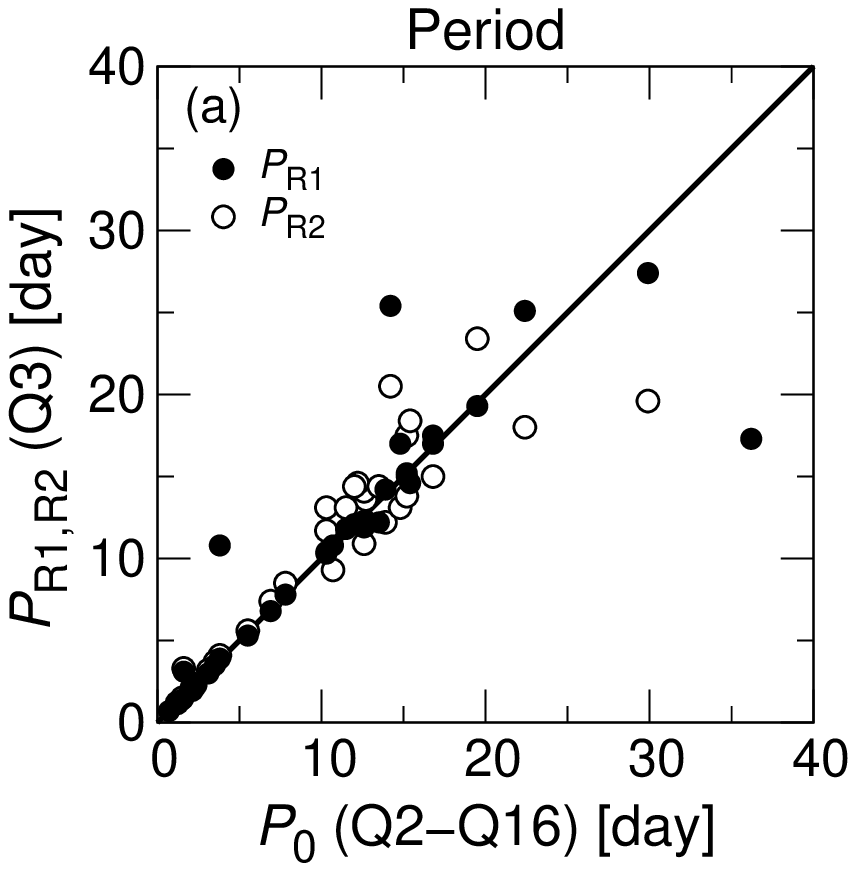}
  \FigureFile(75mm,75mm){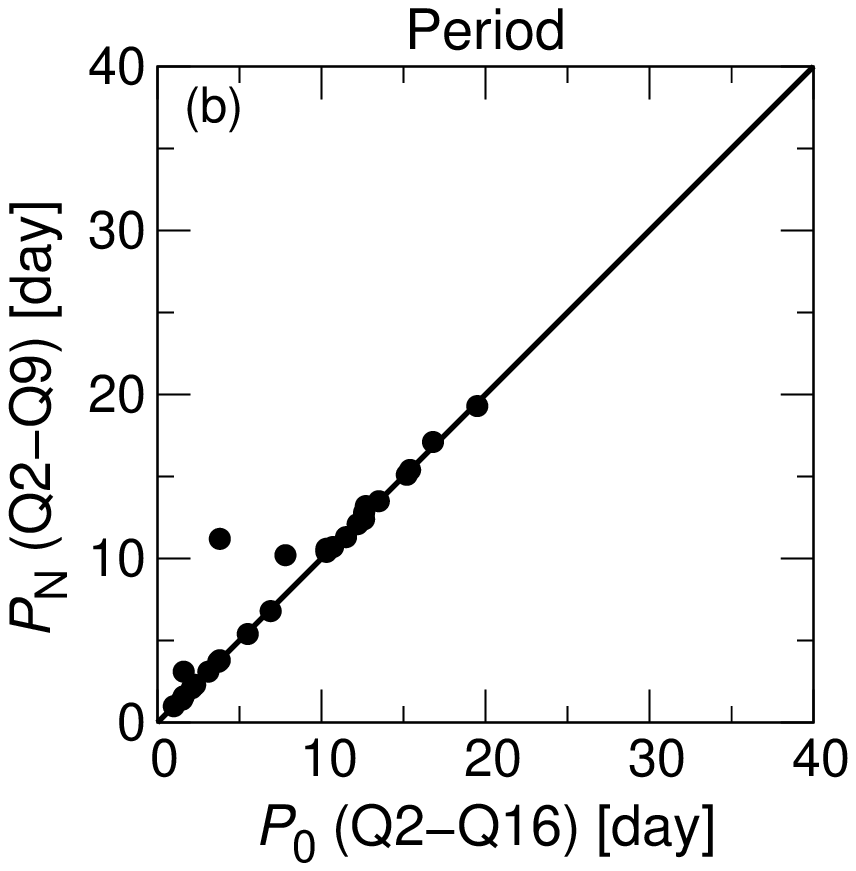}
  \FigureFile(75mm,75mm){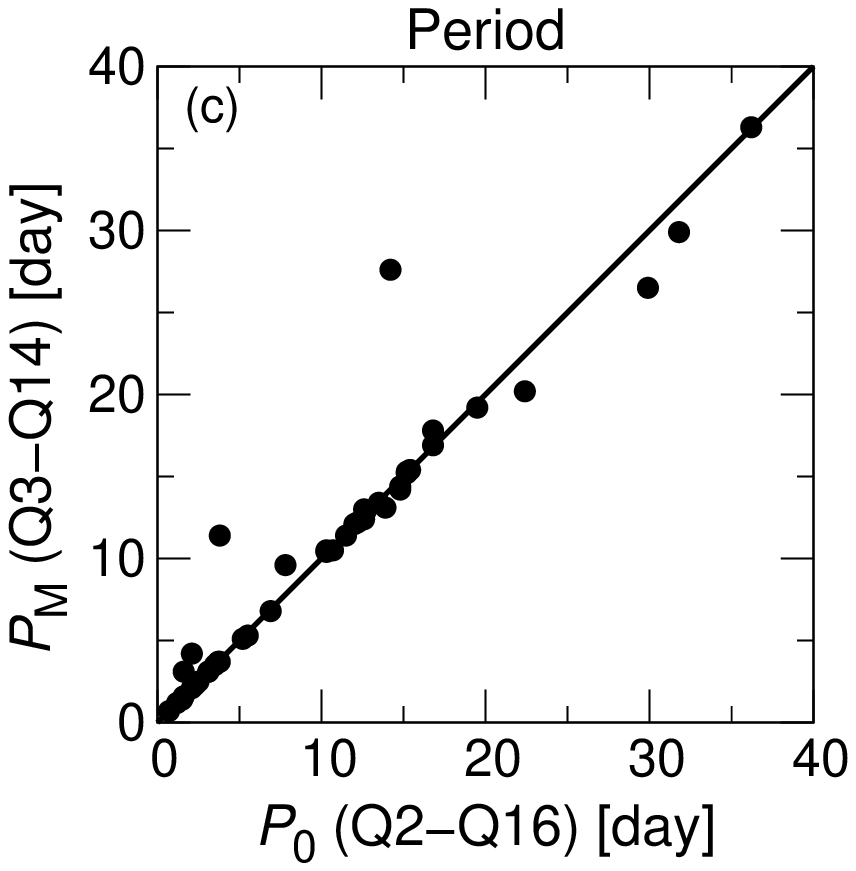}
 \end{center}
\caption{
(a) Comparison of our resultant period value $P_{0}$ estimated using Kepler Quarter 2$\sim$16 data 
with the value estimated in \citet{Reinhold2013} 
using Kepler Quarter 3 data ($P_{\rm{R1}}$ and $P_{\rm{R2}}$).
(b) Comparison of our resultant period value $P_{0}$ with the value estimated in \citet{Nielsen2013} 
using Kepler Quarter 2-9 data ($P_{\rm{N}}$).
(c) Comparison of our resultant period value $P_{0}$ with the value estimated in \citet{McQuillan2014} 
using Kepler Quarter 3-14 data ($P_{\rm{M}}$).
}\label{fig:period-others}
\end{figure}

Two period values ($P_{S}$ and $P_{1}$), which are explained in Section \ref{subsec:tarselc}, are listed in Supplementary Table 1.
We also listed other period values $P_{2}$ and $P_{3}$, which correspond to the second and third peak of the power spectra used for 
estimating $P_{1}$, respectively. 
We compare the above two period value ($P_{\rm{S}}$ and $P_{1}$) in Figure \ref{fig:periodQ17} (a). 
Figure \ref{fig:periodQ17} (a) shows that $P_{1}$ is not so well consistent with $P_{\rm{S}}$ for longer period values ($P>15$ days),
but, in such cases, $P_{2}$ or $P_{3}$ are well consistent with $P_{\rm{S}}$.
We can also see two clear branches in the plotted distribution of the data points in Figure \ref{fig:periodQ17} (a).
These two branches correspond to $P_{\rm{S}}\approx P_{1,2,3}$ and $P_{\rm{S}}\approx (1/2)P_{1,2,3}$, respectively.
The latter branch implies that half periods are sometimes selected 
when we use Kepler data (Q2$\sim$16 data; $\sim$1500 days) longer 
than those in \citet{Shibayama2013} (Q0$\sim$6 data; $\sim$500 days).
We compared these period values, and checked by eye the lightcurves and power spectra shown in Supplementary Figure 1.
We then selected the resultant period value $P_{0}$ from the above three period values ($P_{1}$, $P_{2}$ and $P_{3}$)
by checking the lightcurve and power spectrum of each star by eye.
The selection criteria are described in Section \ref{subsec:tarselc}.
In many cases, we adopt $P_{1}$ as $P_{0}$.
The resultant $P_{0}$ values are listed in Supplementary Table 1 and plotted in Figure \ref{fig:periodQ17} (b). 
In Paper II, we use only $P_{0}$ as the period of the brightness variation, as explained in Section \ref{subsec:tarselc}. 
\\ \\
\ \ \ \ \ \ \
We also listed other period values ($P_{\rm{R1}}$, $P_{\rm{R2}}$, $P_{\rm{N}}$, and $P_{\rm{M}}$) in Supplementary Table 1 for comparison.
$P_{\rm{R1}}$ and $P_{\rm{R2}}$ values are reported by \citet{Reinhold2013}, 
and they only used Kepler Quarter 3 data ($\sim$90 days) for estimating their period values. 
According to \citet{Reinhold2013}, $P_{\rm{R1}}$ correspond to their ``primary" period value,
 while $P_{\rm{R2}}$ are their ``second" period value, which they used for investigating the surface differential rotation.
$P_{\rm{N}}$ values are reported in \citet{Nielsen2013}, 
and they are derived from Kepler Quarter 2-9 data.
$P_{\rm{M}}$ values are reported in \citet{McQuillan2014}, 
and they are derived from Kepler Quarter 3-14 data.
We compare these period values ($P_{\rm{R1, R2}}$, $P_{\rm{N}}$, and $P_{\rm{M}}$) with our resultant values ($P_{0}$) 
in Figures \ref{fig:period-others} (a), (b), and (c), respectively. 
These figures show that our resultant period values ($P_{0}$) and the above other period values ($P_{\rm{R1, R2}}$, $P_{\rm{N}}$, and $P_{\rm{M}}$) 
are consistent for most stars. 

\clearpage

\section{Lightcurves around the observational dates of S11B and S12B}\label{sec:obslc} 
We show the Kepler lightcurves obtained around the period of S11B and S12B observation in Figure \ref{fig:lcS12B}. 
There are no lightcurve data for S13A observation since Kepler ended its general observation mode on 2013 May \citep{Thompson2013d}.
No flare events were detected during the observation period.

\begin{figure}[htbp]
 \begin{center}
 \FigureFile(65mm,65mm){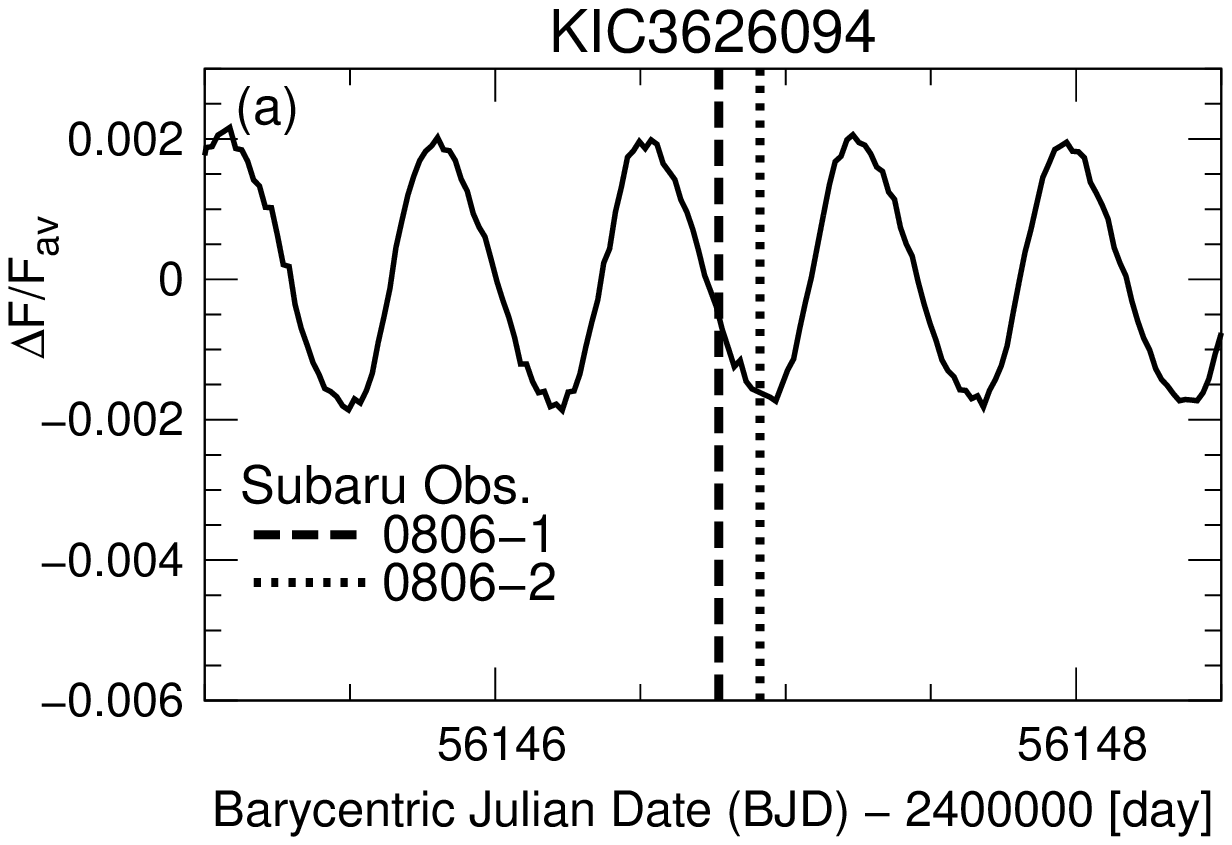}
 \FigureFile(65mm,65mm){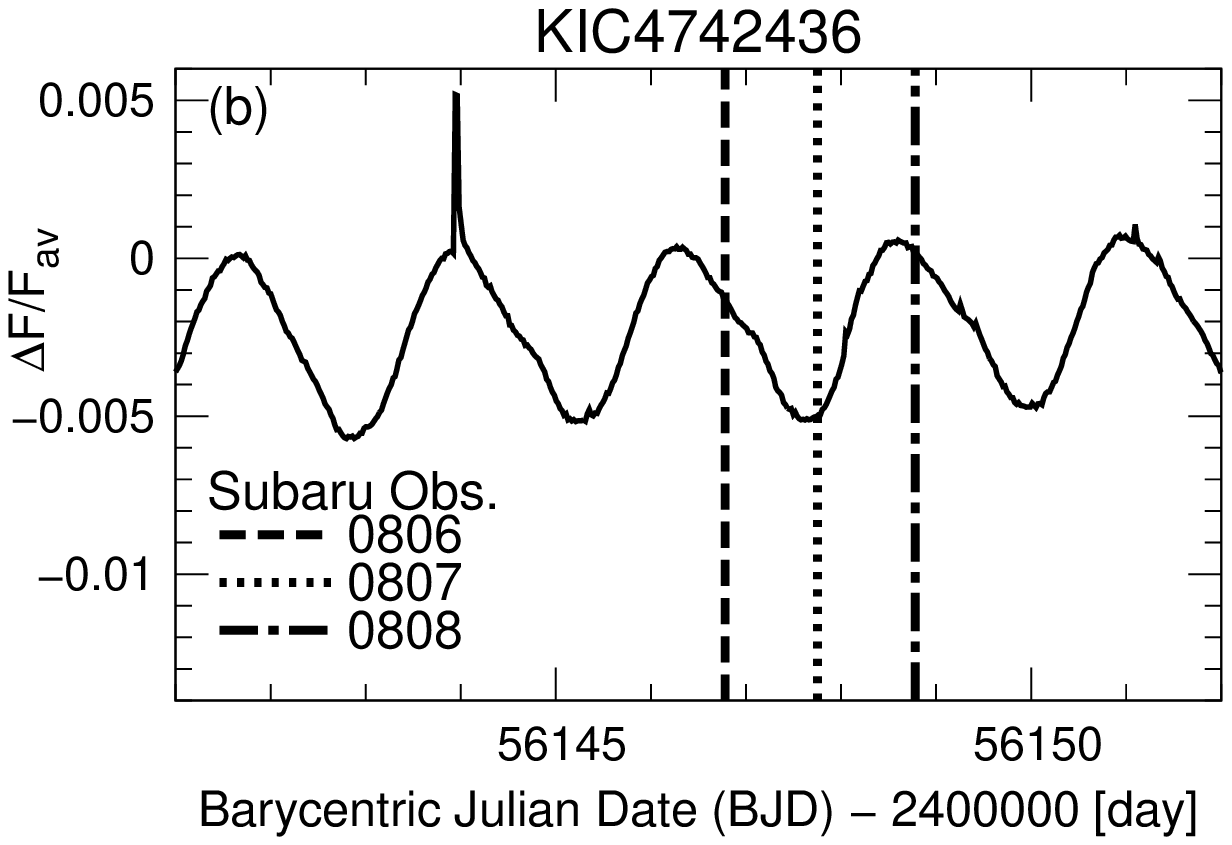} 
 \FigureFile(65mm,65mm){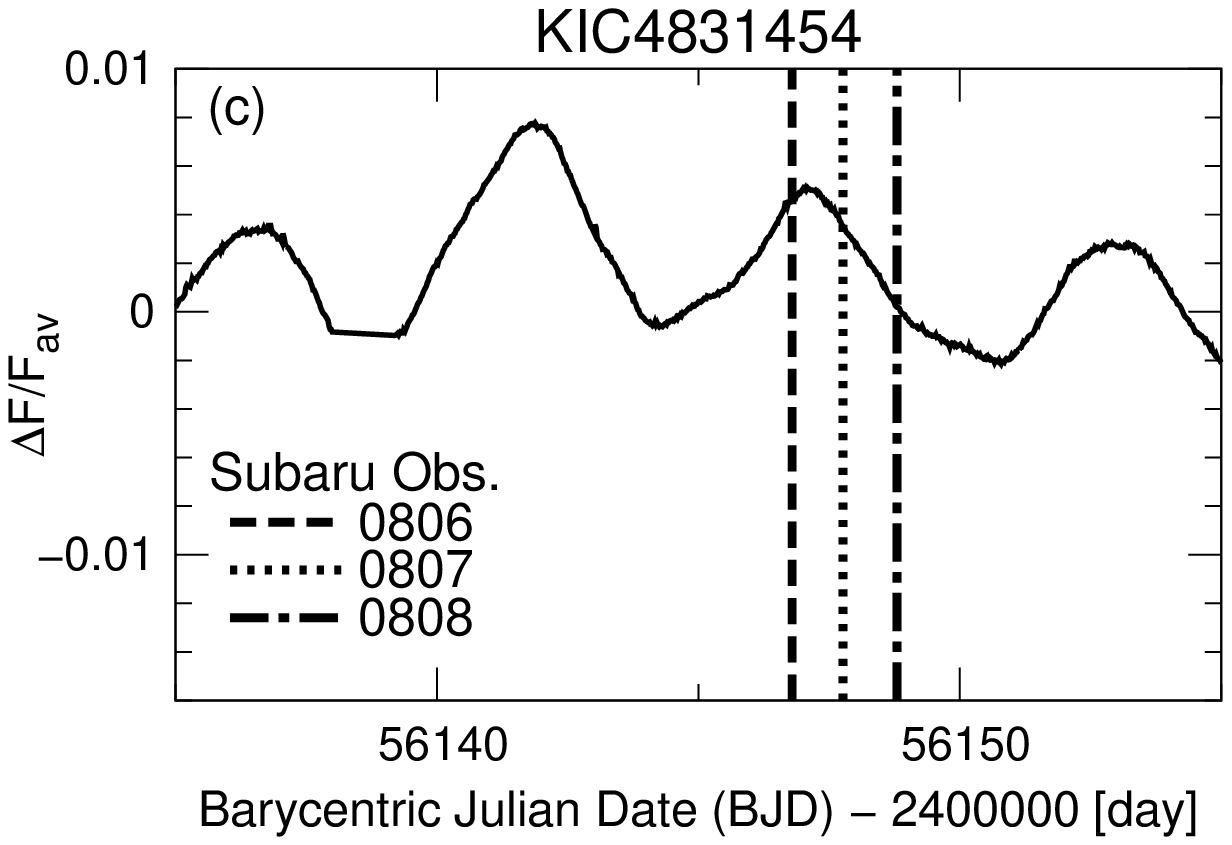}
 \FigureFile(65mm,65mm){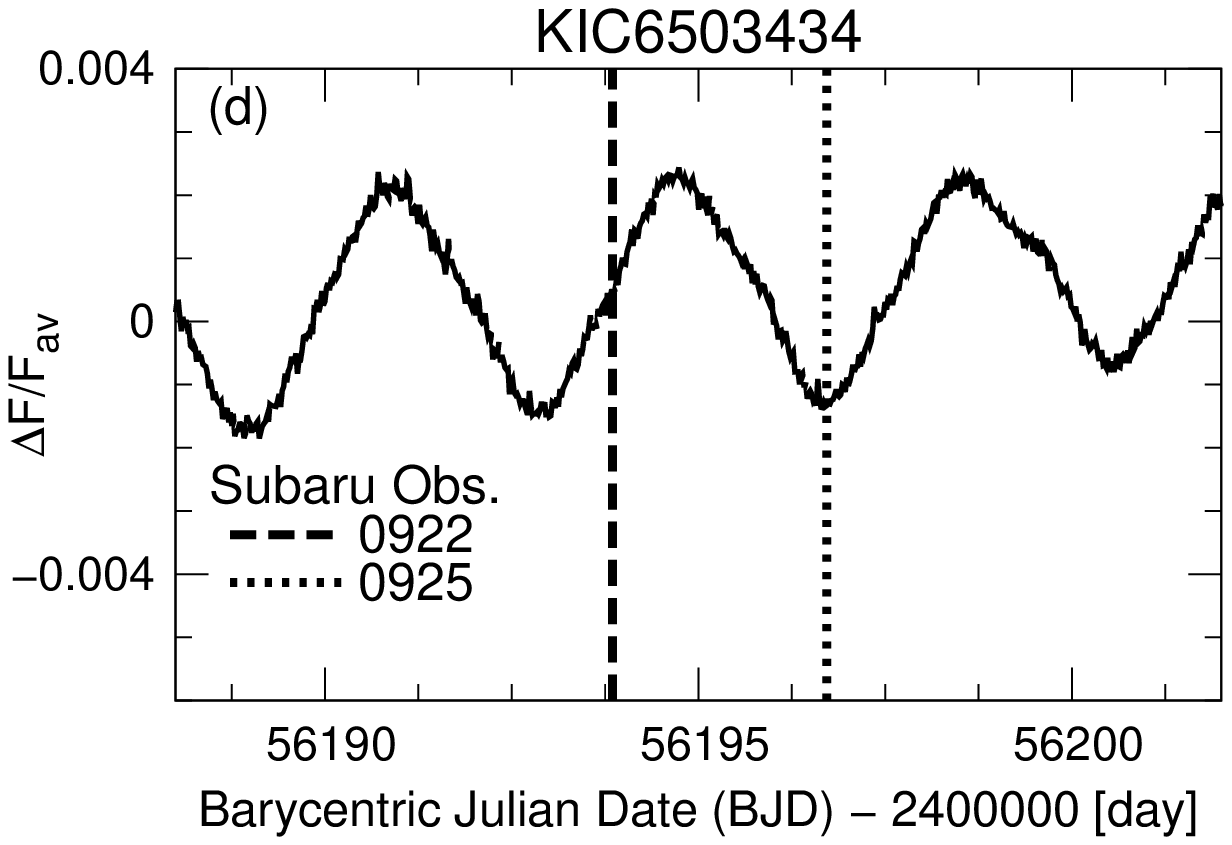} 
 \FigureFile(65mm,65mm){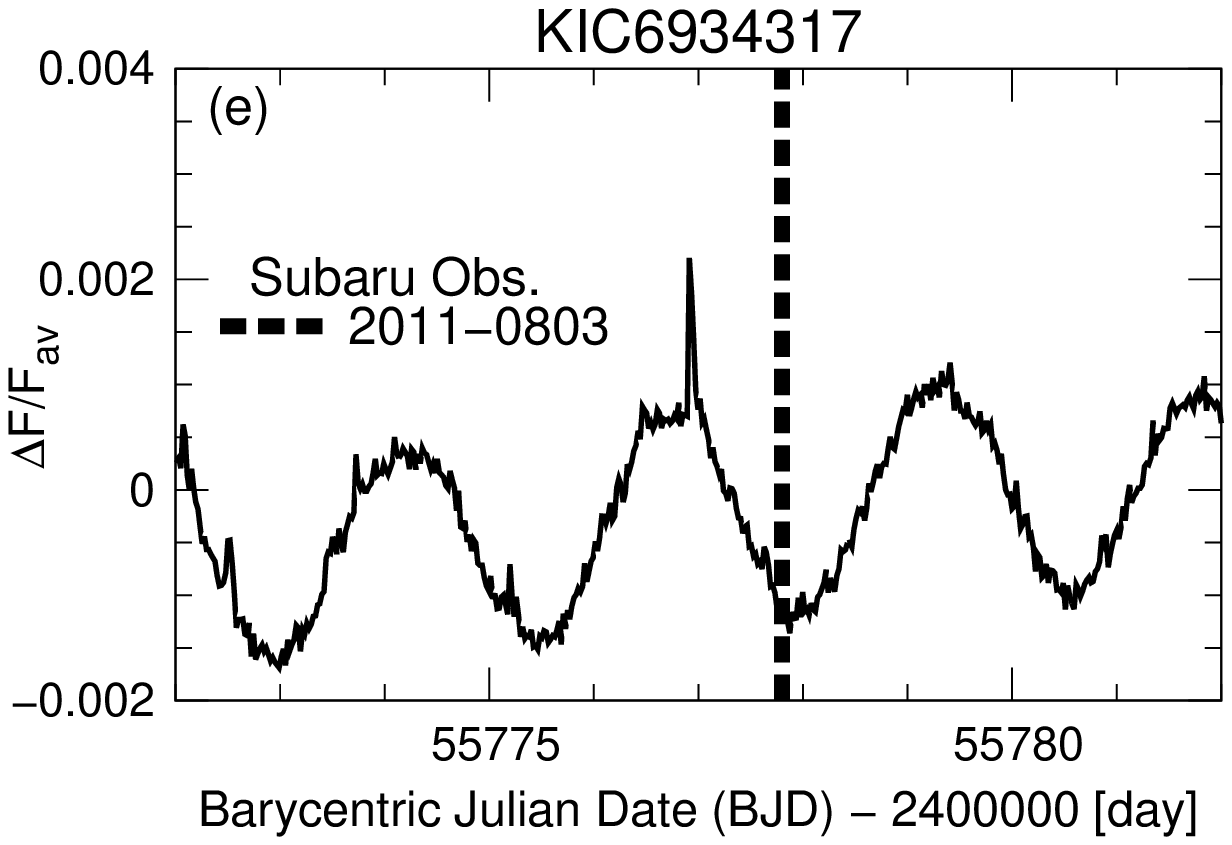}
 \FigureFile(65mm,65mm){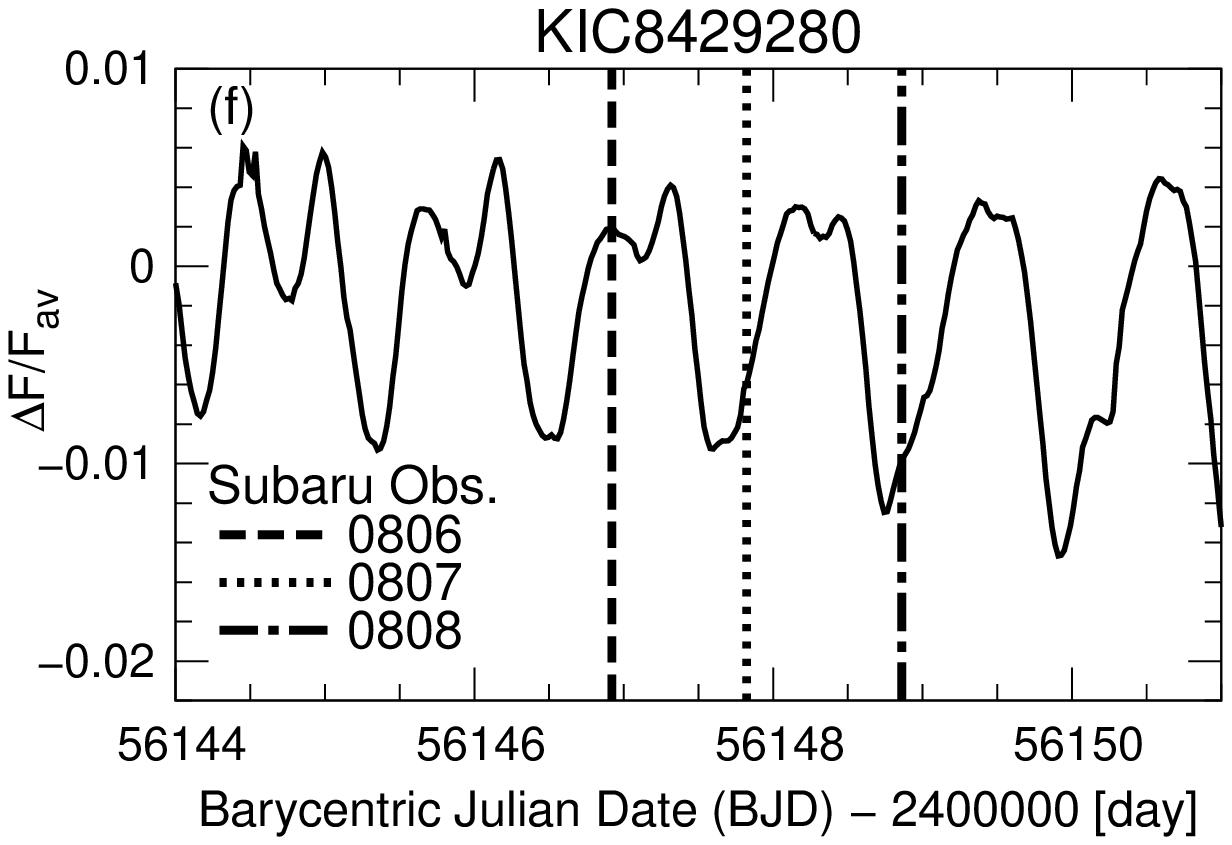}
 \FigureFile(65mm,65mm){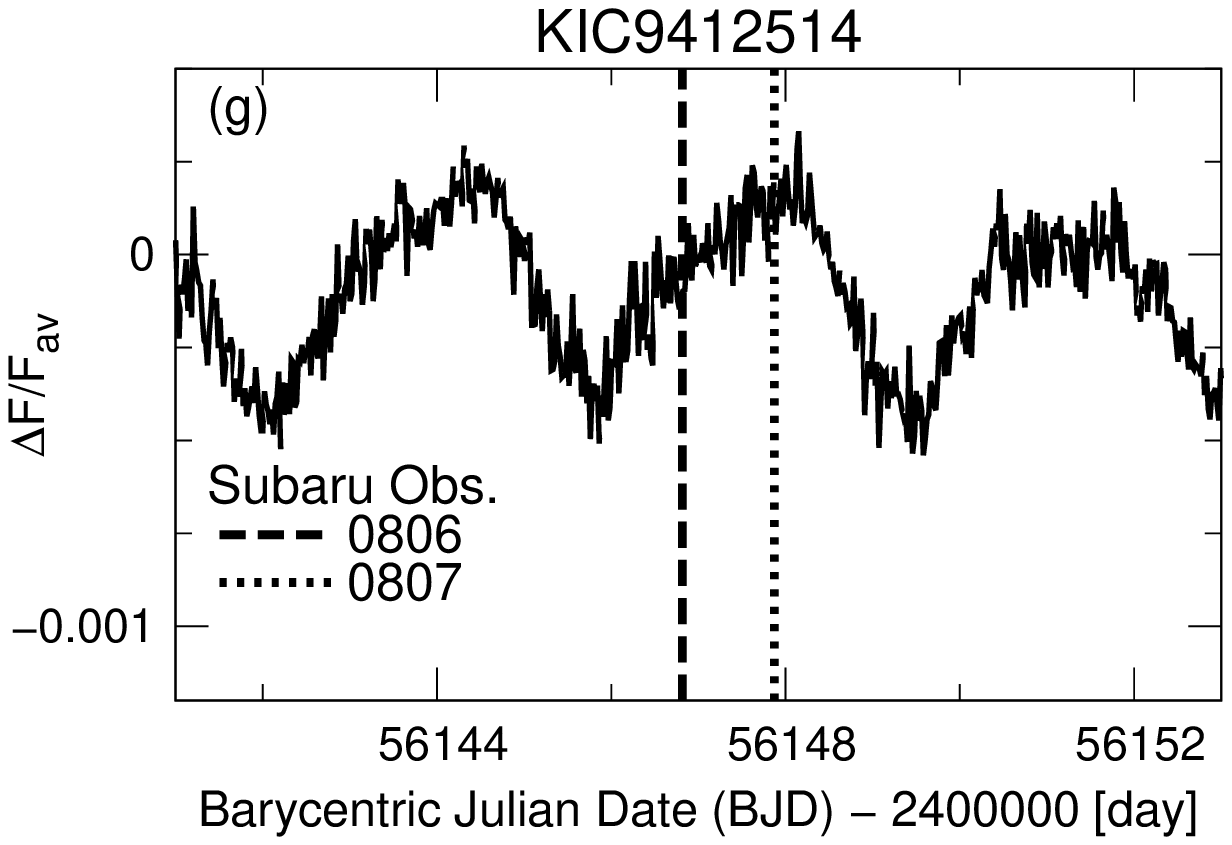}
\end{center}
\caption{
Light curves of the target superflare stars around the observational dates of S11B (2011 August 3) and S12B  (2012 August 6, 7, 8, and September 22, 23, 24, 25). 
Horizontal axes show Barycentric Julian Date, and vertical axes correspond to
stellar brightness normalized by the average one ($F_{\rm{av}}$). 
Vertical dashed and dash-dotted lines show the observational date, respectively.}\label{fig:lcS12B}
\end{figure}
 
 \addtocounter{figure}{-1}
\begin{figure}[htbp]
 \begin{center} 
 \FigureFile(65mm,65mm){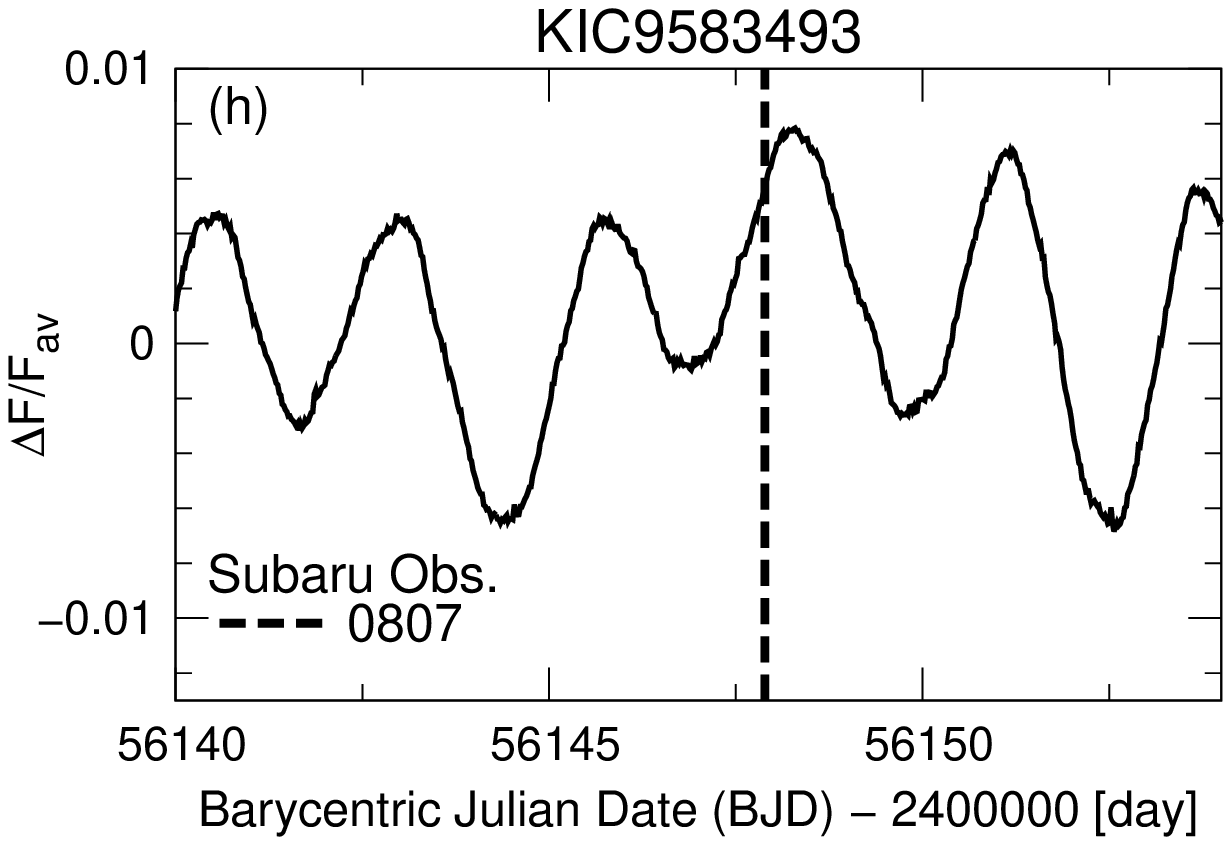}
 \FigureFile(65mm,65mm){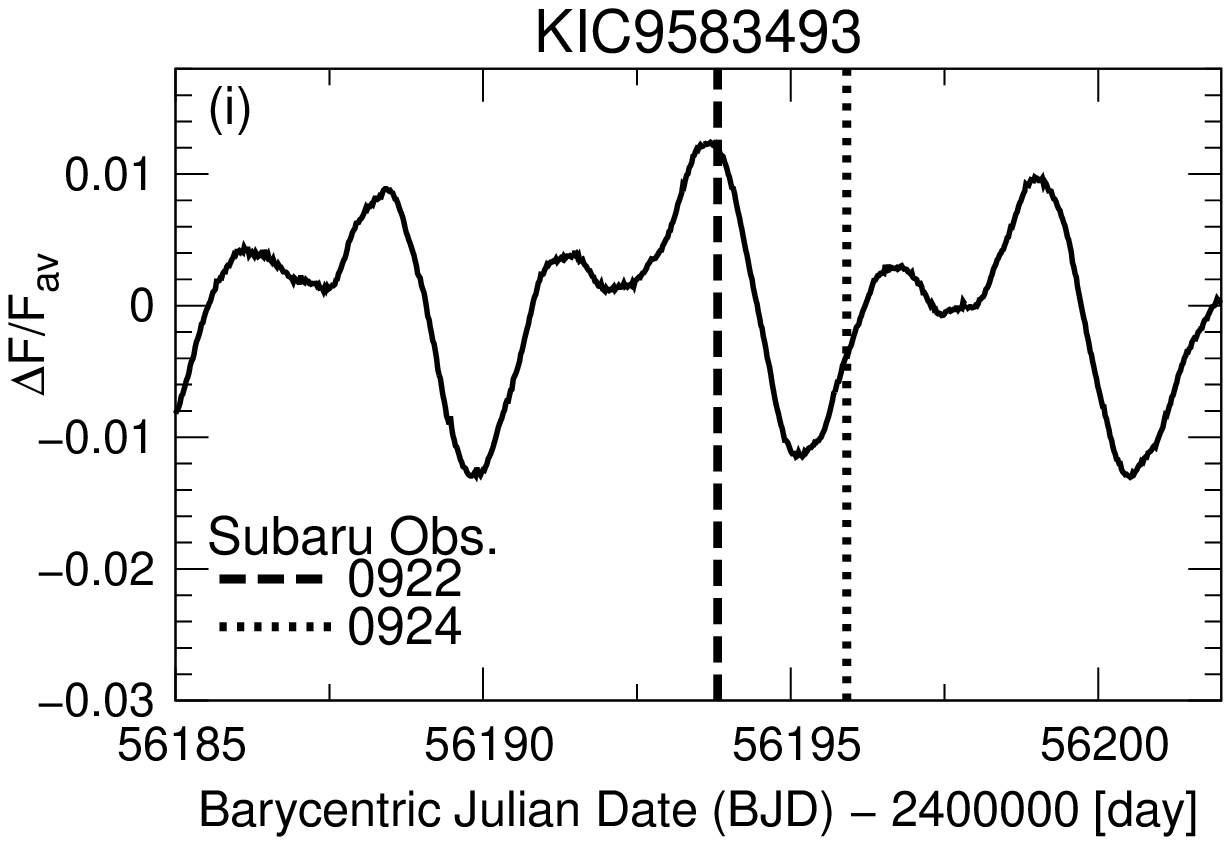}
 \FigureFile(65mm,65mm){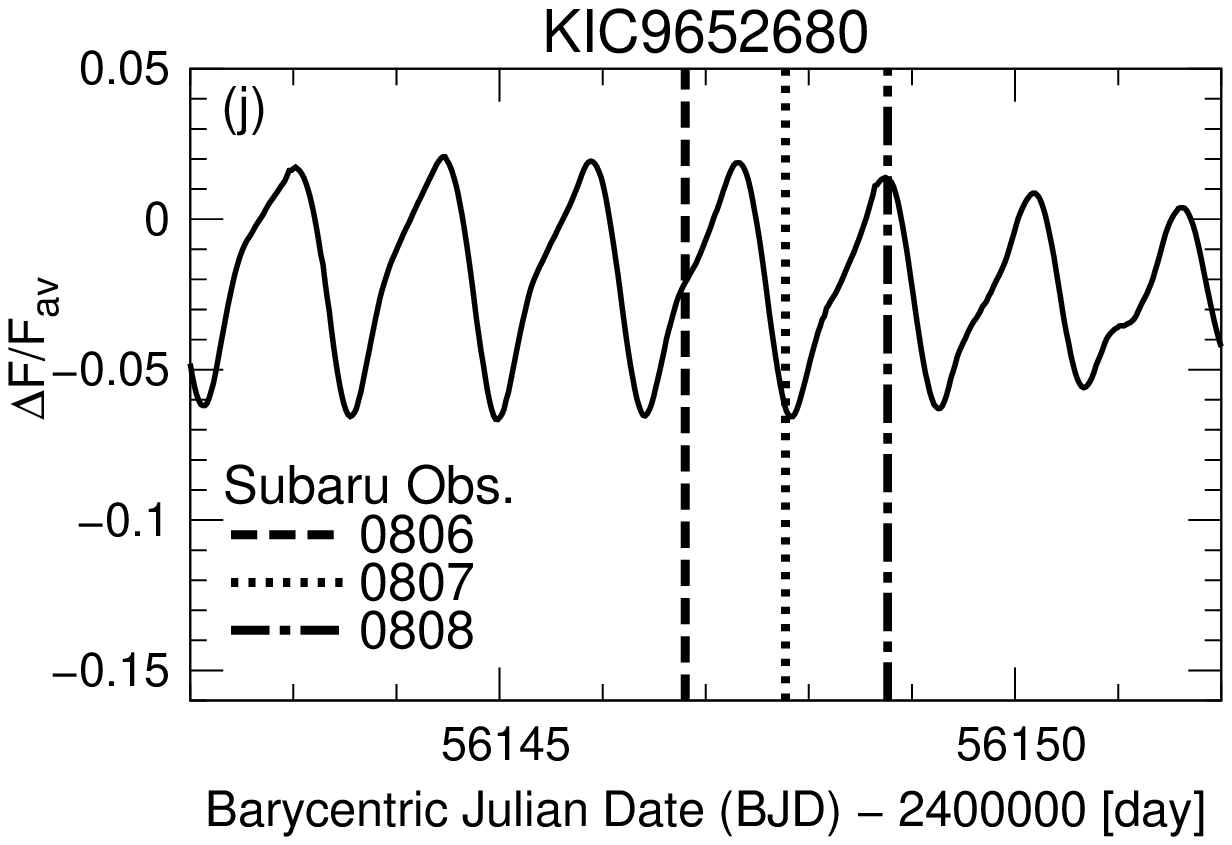}
 \FigureFile(65mm,65mm){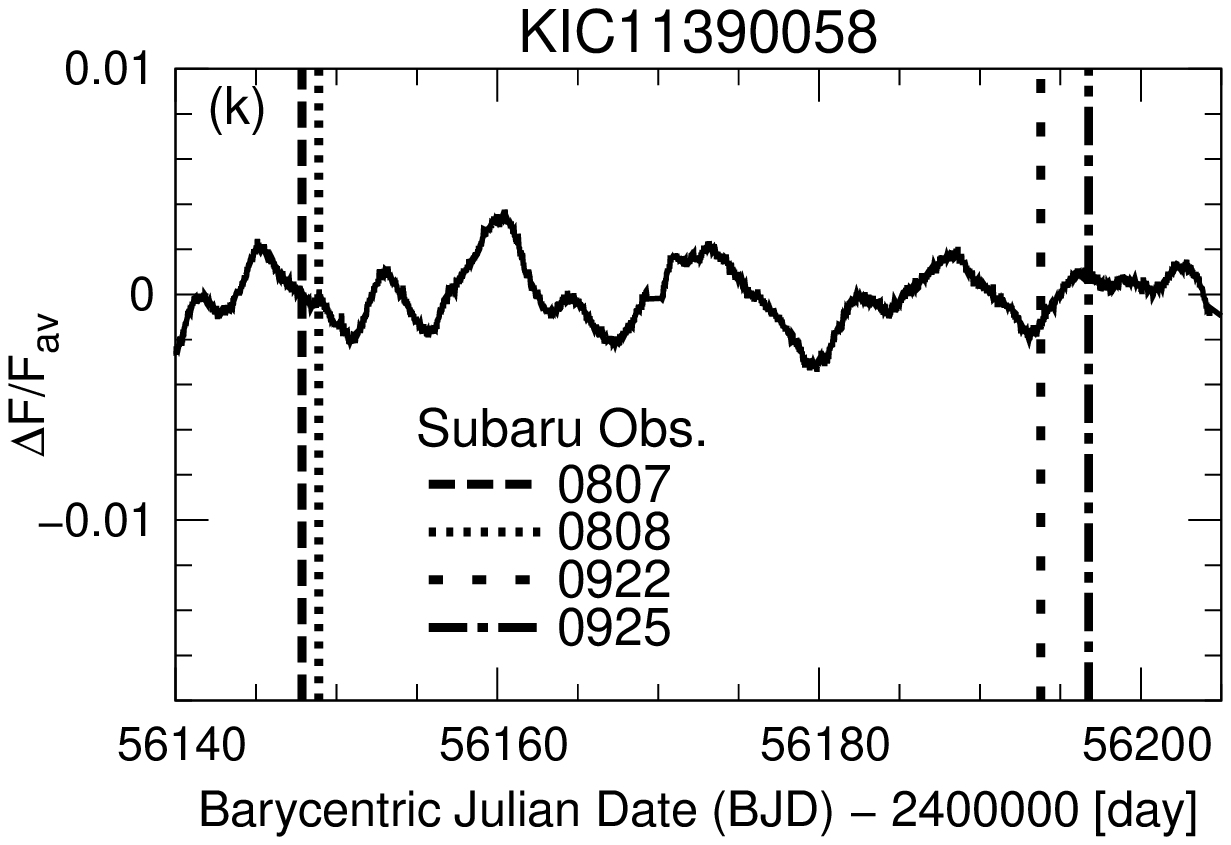}
 \FigureFile(65mm,65mm){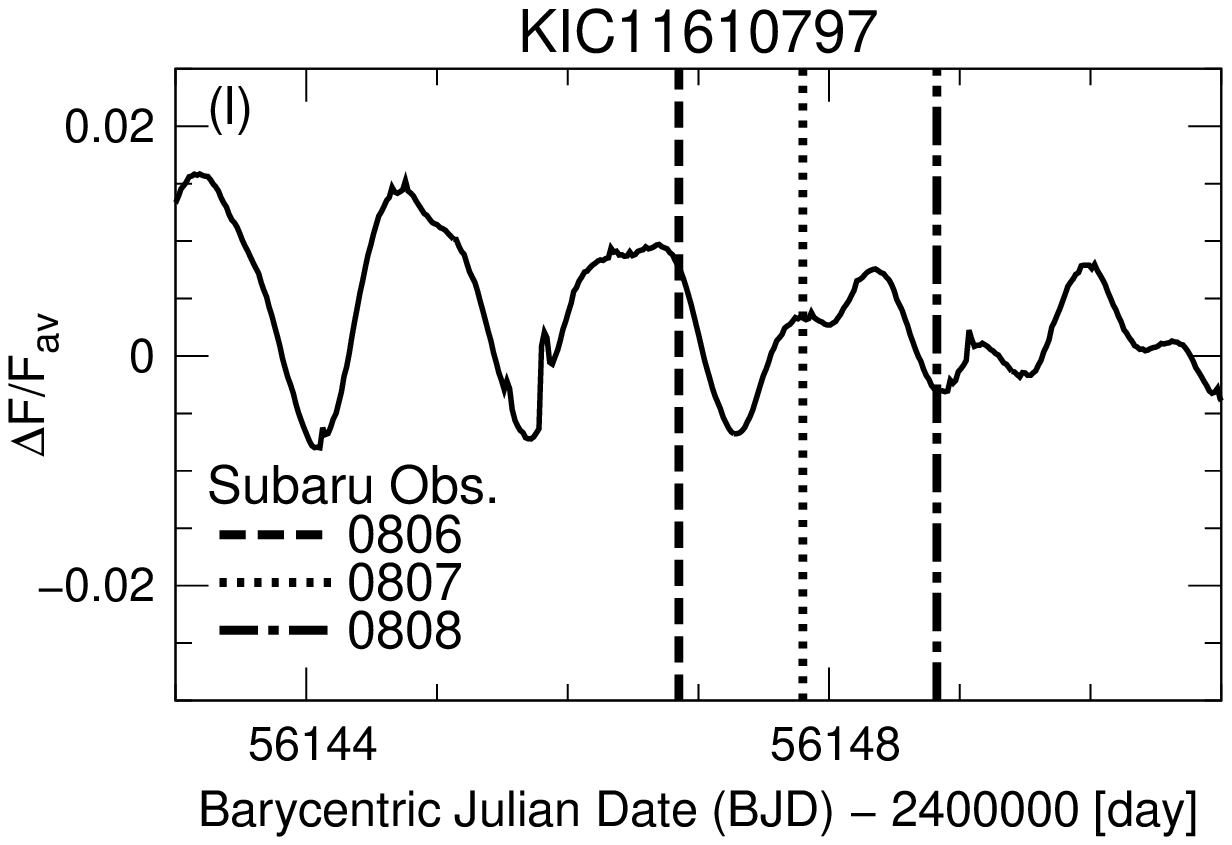}
 \FigureFile(65mm,65mm){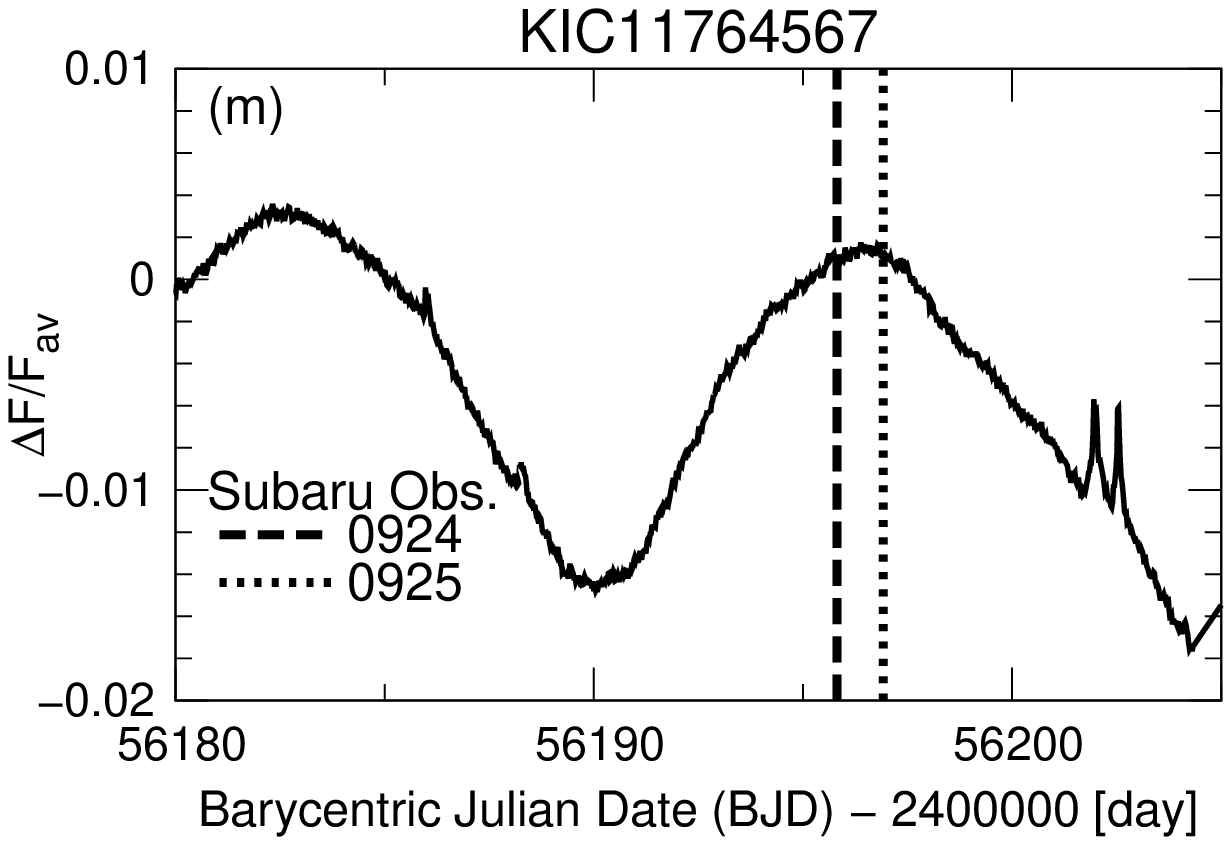}
 \FigureFile(65mm,65mm){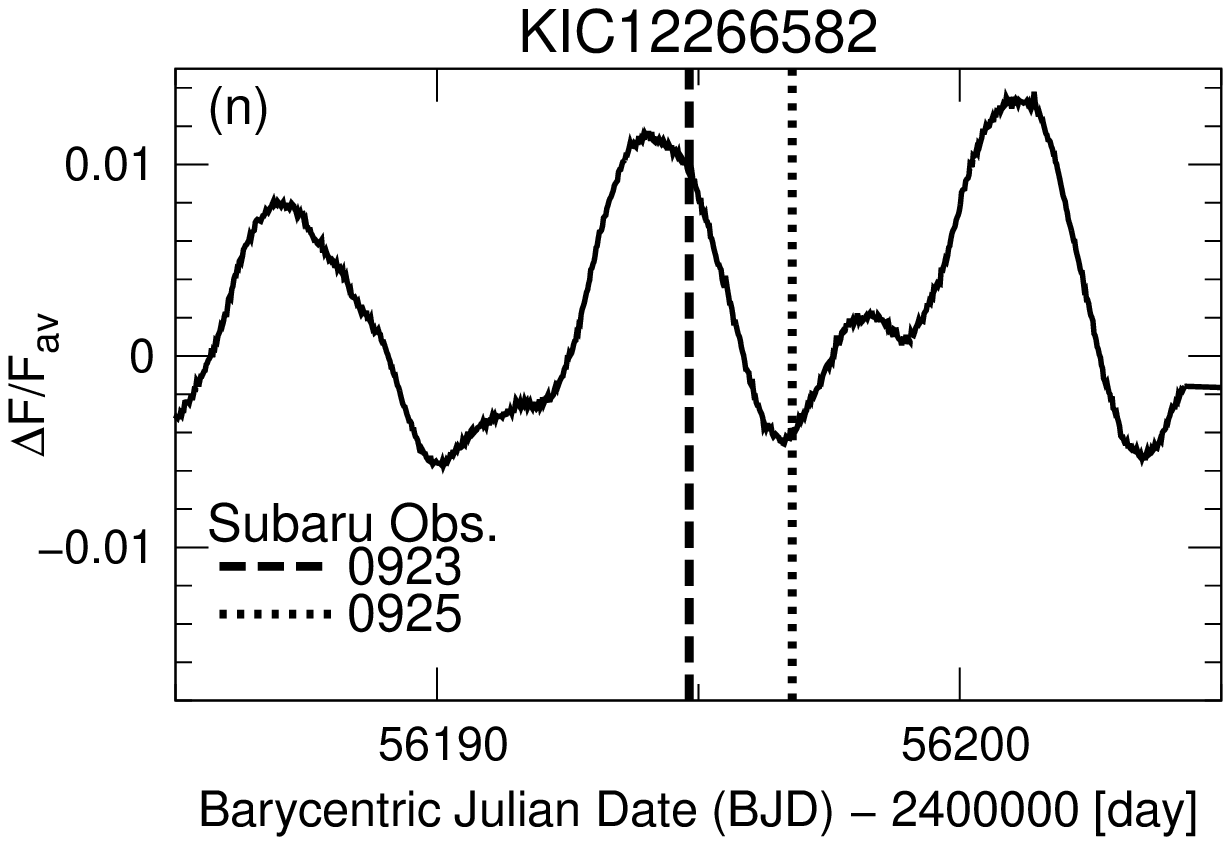}
  \FigureFile(65mm,65mm){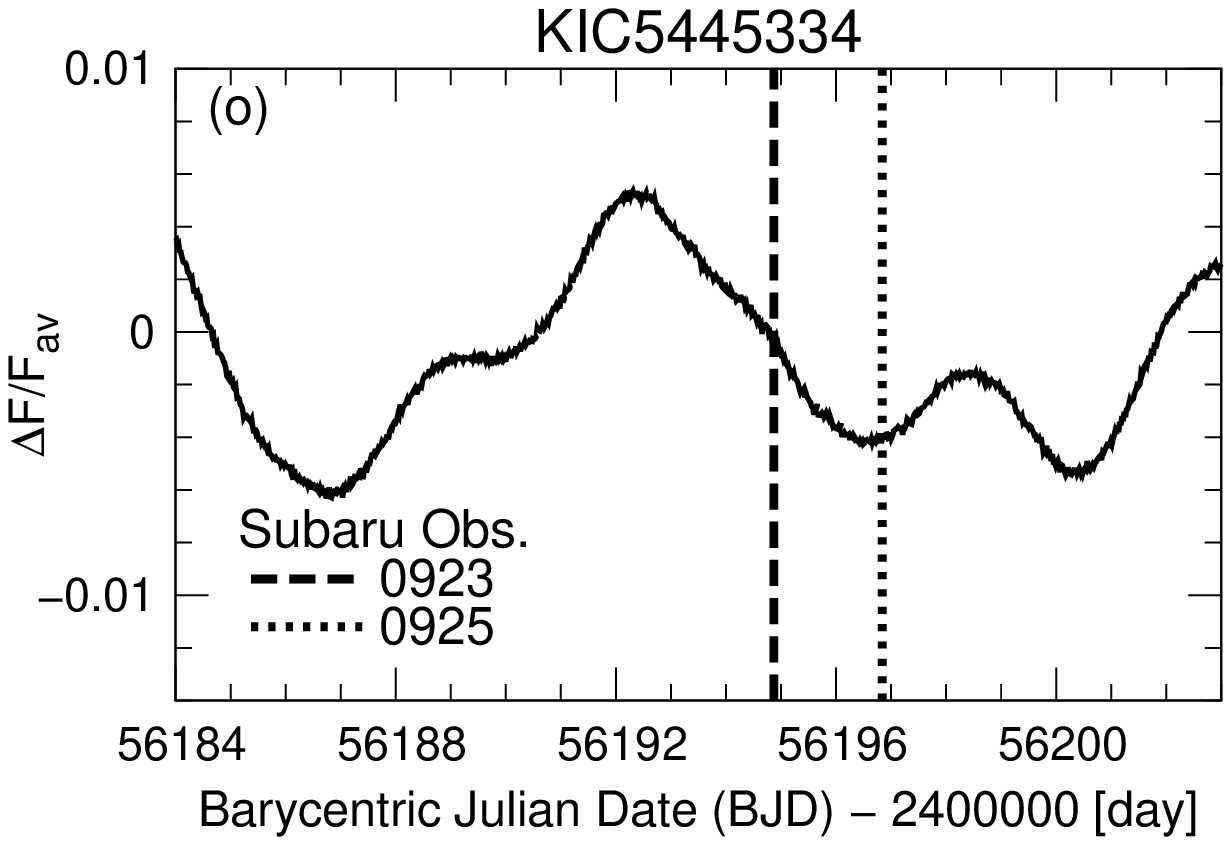} 
 \end{center}
\caption{Continued.}
\end{figure}

\addtocounter{figure}{-1}
\begin{figure}[htbp]
 \begin{center}
 \FigureFile(65mm,65mm){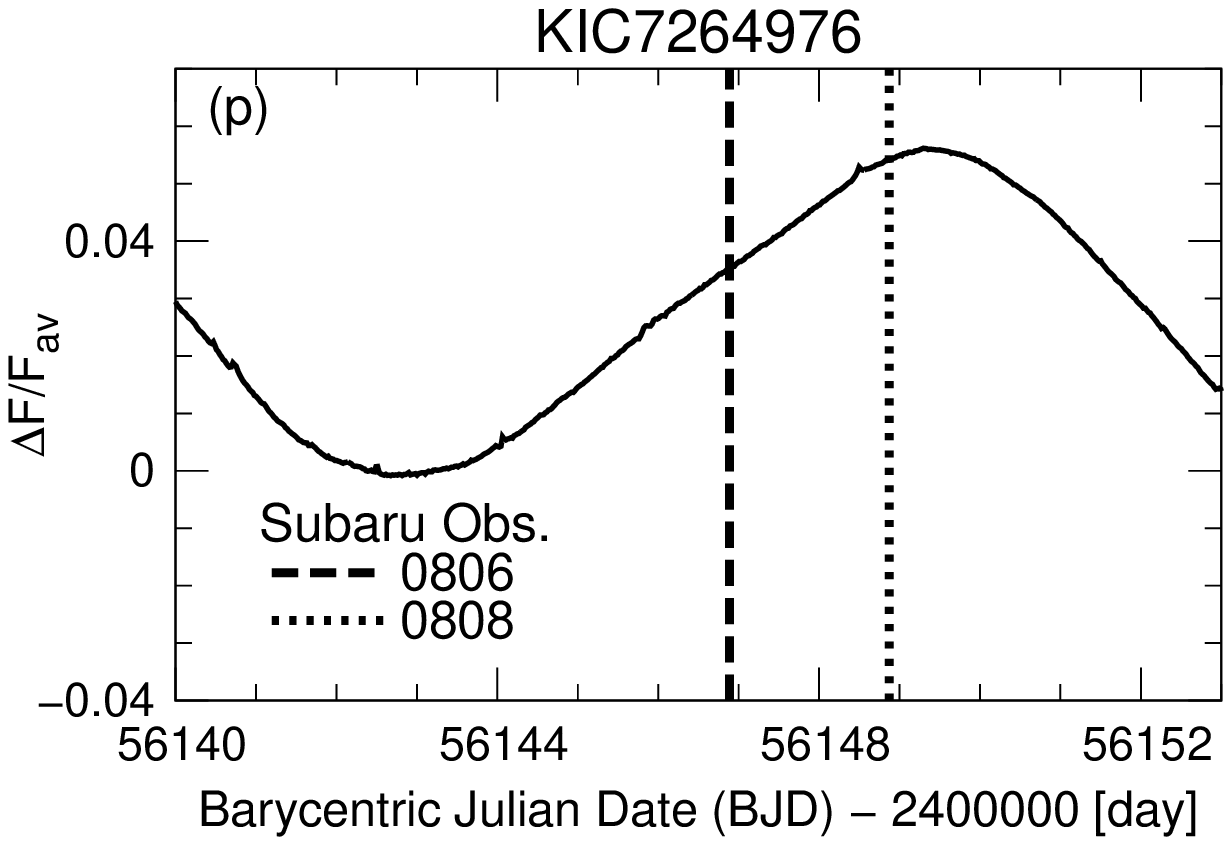}
 \FigureFile(65mm,65mm){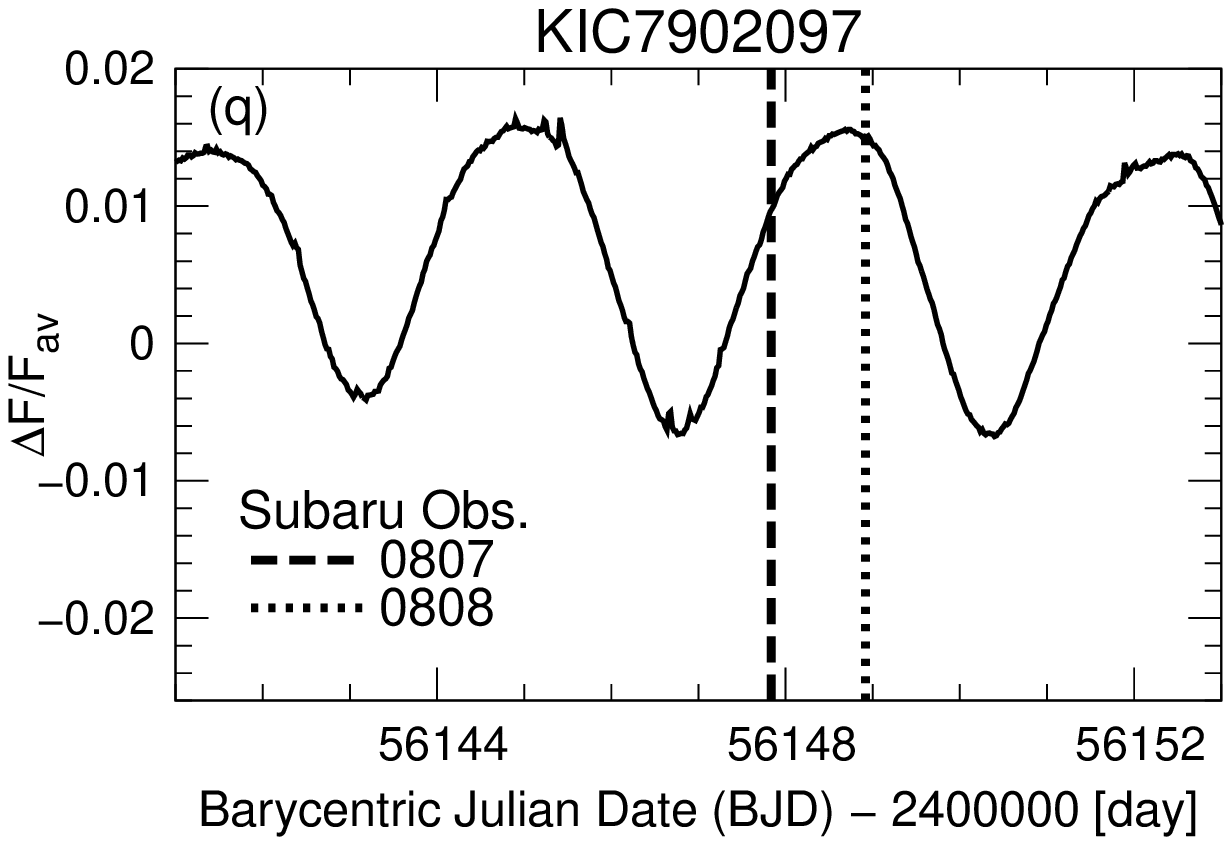} 
 \FigureFile(65mm,65mm){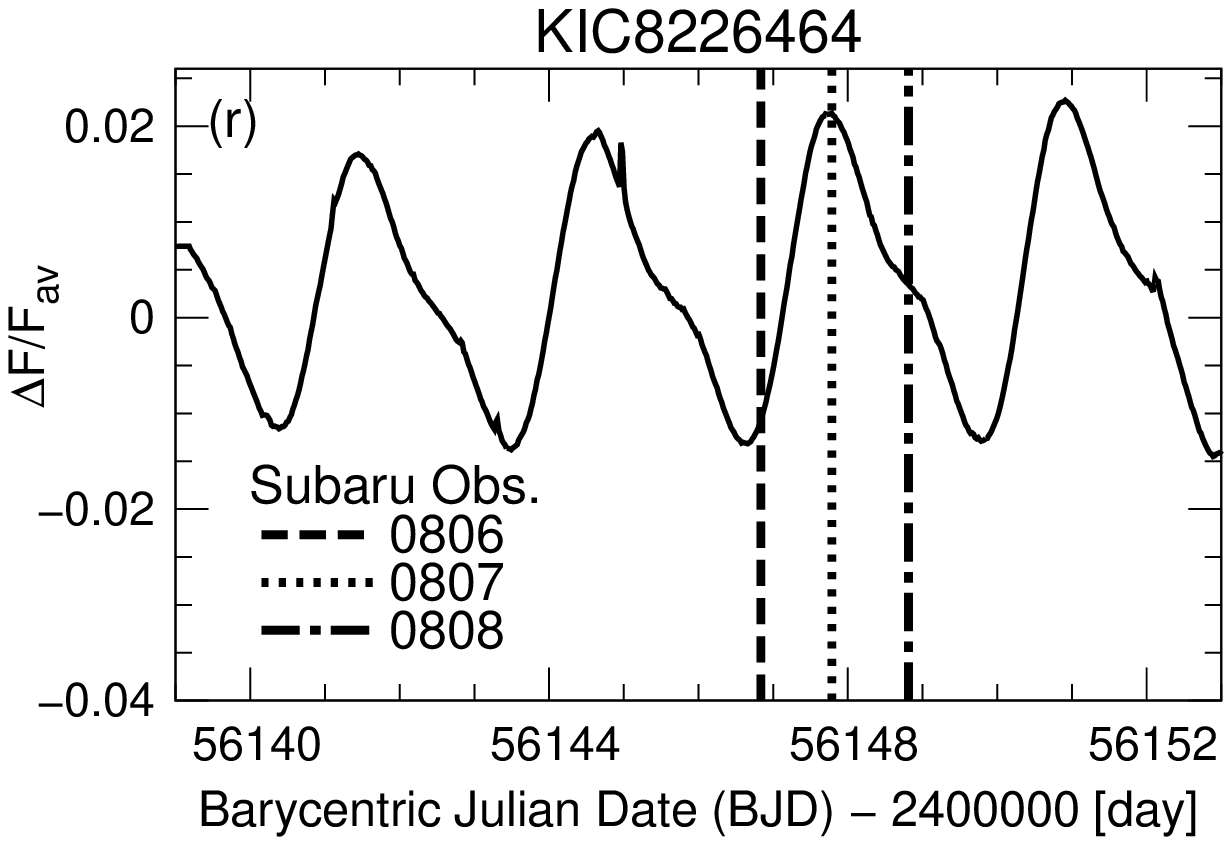}
 \FigureFile(65mm,65mm){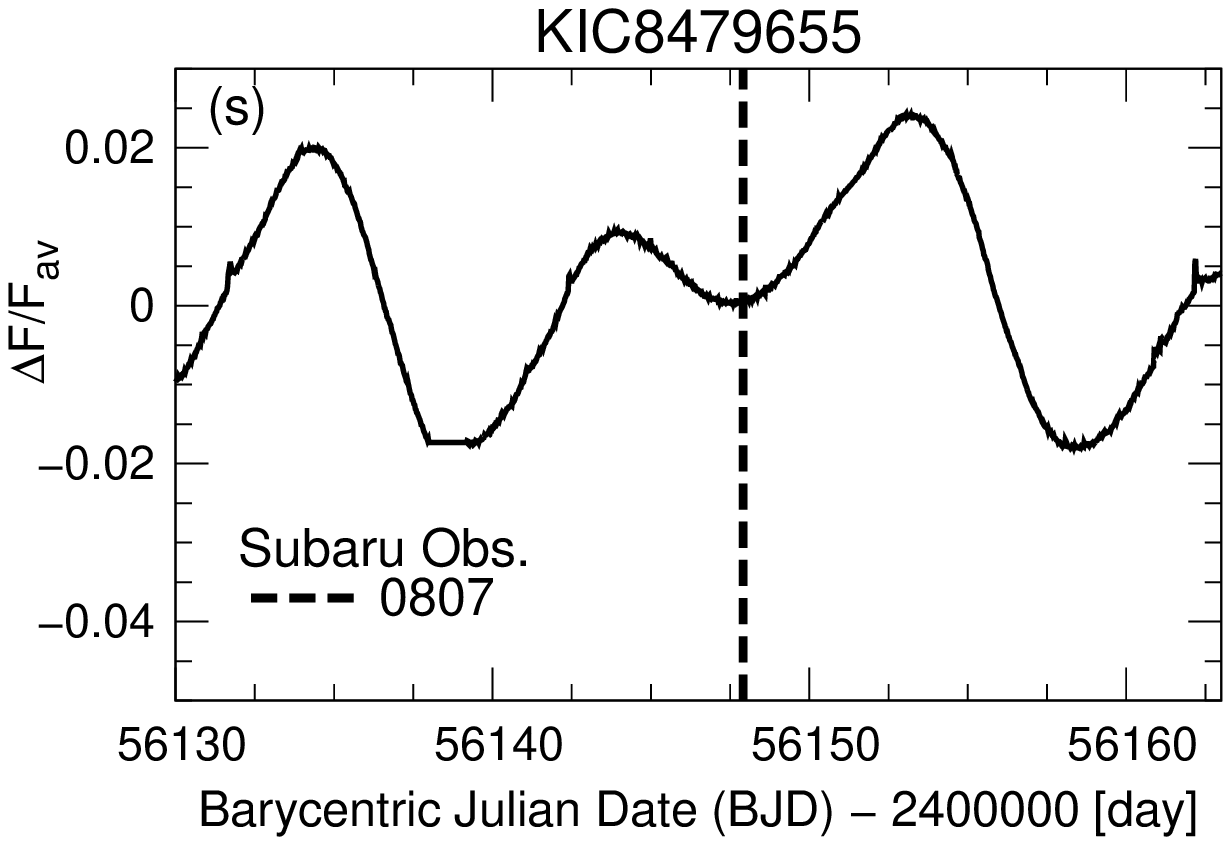}
 \FigureFile(65mm,65mm){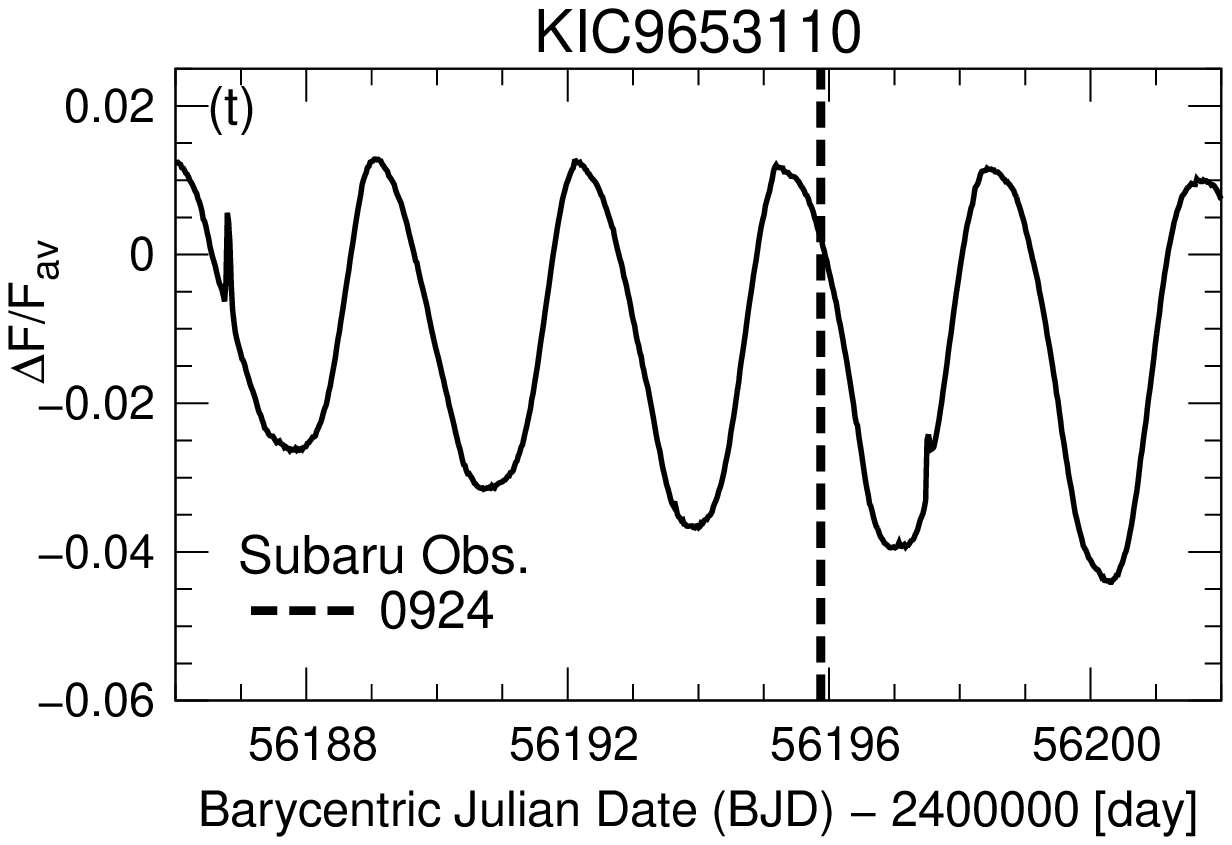}
 \FigureFile(65mm,65mm){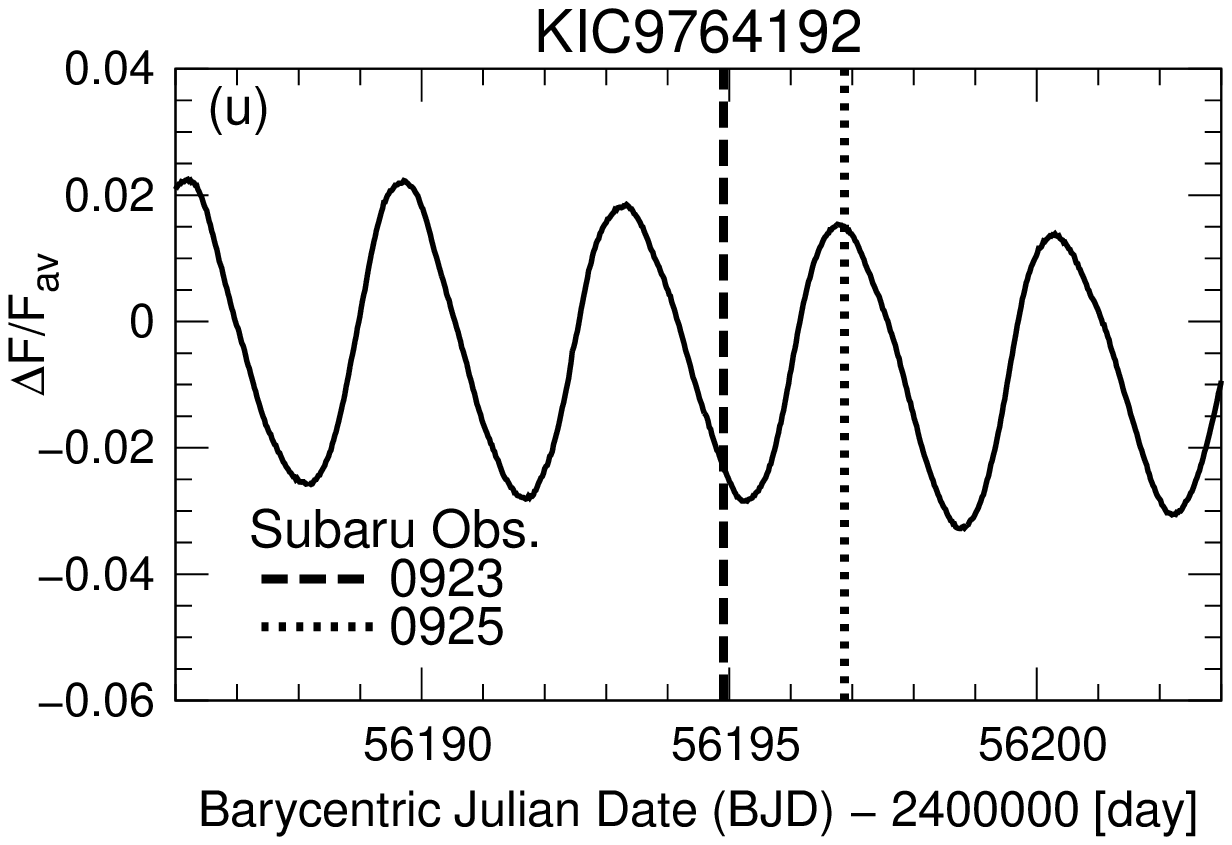} 
 \FigureFile(65mm,65mm){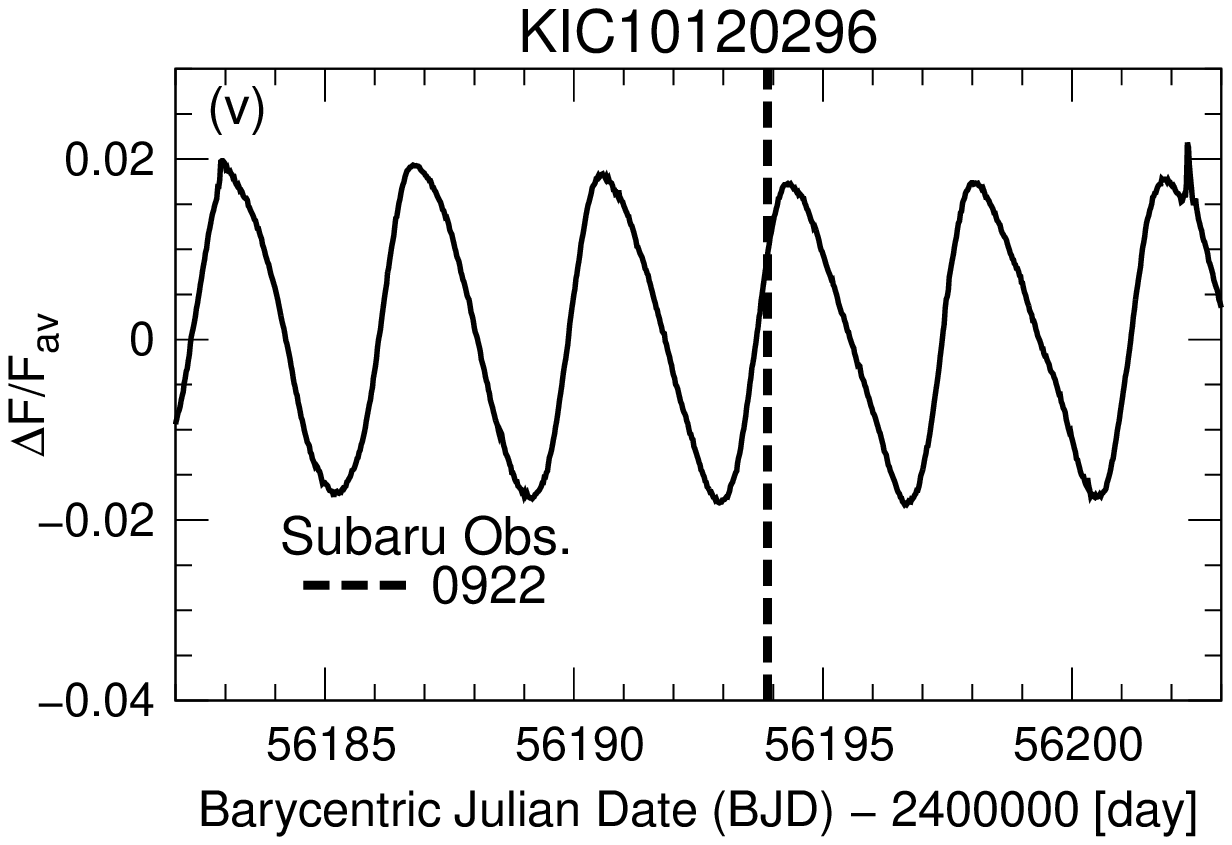}
 \FigureFile(65mm,65mm){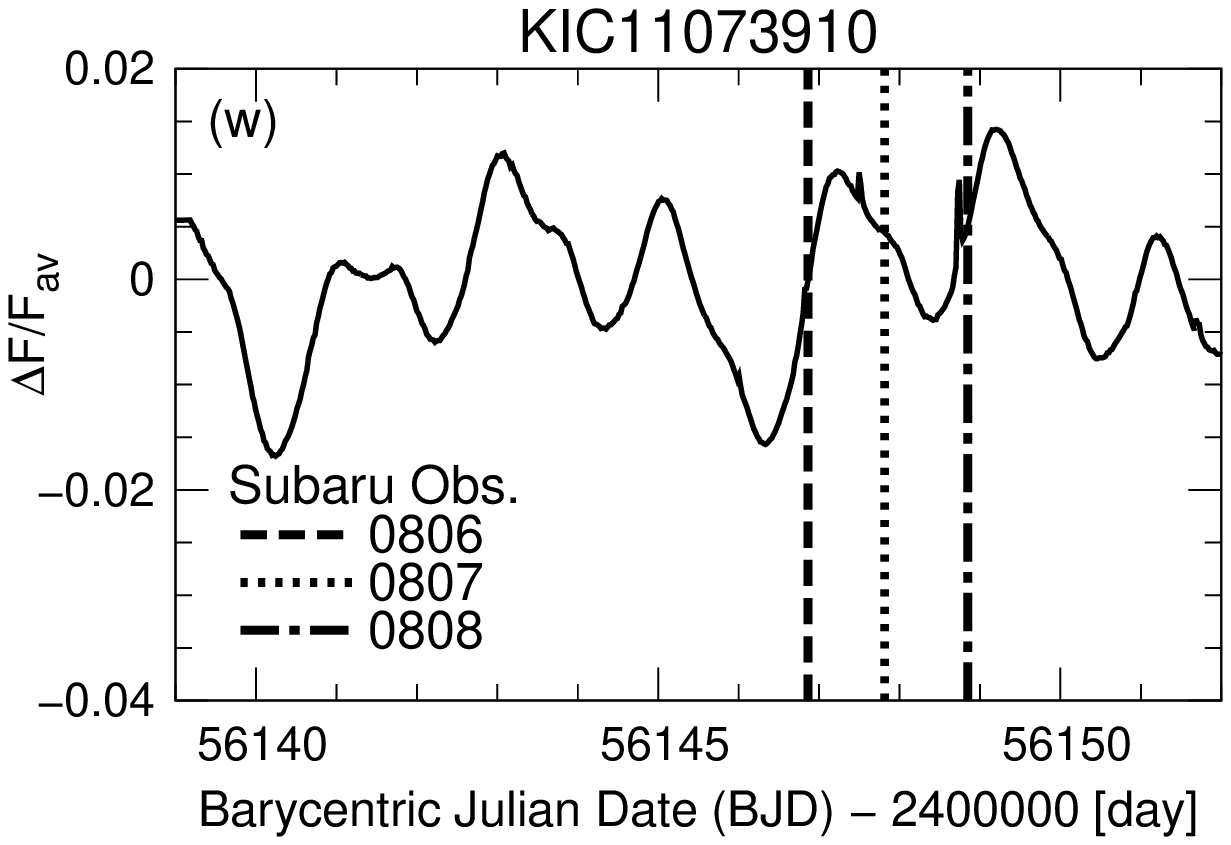}
 \end{center}
\caption{Continued.}
\end{figure}

\clearpage
\newpage

\section{Comparison stars}\label{sec:compConf-ape}
\begin{figure}[htbp]
 \begin{center}
  \FigureFile(80mm,80mm){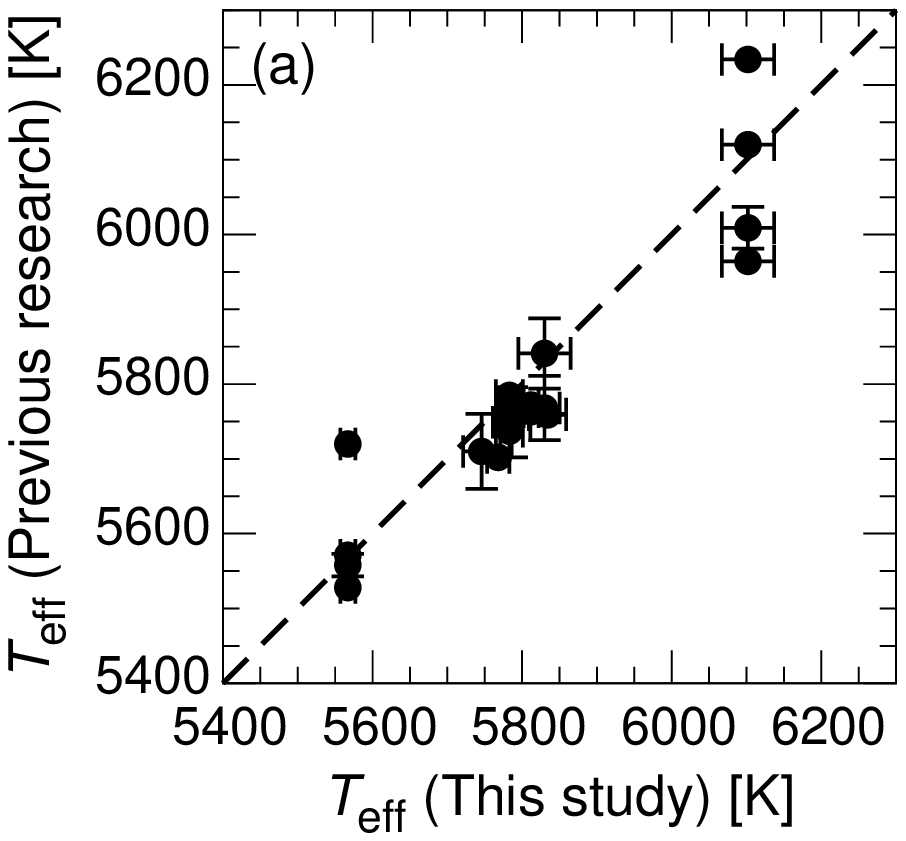}
  \FigureFile(80mm,80mm){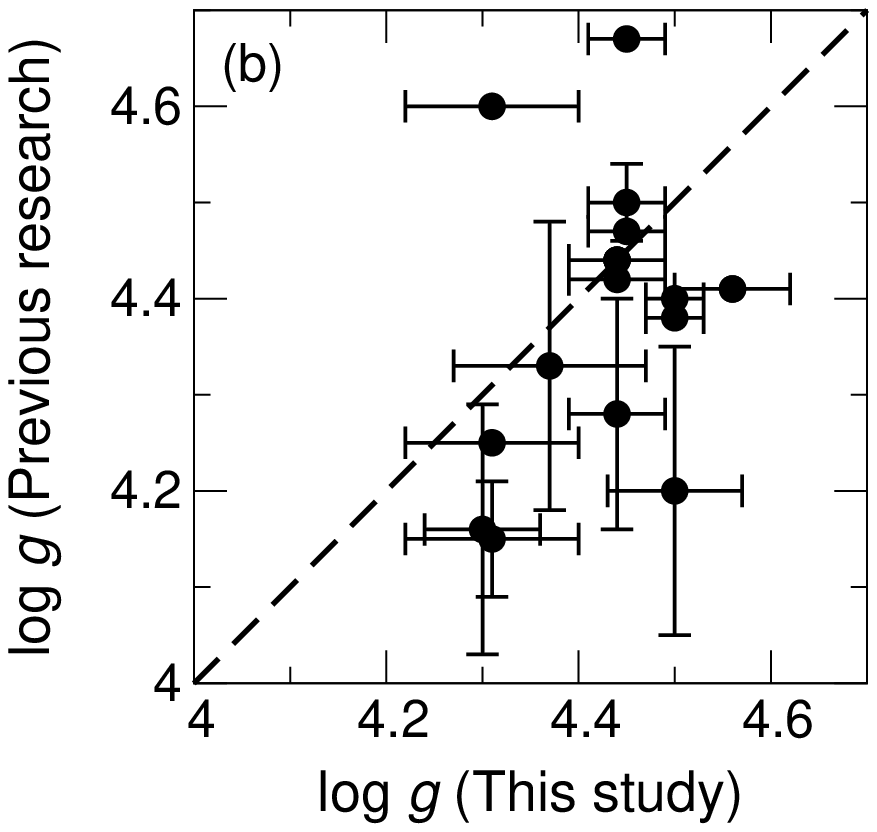}
  \FigureFile(80mm,80mm){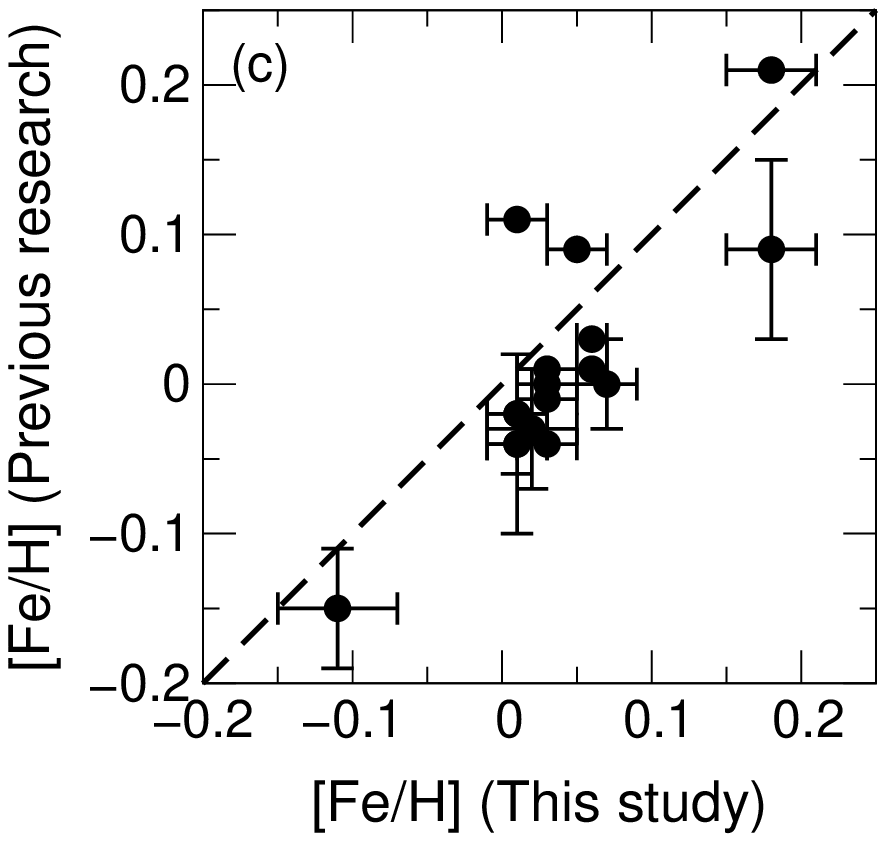}
  \FigureFile(80mm,80mm){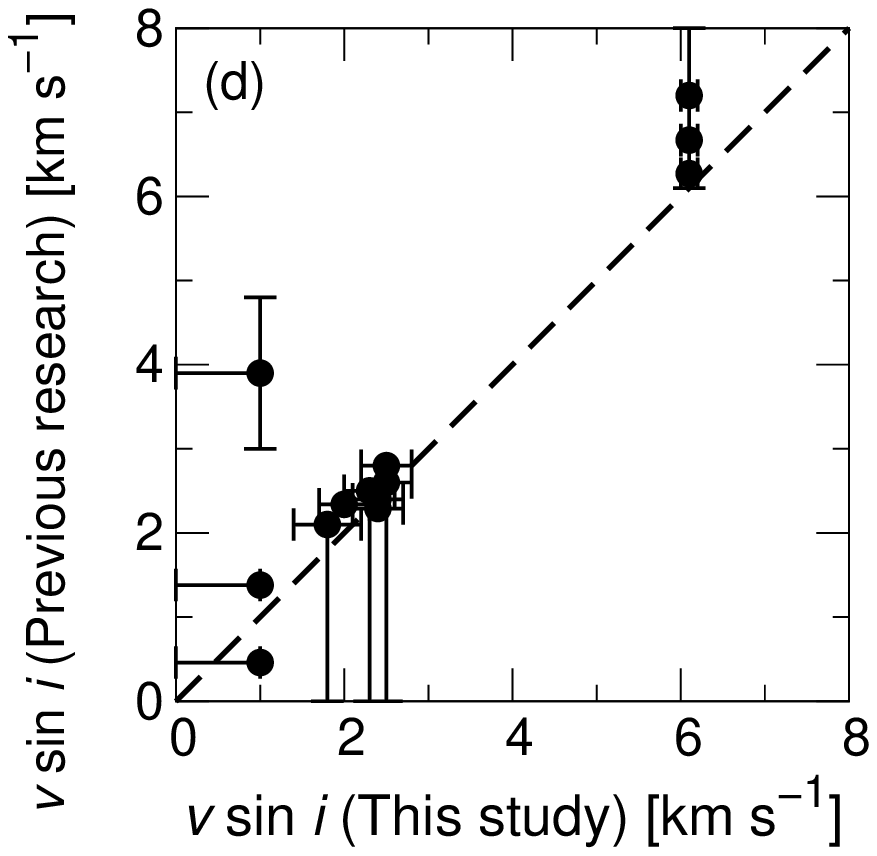}
 \end{center}
\caption{Temperature ($T_{\rm{eff}}$), surface gravity ($\log g$), metallicity ([Fe/H]) and projected rotational velocity ($v \sin i$) 
of comparison stars including Moon, and comparison with previous results listed in Table \ref{tab:compara}.}\label{fig:comppara}
\end{figure}

\begin{figure}[htbp]
 \begin{center}
  \FigureFile(70mm,70mm){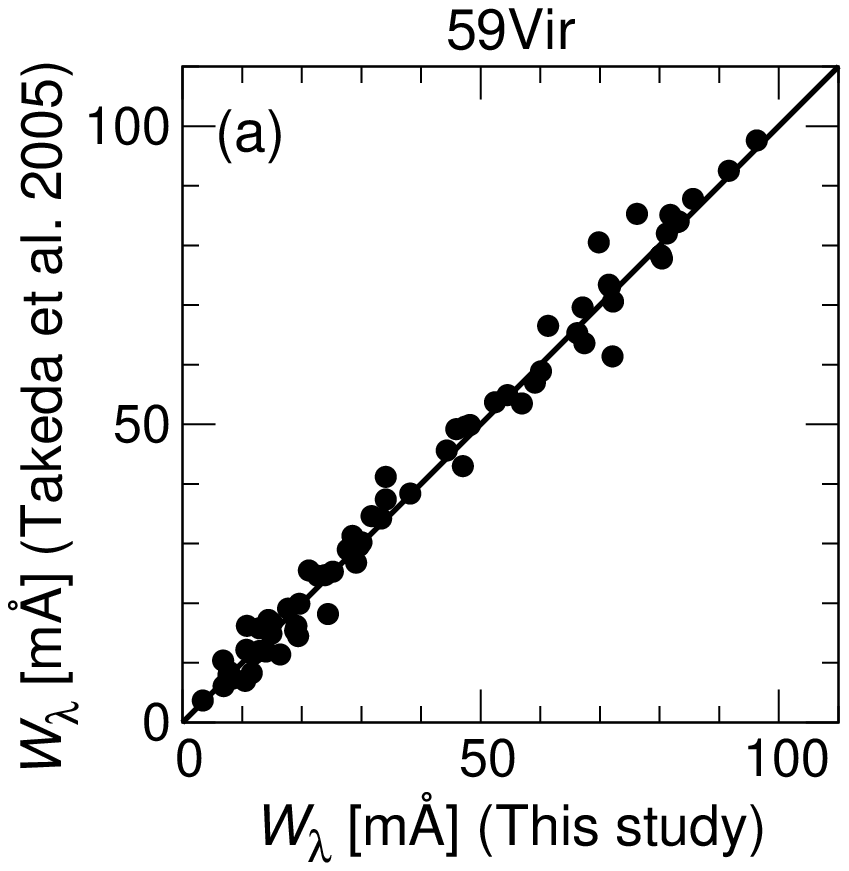}
  \FigureFile(70mm,70mm){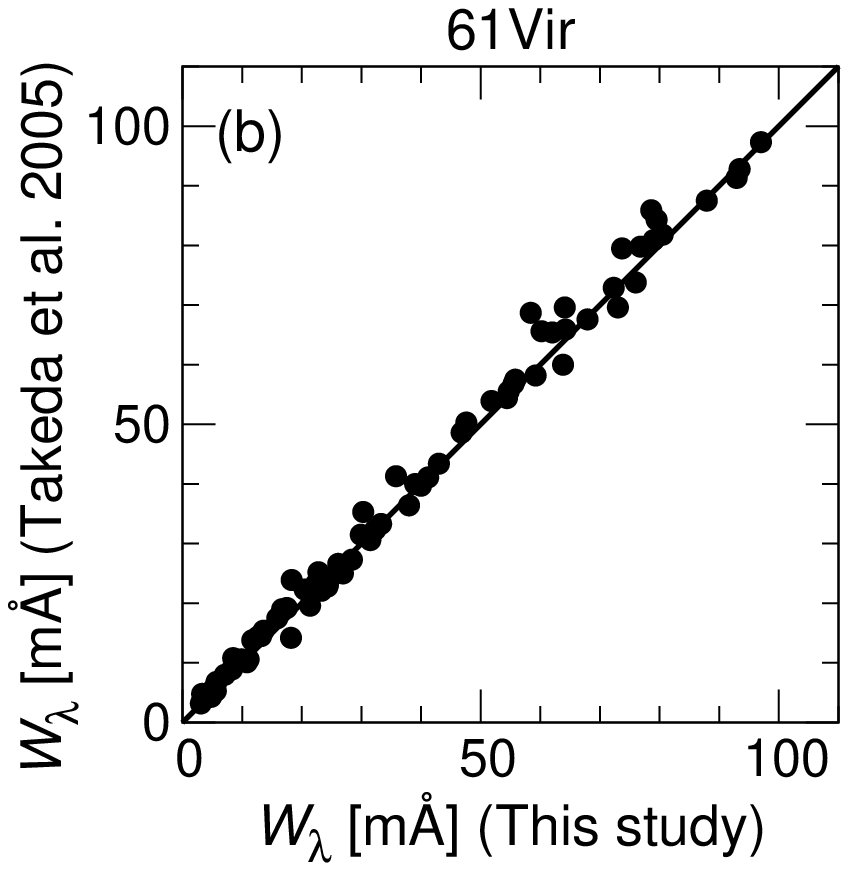}  
  \FigureFile(70mm,70mm){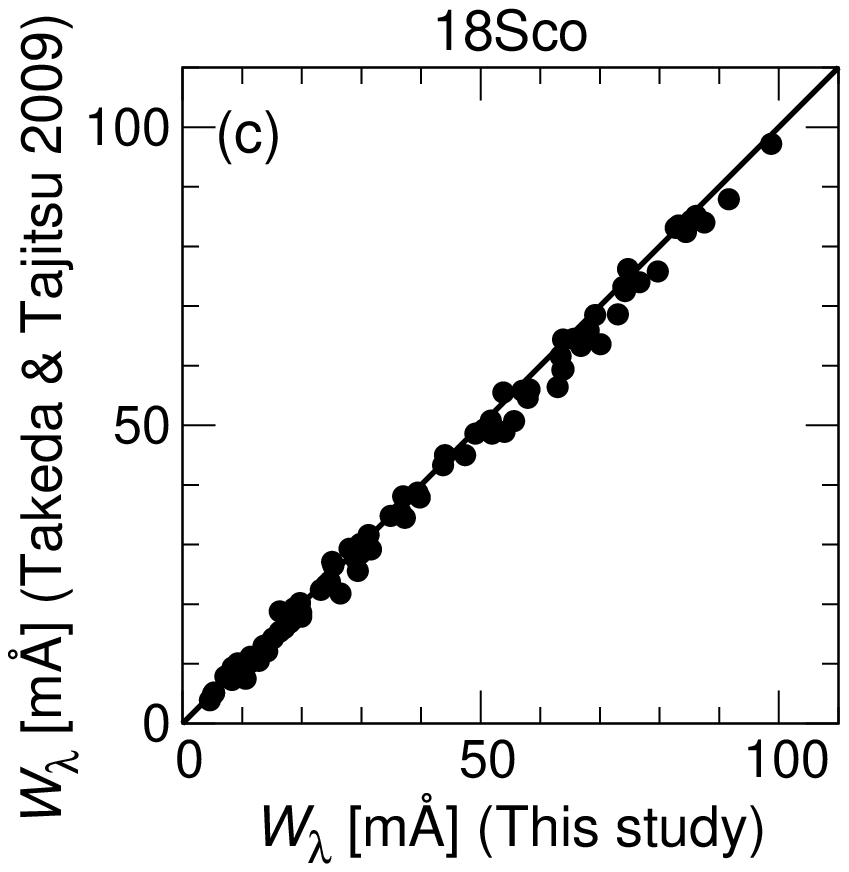}  
  \FigureFile(70mm,70mm){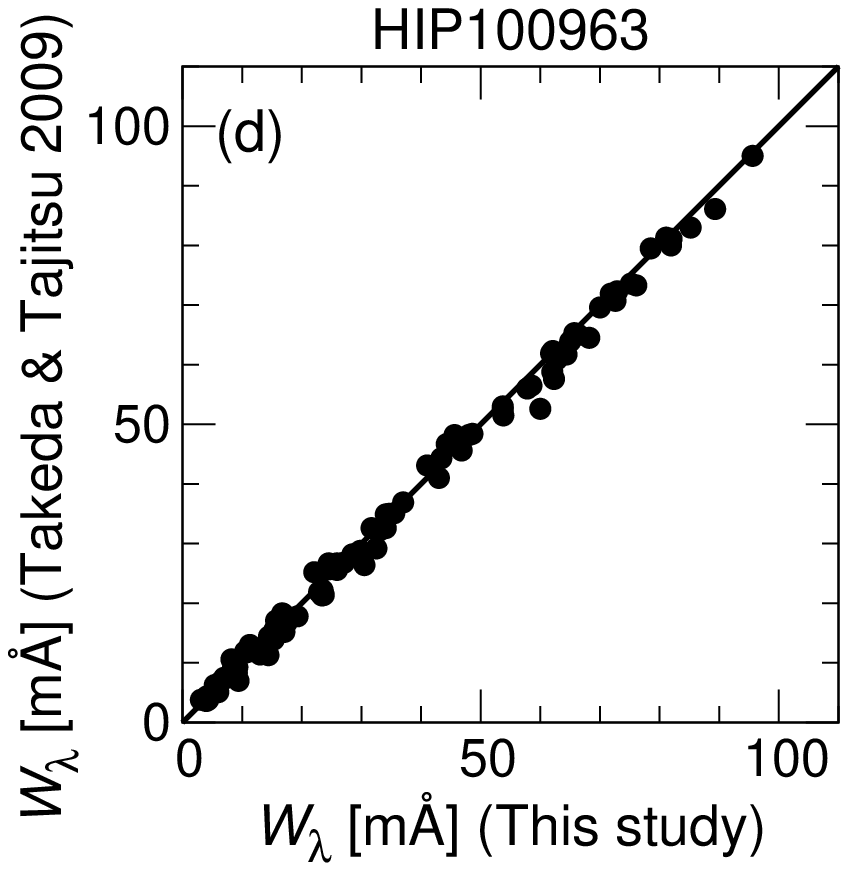}    
  \FigureFile(70mm,70mm){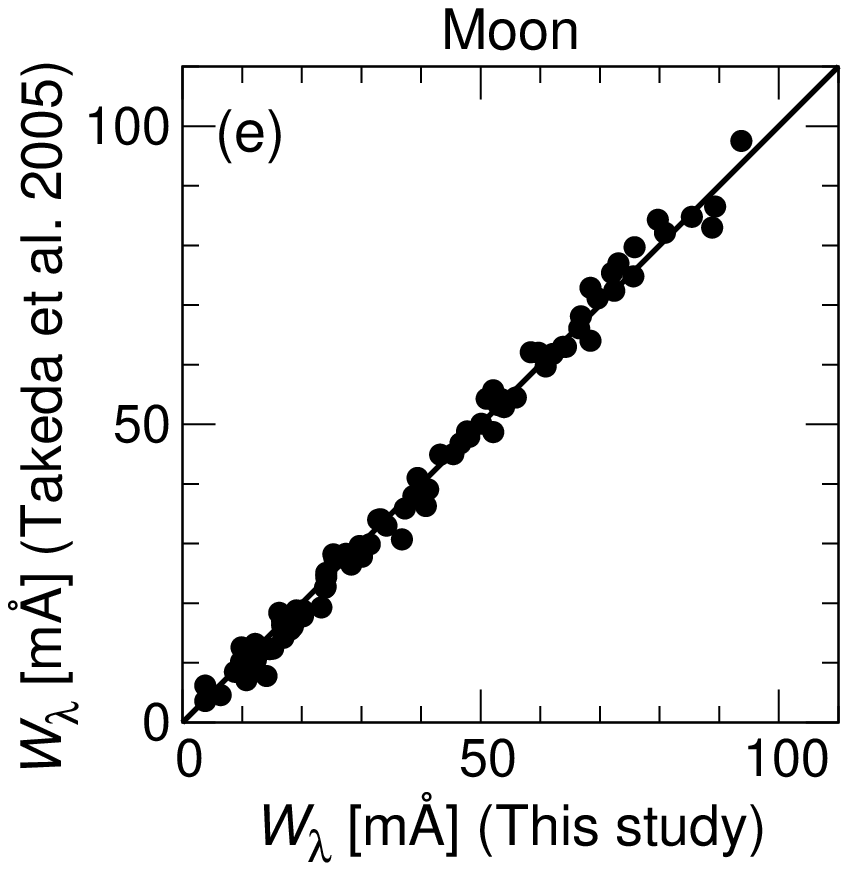}      
  \FigureFile(70mm,70mm){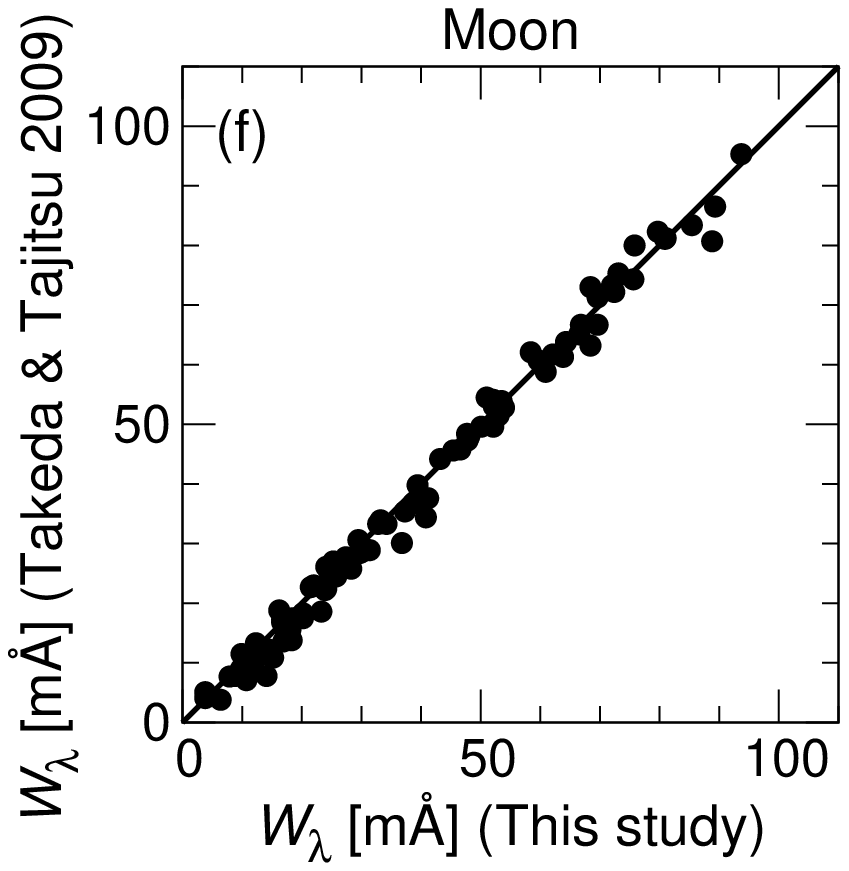}      
 \end{center}
\caption{Comparison of the EW values ($W_{\lambda}$) of Fe I and Fe II lines shown in \citet{Takeda2005} or \citet{TakedaTajitsu2009} with those measured by us.}\label{fig:ewcomp}
\end{figure}

We estimated temperature ($T_{\rm{eff}}$), surface gravity ($\log g$), metallicity ([Fe/H]), projected rotational velocity ($v \sin i$), and stellar radius ($R_{\rm{s}}$) 
of 34 single superflare stars, 10 comparison stars, and Moon (Sun) in Section \ref{subsec:atmos}$\sim$\ref{subsec:AgeR}.
The estimated values of 10 comparison stars and Moon (Sun) are listed in Table \ref{tab:compara}.
The values reported by the previous studies were also listed in Table \ref{tab:compara}.
In Figure \ref{fig:comppara}, we compared our $T_{\rm{eff}}$, $\log g$, [Fe/H], and $v \sin i$ values of comparison stars including Moon 
with those reported by the previous studies listed in Table \ref{tab:compara}.
Our values seem to be in good agreement with the previous values especially for $T_{\rm{eff}}$, [Fe/H], and $v \sin i$ values.
\\ \\
\ \ \ \ \ \ \
When estimating $v\sin i$ values, the choice of macroturbulence is often important, 
as we mentioned at the end of Section \ref{subsec:dis-pa}.
We briefly summarize which formula was used for evaluating the macroturbulence velocity 
in the other studies whose $v\sin i$ values are listed in Table \ref{tab:compara}.
\citet{SNotsu2013} (referred as (2) in Table \ref{tab:compara}) used the same formula 
as we do in this paper (Equation (\ref{eq:macroT})). 
\citet{Anderson2010} (referred as (4)) treated the macroturbulence velocity as one of the free parameters 
in the process of spectral line fitting. 
\citet{Ammler-von2012} (referred as (5)) did not assume the macroturbulence velocity 
since they conducted line profile analyses in Fourier space 
(The detailed explanations are in the 8th paragraph of Section 1 of \citet{Ammler-von2012}).
\citet{Takeda2010} (referred as (7)) assumed $v_{\rm{mt}}=1.5$km s$^{-1}$ since atmospheric parameters 
of most of their target stars are similar to the Sun.
\citet{King2005} (refereed as (9)) did not consider the effect of macroturblence velocity 
and they could only estimate the upper limit of $v\sin i$.
\\ \\
\ \ \ \ \ \ \
Among the previous studies referred in Table \ref{tab:compara}, 
\citet{Takeda2005} and \citet{TakedaTajitsu2009} used the basically same method for deriving $T_{\rm{eff}}$, $\log g$, and [Fe/H] of 59 Vir, 61 Vir, 18 Sco, HIP100963, and Moon
as we have done in Section \ref{subsec:atmos}.
In this method, equivalent width values of Fe I/II lines are used for deriving stellar parameters.
Then, we here compared the measured equivalent width values ($W_{\lambda}$) of Fe I/II lines in this paper 
with those in \citet{Takeda2005} and \citet{TakedaTajitsu2009}.
The results are shown in Figure \ref{fig:ewcomp}.
We can see that these two values are comparable, 
and that our measurement of equivalent width values are consistent with that of the previous studies.
\\ \\
\ \ \ \ \ \ \
Summarizing this section, we confirmed that our resultant atmospheric parameters and measured equivalent width values of comparison stars 
are comparable to the results of previous researches. 
We can say that our result of spectroscopic determination of atmospheric parameters is consistent with that of the previous studies.

\begin{center}
\begin{longtable}{lcccccccc}
\caption{Stellar parameters of comparison stars.}\label{tab:compara}
\hline
Starname &$ v \sin i$ & $T_{\rm{eff}}$ & $ \log g$ & $v_{t}$ &[Fe/H] & $M_{\rm{s}}$~\footnotemark[a] & $R_{\rm{s}}$~\footnotemark[b] & Ref.~\footnotemark[c]  \\ 
 & [km s$^{-1}$] & [K] & [cm s$^{-2}$]& [km s$^{-1}$] & & [$M_{\odot}$] & [$R_{\odot}$]  & \\ 
\hline
\endhead
\hline
\endfoot
\hline 
\multicolumn{8}{l}{\hbox to 0pt{\parbox{170mm}{\footnotesize
    \footnotemark[a] The resultant stellar mass ($M_{\rm{s}}$) value in this column 
is a median between the maximum and minimum values among all the possible $M_{\rm{s}}$ values selected from the isochrone data.
The error value in this column corresponds to these maximum and minimum values. }}}
\\
\multicolumn{8}{l}{\hbox to 0pt{\parbox{170mm}{\footnotesize
    \footnotemark[b] The resultant stellar radius ($R_{\rm{s}}$) value in this column 
is a median between the maximum and minimum values among all the possible $R_{\rm{s}}$ values selected from the isochrone data.
The error value in this column corresponds to these maximum and minimum values.}}}
\\
\multicolumn{1}{l}{\hbox to 0pt{\parbox{160mm}{\footnotesize
 \footnotemark[c] (1) Present work. (2)\citet{SNotsu2013}; (3)\citet{Takeda2005}; (4)\citet{Anderson2010}; }}}
\\
\multicolumn{1}{l}{\hbox to 0pt{\parbox{160mm}{\footnotesize
(5)\citet{Ammler-von2012}; (6)\citet{TakedaTajitsu2009}; (7)\citet{Takeda2010};}}}
\\
\multicolumn{1}{l}{\hbox to 0pt{\parbox{160mm}{\footnotesize
 (8)\citet{Datson2012}; (9)\citet{King2005}.}}}
\endlastfoot
59Vir & 6.1$\pm$ 0.1 & 6102$\pm$ 35 & 4.31$\pm$ 0.09 & 1.33$\pm$ 0.14 & 0.18$\pm$ 0.03 & 1.18$\pm$ 0.03 & 1.27$\pm$ 0.14 & (1)  \\ 
 & 6.27 & 6009$\pm$ 28 & 4.15$\pm$ 0.06 & 1.32$\pm$ 0.09 & 0.09$\pm$ 0.06 &  &  & (2)  \\ 
 &  & 6120 & 4.25 &  & 0.21 & 1.22 &  & (3)  \\ 
 & 6.67 & 6234 & 4.60 &  &  &  &  & (4)  \\ 
 & 7.2$\pm$ 1.1  & 5964 &  &  &  &  &  & (5)  \\ 
61Vir & $<$1.0$\pm$  & 5567$\pm$ 10 & 4.45$\pm$ 0.04 & 0.87$\pm$ 0.10 & 0.01$\pm$ 0.02 & 0.90$\pm$ 0.02 & 0.93$\pm$ 0.03 & (1)  \\ 
 & 1.38 & 5558$\pm$ 15 & 4.50$\pm$ 0.04 & 0.87$\pm$ 0.08 & -0.04$\pm$ 0.06 &  &  & (2)  \\ 
 &  & 5720 & 4.67 &  & 0.11 & 1.02 &  & (3)  \\ 
 & 0.46 & 5571 & 4.47 &  &  &  &  & (4)  \\ 
 & 3.9$\pm$ 0.9  & 5528 &  &  &  &  &  & (5)  \\ 
18Sco & 2.0$\pm$ 0.3 & 5812$\pm$ 10 & 4.50$\pm$ 0.03 & 1.04$\pm$ 0.08 & 0.06$\pm$ 0.01 & 1.03$\pm$ 0.01 & 0.96$\pm$ 0.02 & (1)  \\ 
(HIP79672) &  & 5772 & 4.40 & 0.97 & 0.01 &  &  & (6)  \\ 
 & 2.34 & 5763 & 4.38 & 0.97 & 0.03 &  &  & (7)  \\ 
HD163441 & 2.5$\pm$ 0.3 & 5768$\pm$ 15 & 4.45$\pm$ 0.04 & 0.87$\pm$ 0.11 & 0.05$\pm$ 0.02 & 0.99$\pm$ 0.02 & 0.98$\pm$ 0.04 & (1)  \\ 
 &  & 5702 &  &  & 0.09 &  &  & (8)  \\ 
HD173071 & 2.7$\pm$ 0.2 & 5982$\pm$ 25 & 4.41$\pm$ 0.07 & 0.98$\pm$ 0.14 & 0.18$\pm$ 0.03 & 1.13$\pm$ 0.03 & 1.11$\pm$ 0.07 & (1)  \\ 
 &  & 5875 &  &  & -0.04 &  &  & (8)  \\ 
HIP100963 & 2.4$\pm$ 0.3 & 5834$\pm$ 25 & 4.56$\pm$ 0.06 & 1.07$\pm$ 0.09 & 0.01$\pm$ 0.02 & 1.03$\pm$ 0.02 & 0.93$\pm$ 0.01 & (1)  \\ 
 &  & 5760 & 4.41 & 0.93 & -0.04 &  &  & (6)  \\ 
 & 2.4 & 5759 & 4.41 & 0.98 & -0.02 &  &  & (7)  \\ 
HIP71813 & 2.3$\pm$ 0.3 & 5786$\pm$ 25 & 4.30$\pm$ 0.06 & 1.07$\pm$ 0.09 & 0.01$\pm$ 0.02 & 0.96$\pm$ 0.02 & 1.15$\pm$ 0.08 & (1)  \\ 
 & $<$2.5 & 5749 & 4.16 & 1.22 & -0.02 &  &  & (9)  \\ 
HIP76114 & 1.8$\pm$ 0.4 & 5746$\pm$ 25 & 4.50$\pm$ 0.07 & 0.98$\pm$ 0.11 & 0.02$\pm$ 0.03 & 0.98$\pm$ 0.03 & 0.94$\pm$ 0.05 & (1)  \\ 
 & $<$2.1 & 5710 & 4.20 & 1.35 & -0.03 &  &  & (9)  \\ 
HIP77718 & 2.5$\pm$ 0.3 & 5830$\pm$ 35 & 4.37$\pm$ 0.10 & 0.97$\pm$ 0.17 & -0.11$\pm$ 0.04 & 0.93$\pm$ 0.03 & 1.04$\pm$ 0.11 & (1)  \\ 
 & $<$2.8 & 5841 & 4.33 & 1.18 & -0.15 &  &  & (9)  \\ 
HIP78399 & 2.5$\pm$ 0.3 & 5830$\pm$ 20 & 4.44$\pm$ 0.05 & 1.04$\pm$ 0.09 & 0.07$\pm$ 0.02 & 1.02$\pm$ 0.03 & 1.02$\pm$ 0.05 & (1)  \\ 
 & $<$2.6 & 5768 & 4.28 & 1.32 & 0.00 &  &  & (9)  \\ 
Moon & 2.4$\pm$ 0.3 & 5783$\pm$ 18 & 4.44$\pm$ 0.05 & 0.85$\pm$ 0.13 & 0.03$\pm$ 0.02 & 0.99$\pm$ 0.02 & 1.00$\pm$ 0.05 & (1)  \\ 
(Sun) &  & 5785 & 4.44 & 0.96 & 0.01 & 1.00 &  & (3)  \\ 
 &  & 5737 & 4.42 & 0.95 & -0.04 &  &  & (6)  \\ 
 & 2.29 & 5761 & 4.44 & 1.00 & -0.01 &  &  & (7)  \\ 
 & $<$2.5 & 5777 & 4.44 & 1.25 & 0.00 &  &  & (9)  \\ 
\hline
\end{longtable}
\end{center}

\end{document}